\documentclass[a4paper,12pt]{report}
\usepackage{amsfonts}
\usepackage{amsmath}
\usepackage{amssymb}
\usepackage{graphicx}
\usepackage[utf8]{inputenc}
\usepackage[english]{babel}
\usepackage{ragged2e}
\usepackage{indentfirst}
\usepackage{lipsum}
\DeclareRobustCommand{\rchi}{{\mathpalette\irchi\relax}}
\newcommand{\irchi}[2]{\raisebox{\depth}{$#1\chi$}}
\usepackage{fullpage}
\usepackage{lscape}
\usepackage{slashbox}
\usepackage{wrapfig}
\usepackage{makeidx}
\usepackage{url}
\usepackage[margin=0.8in]{geometry}
\usepackage{pdfpages}
\usepackage{xcolor}
\usepackage[]{hyperref}
\hypersetup{
	colorlinks=true,
	linkcolor={red!90!blue},
	citecolor={cyan!80!black},
	filecolor={red!60!green},      
	urlcolor={magenta!60!black},
	pdfborder = {1 1 1}
}

\newcommand*{\bfrac}[2]{\genfrac{\lbrace}{\rbrace}{0pt}{}{#1}{#2}}

\begin{document}

\begin{titlepage}
	\begin{center}
		{\Huge \textbf{\textsc{The Einstein-Cartan-Dirac (ECD) theory}}}\\[1.1cm]
		\includegraphics[width=0.35\textwidth]{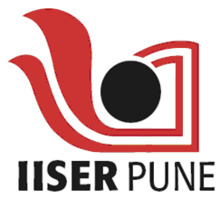}\\[1.0cm]    
		\large A thesis submitted towards partial fulfilment of \\BS-MS Dual Degree Programme\\[1.0cm]
		by\\[1.0cm]
		\Large\textbf{\textsc{Swanand Milind Khanapurkar\\
				(swanand.khanapurkar@students.iiserpune.ac.in)}}\\[1cm]
		\large under the guidance of \\[1cm]
		\Large{\textbf{\textsc{Prof. Tejinder P. Singh }}}\\[0.5cm]
		\large \textsc{Department of Astronomy and Astrophysics (DAA)\\
			Tata Institute of Fundamental Research (TIFR)\\
			Homi Bhabha Road, Mumbai 400005 India}\\[2.5cm]
		%
		\Large \textsc{Indian Institute of Science Education and Research (IISER) Pune}\\[1cm]
		
	\end{center}
\end{titlepage}

\begin{titlepage}
	
	\begin{center}

		{\huge \bfseries An ode to reality}\\[1cm]
		\includegraphics[width=10cm]{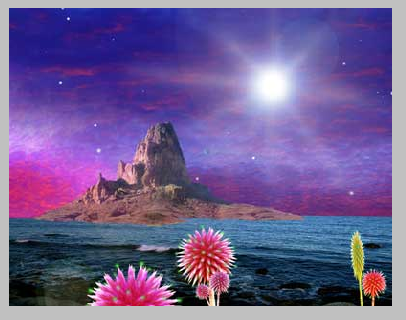}\\
		
		The solidified isle is the realm of physical reality,\\ 
		Whose edges alone are probed by the restless waves of Thought and Reason, \\
		Aided in the foreground by the floral sense of beauty, \\
		Whilst the All-Knowing Sun of Intuition shines brightly above \\
		Illuminating all realms, even those recondite noumenal recesses \\
		Unknown and Unknowable to Thought and Emotion, \\
		Where you reign supreme, Oh Reality! \footnote{Composed by Prof. R. Srikanth. For the source, \href{http://poornaprajna.org/srik/srir.htm}{\fbox{click here}}}\\
		
		\includegraphics[width=10cm]{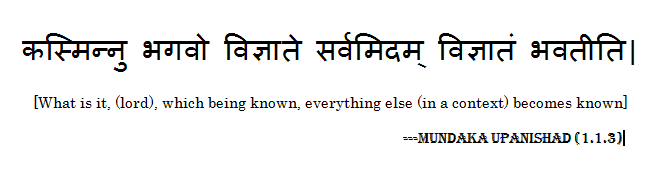} \\
		
		Yet, nature is made better by no mean \\
		But nature makes that mean: So, over that art, \\
		Which you say adds to the nature, is an art \\
		That nature makes. \footnote{An excerpt from `A winter's tale' by William Shakespeare}
		
	\end{center}
	
\end{titlepage}

\begin{titlepage}
	
	\begin{center}
		{\huge \bfseries Dedicated to}\\[1cm]
		
		\textbf{To my parents} \\
		For giving me an immense support during the whole of my journey through life
		\linebreak
		\linebreak
		\textbf{To Prof. Tejinder P. Singh}\\
		For the patience and the faith which he kept in me during the thesis work. 
		\linebreak
		\linebreak
		\textbf{To my friends and mentors at IISER-Pune, TIFR and NIRMAN.}\\
		In the presence of whom, I happen to get (momentarily) relieved from the curse of Sisyphus
		\linebreak
		\linebreak
		\textbf{To the magnificent SUNSET at the shores of TIFR, Mumbai}
		\includegraphics[width=13cm]{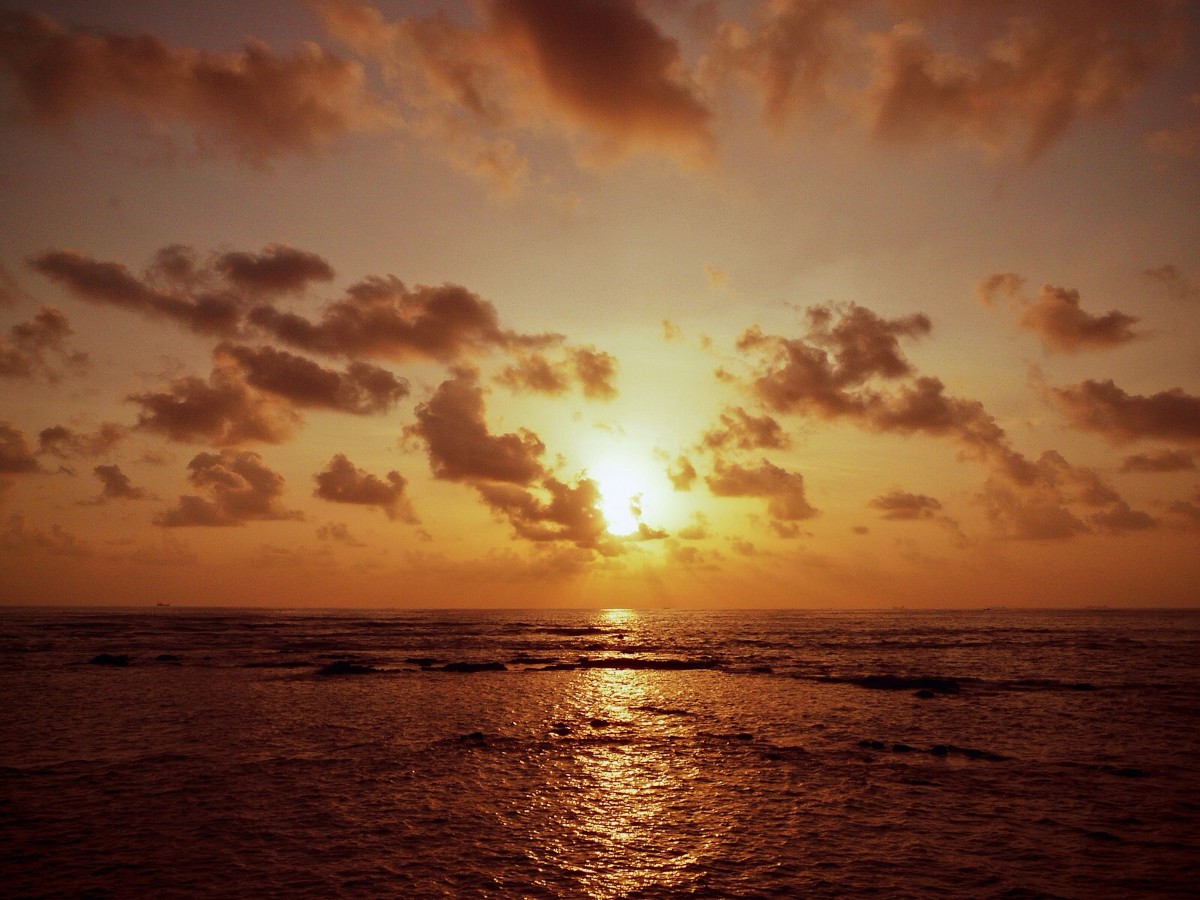}
		\footnote{Photo credits:- Pranab Das's collection. For the source of this photo, \href{https://dayone.me/FOtzMa}{\fbox{click here}}}.
		\linebreak
		-- -- -- -- -- -- -- -- X -- -- -- -- -- -- -- -- -- -- X-- -- -- -- -- -- -- --
		{\huge \bfseries Acknowledgements}\\[0.5cm]
	\end{center}
	Firstly, I express my sincere and deepest gratitude to my supervisor Prof. TP Singh for providing me with an opportunity to work with him in TIFR. His expertise, constant encouragement, understanding and most importantly his affectionate attitude has added considerably to my experience. Without his continual inspiration in the group meetings, it would have not been possible to complete this study. I am grateful to Prof. Rajeev Bhalerao for his guidance as TAC member. I express my sincere thanks to TIFR-VSRP-2017 students Arnab, Vedant and project students Abhinav, Navya and Nehal. It was an enriching experience to work with them. I am indebt to the tax payers of our country who supported me via DST-INSPIRE fellowship. Last but not the least, my head bows down to all those who believe in the open-source movement and share their works/comments/criticisms extensively on INTERNET and promote the culture of collective scientific voyage. 
	\vfill 
\end{titlepage}
\def\baselinestretch{1.3}\selectfont

\begin{titlepage}
	\begin{center}
		{\huge \bfseries Abstract}\\[1.5cm]
	\end{center}
	
	There are various generalizations of Einstein's theory of gravity (GR); one of which is the Einstein- Cartan (EC) theory. It modifies the geometrical structure of manifold and relaxes the notion of affine connection being symmetric. The theory is also called $U_4$ theory of gravitation; where the underlying manifold is not Riemannian. The non-Riemannian part of the space-time is sourced by the spin density of matter. Here mass and spin both play the dynamical role. We consider the minimal coupling of Dirac field with EC theory; thereby calling the full theory as Einstein-Cartan-Dirac (ECD) theory. In the recent works by T.P Singh titled ``A new length scale in quantum gravity \cite{TP_1}", the idea of new unified mass dependent length scale $L_{cs}$ has been proposed. We discuss this idea and formulate ECD theory in both - standard as well as this new length scale. We found the non-relativistic limit of ECD theory using WKB-like expansion in $\sqrt{\hbar}/c$ of the ECD field equations with both the length scales. At leading order, ECD equations with standard length scales give Schr\"{o}dinger-Newton equation. With $L_{cs}$, in the low mass limit, it gives source-free Poisson equation, suggesting that small masses don't contribute to gravity at leading order. For higher mass limit, it reduces to Poisson equation with delta function source. Next, we formulate ECD theory with both the length scales (especially the Dirac equation which is also called hehl-Datta equation and Contorsion spin coefficients) in Newman-Penrose (NP) formalism. The idea of $L_{cs}$ suggests a symmetry between small and large masses. Formulating ECD theory with $L_{cs}$ in NP formalism is desirable because NP formalism happens to be the common vocabulary for the description of low masses (Dirac theory) and high masses (gravity theories). We propose a conjecture to establish this duality between small and large masses which is claimed to source the torsion and curvature of space-time respectively. We therefore call it ``Curvature-Torsion" duality conjecture. In the context of this conjecture, Solutions to HD equations on Minkowski space with torsion have been found and their implications for the conjecture are discussed. Three new works which we have done in this thesis [Non-relativistic limit of ECD theory, formulating ECD theory in NP formalism and attempts to find the solution to non-linear Dirac equation on $U_4$] are valid for standard theory and also the theory with $L_{cs}$. The conjecture to establish the Curvature-torsion duality is formulated in the context of idea of $L_{cs}$. 
	\vfill 
\end{titlepage}

\tableofcontents
\listoffigures

\chapter{Introduction}

\section{Introducing four broad ideas to establish the grounds for this thesis}
\subsection{Einstein-Cartan theory}\label{sec:ECintro}

\par Einstein's theory of gravity, more commonly called as ``General theory of relativity" (GR), published in 1915 is one of the most important works of $20^{th}$ century. It has been described as the most beautiful of all the existing physical theories \cite{landau_2}. In GR, Gravity is described as a geometric property of space-time continuum; thereby generalizing special relativity and Newton's law of universal gravitation. In GR, background space-time is Riemann manifold (denoted as $V_4$) which is torsion less. Affine connection coincides uniquely with levi-civita connection and geodesics coincides with the path of shortest distance. 

There are few possible modifications of Einstein's theory of gravity (GR) [consistent with the principle of equivalence]; one of which is the Einstein- Cartan (EC) theory. It modifies the geometrical structure of manifold and relaxes the notion of affine connection being symmetric. The theory is also called $U_4$ theory of gravitation; where the underlying manifold is not Riemannian. The non-Riemannian part of the space-time is sourced by the spin density of matter. Mass and spin both play the dynamical role. Torsion, as an antisymmetric part of the affine connection was introduced by Elie-Cartan (1922) \cite{Cartan_1922}. In May 1929, He wrote a letter to Albert Einstein suggesting that his studies on torsion might be of physical relevance in General Relativity. The local Minkowskian structure of space-time (which is the essential constraint on manifold if it has to describe physically plausible space-time) is not violated in the presence of torsion. So a manifold with torsion and curvature [with an essential constraint that non-metricity = 0 \cite{hehl_RMP}] can very well describe physical space-time. It is called Riemann-Cartan ($U_4$) manifold. Since the works of E.Cartan, many people like D. Scima, Kibble, F.Hehl, Trautman etc. have studied the theories of gravity on a Riemann-Cartan space time $U_4$ over last century. The basic framework of EC theory was laid down by D.Scima (1962, 1964) \cite{sciama_GR}, \cite{sciama_physica_structure_of_GR} and Kibble (1961) \cite{kibble_GR}. Hence the theory is called Einstein-Cartan-Scima-Kibble (ECSK) theory. Modern review on the subject of ECSK theory is by F.Hehl et.al (1976) \cite{hehl_RMP}. It is titled ``General relativity with spin and torsion: Foundations and prospects". In a recent work of Trautman \cite{trautman_ECT}, he suggests, ``It is possible that the Einstein–Cartan theory will prove to be a better classical limit of a future quantum theory of gravitation than the theory without torsion''. It is worth asking the question that why don't we observe torsion in the universe around us. We note that torsion becomes comparable to curvature only at length scales smaller than the Einstein-Cartan radius $r_c = (\lambda_c L_{pl}^2)^{1/3}$ and at densities higher than $m/r_c^3$ where $\lambda_c $ and $ L_{pl}$ are Compton wavelength and plank length respectively.  For nucleons, the Einstein-Cartan radius is about $10^{-27}$ cms, and the density above which torsion becomes important is about $10^54$ gms/cc \cite{mjlake}. These scales are beyond current technology, and since GR is in excellent agreement with observations, it is said that torsion can be safely neglected in today’s universe. Literature on Einstein-Cartan theory in the context of cosmology and early universe can be refereed in \cite{poplawski_cosmo1}, \cite{poplawski_cosmo2} and the references therein. \cite{poplawski2010nonsingular} 

When we minimally couple Dirac field on $U_4$, we get Einstein-Cartan-Dirac (ECD) theory. There are 2 independent geometric fields (metric, torsion) in this theory and one matter field $\psi$. We get 3 equations of motion. Dirac equation on $U_4$ becomes non-linear and is then called Hehl-Datta equation \cite{Hehl1971}. Einstein-Cartan theory and its coupling with Dirac field has been discussed in details in chapter (\ref{chp:ECD theory}). $U_4$ theory has also been discussed in details in book by Gasperini \cite{gasperini}. We have used some results from this book.  

\subsection{The Schr\"{o}dinger-Newton equation}\label{sec:SN eq}

The Schr\"{o}dinger equation  describes the evolution of the wave-function over time. Born's probability rule gives a connection between the wave-function and the physical world. However the process of wave-function collapse is one of instantaneous nature and its mechanism is not explained via any acceptable theory. Broadly, this is often called ``Quantum-measurement problem". A brief review of various interpretations which revolve around this problem can be looked up in section I.B of \cite{bassi2013models} and the references therein.

The Schr\"{o}dinger-Newton equation came first into the discussions within the scientific literature due to  Ruffini and Bonazzola in their work \cite{SN_eqn_originalidea}. Diosi et.al in their works \cite{SN_diosi} proposed this equation as a model of wave function collapse; more specifically as a model of gravitational localization of macro objects. Roger Penrose developed this idea further and proposed that Schr\"{o}dinger-Newton equation describes the basis states for the scheme of gravitationally induced wavefunction collapse. This can be looked up in his works \cite{SN_penrose_1}, \cite{SN_penrose_2}. In deriving Schr\"{o}dinger-Newton equation, we primarily observe the self-gravity of a quantum mechanical object; that is we observe the modification of Schr\"{o}dinger's equation due to the gravity of the particle for which the equation is being written. Here, matter is taken to be of quantum nature while gravity is still treated classically. Here we assume the fact that, to leading order, the particle produces a classical potential satisfying the Poisson equation, whose source is a density proportional to the quantum probability density. \\
\begin{equation}\label{2}
	\nabla^{2} \phi = 4\pi G m \vert \psi \vert ^{2}
\end{equation}
The Schr\"{o}dinger equation is then modified to include this potential and we get the Schr\"{o}dinger-Newton equation,
\begin{align}
	i\hbar \frac{\partial \psi(\textbf{r},t)}{\partial t} &= - \frac{\hbar^2}{2m} \nabla^{2}\psi(\textbf{r},t) + m\phi\psi \label{3} \\
	i\hbar \frac{\partial \psi(\textbf{r},t)}{\partial t} &= - \frac{\hbar^2}{2m} \nabla^{2}\psi(\textbf{r},t) - Gm^{2} \int \frac{\vert\psi(\textbf{r}',t)\vert^{2}}{\vert \textbf{r} - \textbf{r}'\vert} d^{3}r' \psi(\textbf{r},t) \label{4}
\end{align}
Equations (\ref{2}), (\ref{3}) and (\ref{4}) together is called ``Schr\"{o}dinger-Newton" system of equations. By many people, this system of equations was taken as hypothesis to be put to test by experiments, whether there are any observational consequences (Ex. in molecular interferometry etc.) Work by Giulini et.al \cite{giulini_SN_gravity_despersion} analyzed the quantitative behavior of Gaussian wave packets moving according to Schr\"{o}dinger-Newton equation and proved that wave packets disperse due to their own gravitational field significantly at mass scales around $10^{10}$u (for a width of 500nm.)   This is just $10^3$ orders of magnitude more than masses which are envisaged in the future molecular interferometry experiments. Some works \cite{SN_with_QG1}, \cite{SN_with_QG2} propose that this equation sheds some light on the question of necessity of quantum gravity. 

Main paper of our interest in this thesis is \cite{guilini_grosardt}. Its a recent study by Guilini and Grossardt aimed at knowing whether this equation can be understood as a consequence of known principles and equations. They found that Schr\"{o}dinger-Newton equation is the non-relativistic limit of self -gravitating Klein-Gordon and Dirac fields. Here the gravity is the classical gravity described by GR (on $V_4$ manifold). 

\subsection{Tetrad formalism, Spinor formalism, Newman-Penrose (NP) formalism}

\underline{\textbf{1) Tetrad formalism in GR}}\\
The usual method in approaching the solution to the problems in General Relativity was to use a \textbf{local coordinate basis} $\hat{e}^{\mu}$ such that $\hat{e}^{\mu} = {\partial_{\mu}}$. This coordinate basis field is covariant under General coordinate transformation. However, it has been found useful to employ non-coordinate basis techniques in problems involving Spinors. This is the tetrad formalism which consists of setting up four linearly independent basis vectors called a `tetrad basis' at each point of a region of spacetime; which are covariant under local Lorentz transformations. [One of the reason of using tetrad formalism for spinors is essentially this fact that transformation properties of spinors can be easily defined in flat space-time]. Tetrads are basically basis vectors on local Minkowski space. Detail account of tetrad formalism in GR can be found in Appendix  [\ref{Tetrad_formalism_CD_for_spinors}].

\underline{\textbf{2) $SL(2,\mathbb{C})$ Spinor formalism}}\\
4-vector on a Minkowski space can be represented by a hermitian matrix by some transformation law. Unimodular transformations on complex 2-Dim space induces a Lorentz transformation in Minkowski space. Unimodular matrices form a group under multiplication and is  denoted by$SL(2,\mathbb{C})$ - special linear group of 2 x 2 matrices over complex numbers. By a simple counting argument, it has six free real parameters corresponding to those of the Lorentz group. The levi-civita symbol $\epsilon_{AB'}$ acts as metric tensor in $\mathbb{C}^2$, which preserves the scalar product under Unimodular transformations. Spinor $P^{A}$ of rank 1 is defined as vector in complex 2-Dim space subject to transformations $\in SL(2, \mathbb{C})$. Similarly higher rank spinor are defined. Analogous to a tetrad in Minkowski space, here we have a spin dyad (a pair of 2 spinors $\zeta_{(0)A}$ and $\zeta_{(1)A}$) such that $\zeta_{(0)A}\zeta_{(1)}^A =1$.

\underline{\textbf{3) Newman-Penrose (NP) formalism}}\\
NP formalism was formulated by Neuman and Penrose in their work \cite{NP_original_paper}. It is a special case of tetrad formalism; where we choose our tetrad as a set of four null vectors viz. 
\begin{equation}
	e_{(0)}^{\mu} = l^{\mu},~~~ e_{(1)}^{\mu} = n^{\mu},~~~ e_{(2)}^{\mu} = m^{\mu}, ~~~e_{(3)}^{\mu} = \bar{m}^{\mu}
\end{equation}
$l^{\mu}, n^{\mu}$ are real and $m^{\mu}, \bar{m}^{\mu}$ are complex. The tetrad indices are raised and lowered by flat space-time metric 
\begin{equation}
	\eta_{(i)(j)} = \eta^{(i)(j)} = \begin{pmatrix}
		0 & 1& 0 & 0 \\
		1 & 0& 0 & 0 \\
		0 & 0& 0 & -1 \\
		0 & 0& -1 & 0 \\
	\end{pmatrix}
\end{equation}
and the tetrad vectors satisfy the equation $g_{\mu\nu} = e_{\mu}^{(i)}e_{\nu}^{(j)}\eta_{(i)(j)} $.
In the formalism, we replace tensors by their tetrad components and represent these components with distinctive symbols. These symbols are quite standard and used everywhere in literature. A brief review of NP formalism can be found in chapter (\ref{Np/ECDinNP}).

Now, it can be shown that there is a natural connection between spin dyads and null tetrads \cite{Chandru}, \cite{SVD_geometry_fields_cosmology}. A null tetrad can be associated with a spin dyad by certain identification. This connection is explained in details in Appendix [\ref{app:NP-SL2C}]. Equations of motion involving spinorial fields (Ex. Dirac field) can be expressed in NP formalism. Dirac equation on $V_4$ has been studied extensively in \cite{Chandru}. Many systems in gravitational physics are also studied in NP formalism \cite{Chandru}. NP formalism happens to be the common vocabulary between physics of quantum mechanical spinor field systems and classical gravitational field systems. 

\subsection{Unified length scale in quantum gravity $L_{cs}$ and curvature-torsion duality} \label{sec:intro_Lcs}
In the recent works of Tejinder P. Singh \cite{TP_1}, \cite{TP_2}, it has been argued, why and how Compton wavelength ($\lambda /\hbar c$) and Schwarzschild radius ($2GM/c^2$) for  a point particle of mass `m' should be combined into one single new length scale, which is called Compton-Schwarzschild length ($L_{CS}$). The idea of $L_{cs}$ is more coherent in the framework of $U_4$. Action principle has been proposed with this new length scale and Dirac equation and Einstein GR equations are shown to be mutually dual limiting cases of this underlying modified action. More details can be looked up in chapter (\ref{chp:introducing L_cs}). It has been proposed that for $m\ll m_{pl}$, the spin density is more important than mass density. Mass density can be neglected and spin density sources the torsion (coupling is through $\hbar$). Whereas, $m\gg m_{pl}$, mass density dominates spin density. spin density can be neglected and as usual, mass density sources the gravity (coupling is through G). In this manner there exists a symmetry between small mass and large mass in the sense that small mass is the source for torsion and large mass is the source for gravity. \cite{TP_2}. Since both small masses and large masses give same $L_{cs}$ (which is the only free parameter in the theory), there is a sort of duality between solutions to small masse and that of large mass. We call such a duality ``Curvature-torsion" duality. We will explain this duality more in chapter (\ref{chp:duality conjecture}) and \cite{GRFessay2018}

\section{Goals and objectives of the Thesis}
\subsection{Finding Non-relativistic limit of ECD theory}\label{sec:intro_NR}

As discussed in the section (\ref{sec:SN eq}), recent work by Giulini and Großardt \cite{guilini_grosardt} derived the non-relativistic limit of self-gravitating Klein-Gordan and Dirac fields. They used WKB-like expansion of Dirac Spinor and metric in (1/c) (as discussed in \cite{kiefer_Singh_1991}) and found that, at leading order, the non-relativistic limit gives Schr\"{o}dinger-Newton equation. This work considers:\\
\# Spherically symmetric gravitational fields\\
\# Background space-time is Riemannian ($V_4$)\\

As a sequel to this study and to the study by TP Singh \cite{TP_2}(where ECD equations are modified with the unified length scale $L_{cs}$), we aim for the following:\\

\noindent \# Consider the generic metric (with no assumptions of symmetry) and find the non-relativistic limit of Einstein-Dirac system. This would generalize their work. We hypothesize that It will also be possible to find the underlying role of symmetry in the metric (in the context of non-relativisic limit).\\
\# If we consider gravitational theories with torsion; especially Einstein-Cartan-Dirac (ECD) theory discussed in section (\ref{sec:ECintro}), it is worthwhile seeing whether the effects of torsion (viz. non-linearity in Dirac equation and correction to the gravity equation by spin-density) modify the Schr\"{o}dinger-Newton equation in its non-relativistic limit. If it doesn't, next question we can ask is - At what order in 1/c, does effects due to torsion start getting manifested in the non-relativistic limit. This is important from the point of view of experimental studies in the detection of torsion and also to study the implications of the ECD theory at low energies.\\
\# Find the non-relativistic limit of ECD equations with modified length scale $L_{cs}$. We wish to analyze the underlying limit at leading order for limiting cases of large mass and small mass.

\subsection{Formulating ECD theory in NP formalism}

Dirac equation has been studied extensively in NP formalism  on $V_4$. It's detail account can be seen in this celebrated book ``The mathematical theory of black holes" By S. Chandrasekhar \cite{Chandru}. We wish to formulate ECD theory in NP formalism. More specifically;\\ 
\# We know that Contorsion tensor is completely expressible in terms of components of Dirac spinor. We want to find an explicit expression for Contorsion spin coefficients (in Newman-Penrose) in terms of Dirac spinor components. We will express this in both length scales - standard and unified length scale $L_{cs}$\\
\# Dirac equation on $V_4$ is presented in equation (108) of \cite{Chandru}. We aim to modify these equations on $U_4$. We will express this in both length scales - standard and unified length scale $L_{cs}$

\underline{There are 2 independent reasons for doing this}:\\
1)  Many gravitational systems in the literature (especially having some specific symmetries explained in details in chapter (\ref{Np/ECDinNP}) are formulated in NP formalism. But the space-time background in all those cases don't have torsion ($V_4$). It is worthwhile seeing the change in equations when we have torsion in the picture. Most of the important and physically relevant geometrical objects/ identities (Ex. Riemann curvature tensor, Weyl tensor, Bianchi identities, Ricci identities etc.) on $U_4$ have been formulated in NP formalism in the work \cite{jogia_Griffiths}. In the context of ECD theory, however, there are 2 important aspects which are not yet accounted viz. Dirac equation on $U_4$ (Hehl-Datta equation), canonical EM tensor etc. Some works \cite{Timofeev2016}, \cite{zecca_NP}, \cite{zecca_NP1} attempt to do that but have not provided explicit corrections to standard NP variables due to torsion. Also, there are notational and sign errors in them. We wish to modify the equations/ physical objects as a sequel to Chandra's work in \cite{Chandru} which is on $V_4$. In the case of vanishing torsion, our equations/ formulations should boil down to standard equations on $V_4$ as given in \cite{Chandru}. With this objective, we formulate the equations of ECD theory (which has 3 primary equations on $U_4$ - Dirac equation, Gravitational equation relating Einstein's tensor and canonical EM tensor, Algebraic equation relating torsion and spin) with standard length scale. \textbf{Especially we would like to analyze the Contorsion spin coefficients and thereby use Chandrasekhar's approach to modify Dirac equation.} \\
2) As explained in section (\ref{sec:intro_Lcs}), the idea of $L_{cs}$ in the context of $U_4$ theory provides a symmetry between small and large mass. There is a duality in the solution to large and small mass (we attempt ton establish it through a conjecture explained in next section). Dirac theory dominates for small masses and gravity dominates for large masses. In order to establish such a duality, its desirable to have a common mathematical language (provided by NP formalism) for dealing with both the domains \cite{TP_1}. To this aim, we formulate the ECD theory in NP formalism with unified length scale $L_{cs}$.

\subsection{Testing Curvature-torsion duality conjecture}

As discussed in section (\ref{sec:intro_Lcs}), the idea of $L_{cs}$ proposed in \cite{TP_1} hints at a symmetry between small and large masses. Solution to small mass is dual to the solution to large mass in the sense that both have same $L_{cs}$ which is the only free parameter in the theory. The motivation for such a ``curvature-torsion" duality has been discussed in \cite{TP_2}. However, we need to make this duality, both qualitatively and mathematically, more evident. To this aim, we propose a conjecture called ``Curvature-torsion duality conjecture" in chapter (\ref{chp:duality conjecture}). Further, this chapter discusses the ways in which such a conjecture can be put to a test. After going through arguments presented in this chapter, we find that if a  solution to ECD equations on Minkowski space with torsion exists, which make a tensor ``T-S"  (defined in \ref{chp:duality conjecture}) vanish, existence of such a solution supports the conjecture. So, the last few sections of this chapter are devoted at finding solutions to Hehl-Datta equations on Minkowski space with torsion and test the duality conjecture. A more detailed account of curvature-torsion duality as an idea can be looked up in \cite{GRFessay2018}.

\section{Brief outline of the Thesis}

\fbox{Chapter \ref{chp:ECD theory} and \ref{chp:introducing L_cs}} are theory chapters. In \fbox{chapter \ref{chp:ECD theory}}, we have explained Einstein-Cartan-Dirac theory in details starting from first principles. \fbox{Chapter \ref{chp:introducing L_cs}} discusses the idea of unified length scale called Compton-Schwarzschild length scale ($L_{cs}$) in the theories which attempt to unify quantum mechanics with gravity. This chapter is mainly based on \cite{TP_1} and \cite{TP_2}. \fbox{Chapter \ref{chp:NR limit}} is dedicated at finding Non-relativistic limit of ECD theory with standard as well as unified length scale. One can directly go to summary section \fbox{\ref{sec:summary_NR}} of this chapter to know some new results. In \fbox{Chapter \ref{Np/ECDinNP}}, we have formulated the ECD theory in NP formalism with standard as well as unified length scale. One can find its summary in \fbox{ section \ref{sec:summary_ECDNP}}. In  \fbox{chapter \ref{chp:duality conjecture}}, we attempt to establish a duality between curvature-torsion via a conjecture and solve ECD equations on Minkowski space (metric flat) with torsion.  \fbox{Chapter \ref{chp:discussions}} is reserved fro presenting conclusions, outlook and future plans. All the important calculations relating to Non-relativistic limit of ECD equations can be looked up in \fbox{Appendix \ref{app:NR}}. ECD equations in NP formalism in \fbox{Appendix \ref{app:ECDinNP}}.

\chapter{Einstein-Cartan-Dirac (ECD) theory}\label{chp:ECD theory}

\begin{center}
	\includegraphics[width=10cm]{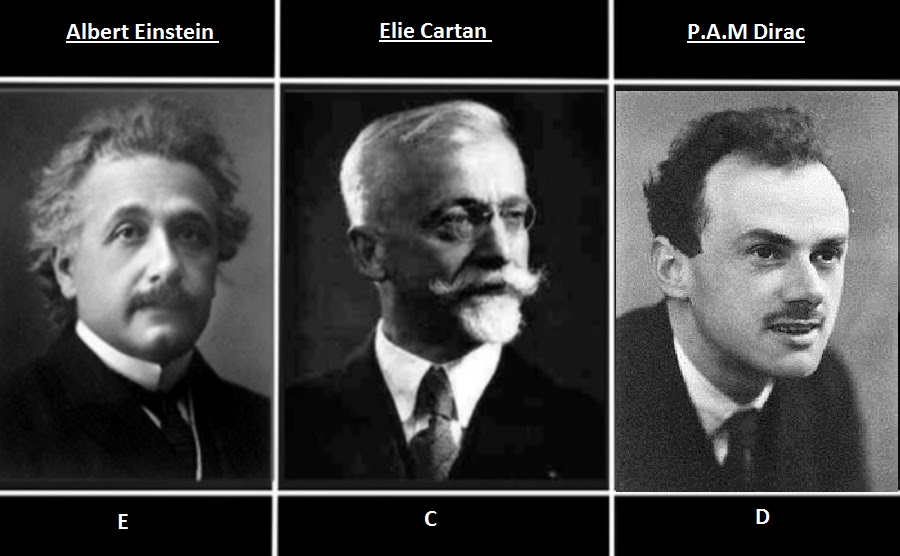}
\end{center}

\section{Brief Review of classical theories of gravity}

Huge strides were made in the European world of 13/14/15 and 16th century about nature of motion seen in the physical world. It took ingenious arguments and efforts of Aristotle, Keplar, Ibn Sina, T. Brahe, Copernicus, Galileo, Leibniz etc. to come up with a coherent, highly falsifiable, internally consistent (that is requiring no additional assumption beyond physical observables), highly predictable and reproducible model/ ontology of the nature of motion. the idea of "conservation of momentum" was an important paradigm shift in our thinking about the ontology of motion. The law also gave a  mathematically characterizable notion to the inertia. It stated that the product of "That property of matter which characterizes inertia" (called inertial mass $M_i$) and "velocity" remains conserved and such a hypothesis (extensively supported by empirical evidences) is sufficient for any type of motion to take place as such; abandoning the idea of ``unmoved mover" of Aristotle. Issac Newton in $16^{th}$ gave (the then universal) law of gravitation. The masses which appear in this law is the attribute of ``gravitational mass $M_g$". This formalism triggered the huge developments in classical physics. Surprisingly, $M_i$ and $M_g$ happened to be numerically exactly the same. It suggested that ``acceleration imparted to a body by a gravitational field is independent of the nature of the body". This motivated Einstein to generalize his special theory of relativity to include general coordinate transformations and non-inertial observers. He found that equivalence between inertia and gravity naturally leads to his theory of gravity called as general theory of relativity (GR). It is a classical theory of gravity. In GR, space-time is curved and the amount of curvature is determined by the Energy distribution on space-time. GR can be summed up in the following equation:

\begin{equation}\label{GRmaineq}
	G_{\mu\nu} = KT_{\mu\nu}
\end{equation}
Where $G_{\mu\nu}$ is Einstein's tensor which characterizes the curvature of space-time manifold and $T_{\mu\nu}$ characterizes the energy distribution on space-time. Gravity is described as a geometric property of space-time continuum. In GR, background space-time is Riemann manifold (denoted as $V_4$) which is torsion less. Affine connection coincides uniquely with levi-civita connection and geodesics coincides with the path of shortest distance. It is also called $V_4$ theory of gravitation. Max Born describes GR as (in his own words) ``GR seemed and still seems to me at present to be the greatest accomplishment of human thought about nature; it is a most remarkable combination of philosophical depth, physical intuition and mathematical ingenuity. I admire it as a work of art." GR has survived 100 years of challenges, both by experimental tests and by alternative theories. It is the basis for the Standard Model of physical cosmology. The review of GR and cosmology w.r.t its unsolved problems and future directions can be looked up in \cite{GR_cosmo_future}

\section{Field theory for first quantized Dirac-field}

Under the coordinate transformations, $x\rightarrow x' =\Lambda x $, the field $\phi$ can transform actively or passively as $\phi\rightarrow\phi'$. Active transformation of a generic field is governed by the equation: $\phi'(x) = L_{\Lambda}\phi(\Lambda^{-1}x)$ where $L_{\Lambda}$ are the elements of representation of a group of rotations [e.g. if $\phi$ real scalar field, then $L_{\Lambda}$ = $\mathbb{I}$, if $\phi$ is real vector field on 3D space, then $L_{\Lambda} = R$ where R represents a 3x3 orthogonal matrix. If $\phi$ is vector field on 4D space-time, then $L_{\Lambda} = \Lambda$ where $\Lambda$ represents a 4x4 matrix of Lorentz transformation. $\phi$ is spinor field on 4D space-time, then $L_{\Lambda} = S[\Lambda]$ where $S[\Lambda]$ is a spinor representation of Lorentz group]. We denote real tensor fields by $\phi$ and spinor fields by $\psi$. We define 2 types of variations - functional variation and total variation and adopt following notation henceforth \cite{book:17026} \\
*Functional variation in $\phi$: $\delta \phi=\phi^{\prime}(x^{\mu})-\phi(x^{\mu})$  and \\
*Total variation in $\phi$ : $\Delta \phi=\phi^{\prime}(x^{\prime\mu})- \phi(x^{\mu})= \delta \phi+(\partial_{\mu}\phi)\delta x^{\mu}$.

\subsection{Generalized Noether theorem and conserved currents}
Let $\phi(x^{\mu}) $ traces out 4-D region $R$ in a 5-D space ($\phi$,x,y,z,t). Initial and final space-like hyperspace; sliced at times t = $t_1$ and t = $t_2$ forms a boundary $\partial R$ of region $R$. Under the condition that the variation of $\phi$ and $x^{\mu}$ vanish on the boundary $\partial R$ we get, the \textsl{Euler-Lagrange} equation of motion for this field $\phi$ as follows:

\begin{equation}\label{eq:ELequation}
	\frac{\partial L}{\partial \phi}=\partial_{\mu}\pi^{\mu};~~~~ \pi^{\mu} = \left(\frac{\partial L}{\partial (\partial_{\mu}\phi)}\right)
\end{equation}
Now we vary action S on a classical trajectory and state Noether's theorem as follows: Suppose action is invariant under a group of transformations on $x^{\mu}$ and $\phi$ [whose infinitesimal version is given by $\Delta x^{\mu}=X^{\mu}_{\nu}\delta \omega^{\nu}$ and
$\Delta \phi =\Phi_{\nu}\delta \omega^{\nu}$ and which are characterized by infinitesimal parameter $\delta \omega^{\mu}$], then there exist one or more conserved quantities which remain invariant under the transformations. For Lagrangian, the condition is that it should either remain invariant or at the most change by total derivative. We will exploit this freedom om Lagrangian later, We will now establish this theorem mathematically. Variation of action over classical trajectory yields:

\begin{equation}\label{eq:noether}
	\delta S = \int_{\delta R} \Big[\pi^{\mu}\Phi_{\nu}-\Theta^{\mu}_{k}X^{k}_{\nu}\Big]\delta \omega^{\nu}\, d\sigma_{\mu}; ~~~~  \Theta_{\nu}^{\mu}=(\pi^{\mu}\partial_{\nu}\phi-L\delta_{\nu}^{\mu})
\end{equation}
Now, if the transformations make $\delta S = 0$ and since $\delta \omega^{\nu}$ is arbitary, we can write equation \ref{eq:noether} as follows:

\begin{equation}
	\int_{\delta R}J_{\nu}^{\mu}d\sigma_{\mu}=0; ~~~~~~~ J_{\nu}^{\mu} = \Big[\pi^{\mu}\Phi_{\nu}-\Theta^{\mu}_{k}X^{k}_{\nu}\Big]
\end{equation}
Using Gauss's theorem; 
\begin{equation}
	\int_{\delta R}J_{\nu}^{\mu}d\sigma_{\mu}=0 \Longrightarrow \int_{R} \partial_{\mu}J_{\nu}^{\mu}d^4x=0 \Longrightarrow \partial_{\mu}J_{\nu}^{\mu} = 0 
\end{equation} 
We therefore have a conserved and divergence-less current $J_{\nu}^{\mu}$ whose existence follows from the invariance of action under the given (generic) set of transformations. Integrating above equation over $t=const$ hyperspace and by using \textsl{Gauss's theorem} we get
\begin{equation}\label{eq:Noether charge}
	\frac{\partial Q_{\nu}}{\partial t}=0 \qquad \left(Q_{\nu}=\int_{V} J^{0}_{\nu}d^{3}x\right)
\end{equation} where $Q_{\nu}$ is \textsl{Noether's Charge}.\\

\subsection{Noether's theorem applied to Real Tensor and spinor fields} 
\underline{\textbf{Ex.1: Translational invariance for real tensor fields}} \\
Under the requirement that the laws of physics are to be translationally invariant i.e., using $\Phi_{\mu}=0$ and 
$X_{\nu}^{\mu}=\delta_{\nu}^{\mu}$ we get $J_{\nu}^{\mu}=-\Theta_{\nu}^{\mu}$; which, using \ref{eq:Noether charge} gives conserved four-momentum of the field 
\begin{align}
	Q_{\nu} =\int \Theta_{\nu}^{0} d^{3}x;~~~\
	Q_0 = \int_V \Big( \frac{\partial L}{\partial \dot{\phi}}\dot{\phi} -L \Big)d^3x = \int_V\mathcal{H}d^3x = \mathbb{H} =P_0; ~~~~ Q_i = \int_V \Big( \frac{\partial L}{\partial \dot{\phi}}\partial_i\phi \Big)d^3x = P_i
\end{align}
Where, $\mathcal{H}$ is the Hamiltonian density and $\mathbb{H}$ is the Total Hamiltonian of the system. Also, $Q_{\mu} = P_{\mu}$ and the fact that $\partial_t(Q_{\mu}) = 0$ suggests that invariance of translations conserve the 4-Momentum $P_{\mu}$. Here the conservation law is $\partial_{\mu}\Theta_{~\nu}^{\mu} = 0$. We observe that the Noether theorem's claim (Action remaining invariant) doesn't specify $\Theta_{~\nu}^{\mu}$ uniquely. The conservation law specifies $\Theta_{\nu}^{\mu}$ upto addition of divergence of an antisymmetric tensor field 'f' as follows:

\begin{equation}\label{eq:modifying canonical EM by f}
	T^{\mu\nu}=\Theta^{\mu\nu}+\partial_{\lambda}f^{\lambda\mu\nu}; \qquad \left(f^{\lambda\mu\nu}=-f^{\mu\lambda\nu}\right); \qquad \partial_{\mu}T^{\mu\nu} = 0
\end{equation}
Owing to Gauss's divergence theorem, such an addition of 'f' doesn't change the physical observables viz. Energy and Momentum. \\ 
\underline{\textbf{Ex.2: Rotational invariance for real tensor fields}}
We characterize infinitesimal Lorentz transformations by an antisymmetric tensor $\epsilon^{\mu\nu}$ such that $\epsilon^{\mu\nu}=-\epsilon^{\nu\mu}$.
Under a requirement that the action should be invariant under Lorentz group i.e. under the infinitesimal transformations $\Delta \phi = 0$ and $\Delta x^{\mu}= \delta x^{\mu}$; which is given by following equation:
\begin{equation}\label{eq:variation in x}
	\delta x^{\mu}=\epsilon^{\mu}_{~\nu}x^{\nu} = X^{\mu}_{\rho\sigma}\epsilon^{\rho\sigma}; ~~~~~~~~~ X^{\mu}_{\rho\sigma}=\frac{\delta_{\rho}^{\mu}x_{\sigma}-\delta_{\sigma}^{\mu}x_{\rho}}{2}
\end{equation}
using equation (\ref{eq:noether}) to find \textsl{Noether's current}; we obtain a 3 component Noether's current $J^{\mu\nu\sigma}$ as follows:
\begin{equation}\label{eq:angular momentum density and its conservtion}
	J^{\mu\nu\sigma}=\frac{-1}{2}\Big(\Theta^{\mu\nu}x^{\sigma}-\Theta^{\mu\sigma}x^{\nu}\Big); ~~~~~~~ \partial_{\mu}J^{\mu\nu\sigma} = \frac{-1}{2}\Big(\Theta^{\sigma\nu}-\Theta^{\nu\sigma}\Big)
\end{equation}
$\Theta^{\mu}_{~\nu}$ is the EM tensor representing 4-Momentum density. Hence RHS in the above expression represents density of angular momentum. Indeed, as we expect from the analogy with classical mechanics, invariance under Lorentz's rotation conserve the angular momentum of the system. The question now is: Does it remain conserved for any $\Theta^{\mu}_{~\nu}$? As we see in the second equation of equation (\ref{eq:angular momentum density and its conservtion}), only for symmetric $\Theta^{\mu}_{~\nu}$, conservation law seems to hold. We will investigate it in the next section.\\
\underline{\textbf{Ex.3: Rotational invariance for Spinor fields (this is of our interest)}}
We know that a Spinor field  transforms as 
\begin{equation}
	\psi^{\alpha}(x) \longrightarrow \psi'^{\alpha}(x) = S[\Lambda]^{\alpha}_{~\beta}\psi^{\beta}(\Lambda^{-1}x); ~~~~~~~~ S[\Lambda]=1+\frac{\omega_{\mu\nu}S^{\mu\nu}}{2}
\end{equation}
Corresponding functional and total variation in $\psi$ is then given by 

\begin{equation}
	\delta[\psi^{\alpha}(x)] = \Big(\frac{1}{2}\omega_{\mu\nu}S^{\mu\nu}\Big)^{\alpha}_{~\beta}\psi^{\beta}(x) - \partial_{\mu}\psi^{\alpha}(x)\omega^{\mu}_{~\nu}x^{\nu}; ~~~~~~~ \Delta\psi = \frac{1}{2}\omega^{\mu\nu}S_{\mu\nu} \psi = \Psi_{\mu\nu}\omega^{\mu\nu}; ~~~~~~~ \Psi_{\mu\nu} =\frac{1}{2} S_{\mu\nu} \psi
\end{equation} 
And the total variation in $x^{\mu}$ is as given in eqn (\ref{eq:variation in x}). Then, by Noether's theorem, the conserved current is:
\begin{align}
	J^{\mu}_{\nu\sigma} &= \pi^{\mu}\Psi_{\nu\sigma} - \Theta^{\mu}_{~\alpha} X^{\alpha}_{\nu\sigma} \\
	&= \frac{1}{2}\frac{\partial L}{\partial_{\mu} \psi} S_{\nu\sigma} \psi - \frac{1}{2}\Big(\Theta^{\mu}_{~\nu}x_{\sigma} - \Theta^{\mu}_{~\sigma}x_{\nu}\Big)
\end{align}
$\Theta^{\mu}_{~\nu}$ is the EM tensor representing 4-Momentum density. Hence the second term in above expression represents density of orbital angular momentum. Therefore $J^{\mu}_{~\nu\sigma}$ can be recognized as the total angular momentum density of the matter provided the first term represents the intrinsic spin density of matter field. We take $\nu,\sigma$ up, define spin density of the matter by a 3 component tensor $S^{\mu\nu\sigma}$ and rewrite the above equation as follows:
\begin{align}
	J^{\mu\nu\sigma} = S^{\mu\nu\sigma} - \frac{1}{2}\Big(\Theta^{\mu\nu}x^{\sigma} - \Theta^{\mu\sigma}x^{\nu}\Big); ~~~~~~~~~S^{\mu\nu\sigma} = \frac{1}{2}\frac{\partial L}{\partial_{\mu} \psi} S^{\nu\sigma} \psi
\end{align}

\subsection{Symmetrization of EM tensor by Belinfante-Rosenfeld transformation}
We find that, unless $\Theta^{\sigma\nu}$ is symmetric (which need not be the case always), we don't have a truly conserved angular momentum density current. But we know that Noether conserved currents are arbitary upto addition of divergence-less fields (refer equation \ref{eq:modifying canonical EM by f}). We can exploit this possibility to modify $\Theta^{\mu\nu}$ to $T^{\mu\nu}$ such that it is a symmetric tensor. The antisymmetric tensor field $f^{\lambda\mu\nu}$' which makes $T^{\mu\nu}$ symmetric is called Belnfinte tensor $B^{\lambda\mu\nu}$. It respects the fact that $\partial_\mu T^{\mu\nu} = 0$ and the fact that new symmetric tensor $T_{\mu\nu}$ defines the \textbf{same physical observable} (namely, energy-momentum) of the field. 

\begin{equation}
	T^{\mu\nu}=\Theta^{\mu\nu}+\partial_{\lambda}B^{\lambda\mu\nu} \qquad \left(B^{\lambda\mu\nu}=-B^{\mu\lambda\nu}\right)
\end{equation}
Is the existence of such a Balinfante tensor (which makes $T^{\mu\nu}$ symmetric) guaranteed? Following theorem proved by Belinfante in \cite{belinfante} gives necessary and sufficient conditions on the existence of $B^{\lambda\mu\nu}$. [We state the converse of the original theorem statement here]\\  
\underline{\textbf{Theorem A \cite{bandopadhyay_chp3}}}: ~ $\exists$ a symmetric stress-energy tensor [equivalently $\exists$ Belinfante tensor $B^{\lambda\mu\nu}$] iff the anti-symmetric part of the conserved canonical EM tensor is a total divergence.\\
\underline{\textbf{Theorem B \cite{bandopadhyay_chp3}}}: ~ Given a tensor $H^{\lambda\mu\nu}$ such that $\Theta^{[\mu\nu]} = -\frac{1}{2}\partial_{\lambda}H^{\lambda\mu\nu}$, one can explicitly construct a Belinfante tensor $B^{\lambda\mu\nu}$ such that $T^{\mu\nu}=\Theta^{\mu\nu}+\partial_{\lambda}B^{\lambda\mu\nu}$ is symmetric. The explicit construction is as follows:
\begin{equation}
	B^{\lambda\mu\nu} = \frac{1}{2}\Big( H^{\lambda\mu\nu} + H^{\mu\nu\lambda}-H^{\nu\lambda\mu}\Big)
\end{equation}
Such a transformation of $\Theta^{\mu\nu}$ to $T^{\mu\nu}$ is called \textbf{"Belinfante-Rosenfeld transformation"}.\\

Einstein's general theory of relativity requires EM tensor in its field equations to be symmetric. $G^{\mu\nu} = kT^{\mu\nu}$ Here $T^{\mu\nu}$ is called \textbf{'Dynamic EM tensor'} and is constructed as $T^{\mu\nu} = \frac{-2}{\sqrt{-g}} \frac{\partial(\sqrt{-g}\mathcal{L})}{\partial g_{\mu\nu}}$. It is symmetric by construction. We now state an important theorem. 
\underline{\textbf{Theorem C}\cite{bandopadhyay_chp3}}: ~ The symmetric EM tensor obtained by Belinfante-Rosenfeld transformation using Belinfante's tensor on matter field \textbf{is the same as} dynamic EM tensor which appears on the RHS of field equations of general theory of relativity.

\subsection{Applying above machinery to Dirac Lagrangian}

Lagrangian density pf Dirac field in given by 
\begin{equation}\label{L for dirac field_on M4}
	\mathcal{L}_{m} = \frac{i\hbar c}{2}(\overline{\psi}\gamma^a\partial_a\psi - \partial_a\overline{\psi}\gamma^a\psi) - mc^{2}\overline{\psi}\psi
\end{equation}
The EM tensor and its antisymmetric part is given by 
\begin{equation}
	\Theta_{ij} = \frac{i\hbar c}{2}[\overline{\psi}\gamma_i\partial_j\psi -\partial_j\overline{\psi}\gamma_i\psi] ~~~~~~~ \Theta_{[ij]} = \partial_k S^{ijk}
\end{equation}
Belinfante tensor is $B^{\lambda\mu\nu}= -S^{\lambda\mu\nu} + -S^{\mu\nu\lambda} + S^{\nu\lambda\mu}$. Hence, according to B-R transformations, 
\begin{equation}
	\Theta^{\mu\nu} \longrightarrow T^{\mu\nu} = \Theta^{\mu\nu} - \partial_{\lambda}[S^{\lambda\mu\nu} + S^{\mu\nu\lambda} - S^{\nu\lambda\mu}]
\end{equation}
And with the Lagrangian density defined in \ref{L for dirac field_on M4}, the explicit expression for $S^{\lambda\mu\nu}$ is given by:

\begin{equation}\label{spin density1}
	S^{\mu\nu\alpha} = \frac{-i\hbar c}{4}\bar{\psi}\gamma^{[\mu}\gamma^{\nu}\gamma^{\alpha]}\psi
\end{equation}
[Note = Up till now, we have used Latin symbols and Greek symbols interchangeably. We will define an unambiguous convention for their usage later]

\section{Einstein-Cartan (EC) theory: Modifying Einstein's GR to include torsion}

First we ask the question - Why consider a modified theory of gravity when General theory of relativity works out beautifully well and has stood \textbf{all} the experimental tests within the limits of the domain of validity of the theory. To understand this, we must realize that GR was formulated to describe gravitational interactions between macroscopic bodies. It is a classical theory of gravity. It is strongly suspected that at very high energies where the gravitational interaction becomes comparable to other quantum interactions and at very small length scales, the current formulation of gravity would not hold. There were (and still under investigation) many attempts to reconcile gravity with other fundamental interactions. One of the approach to do this is to expand the domains of validity of ordinary GR (validity in terms of micro/macro extent of matter) and to modify it so as to accomodate the new physical principles/ new experiments offered by the expanded domain of validity.

The Einstein-Cartan theory (EC) or also known as Einstein-Cartan-Sciama-Kibble (ECSK) theory [First published in \cite{sciama_GR}, \cite{sciama_physica_structure_of_GR} and extensively reviewed in \cite{hehl_RMP}] is one such attempt which \textbf{"extends" the geometrical principles and concepts of GR  to the certain aspects of micro-physical world}. In ordinary GR, matter is represented by Energy-Momentum tensor, which essentially provides the description of mass density distribution on space-time. However, when we delve into the microscopical scale we see that particles obey the laws of quantum mechanics and special relativity. At such length scales, the 'spin' (along with mass) of the particle has to be taken into account. Just like mass (which is characterized by EM tensor), it is a fundamental \textbf{and independent} property of matter . In macro physical limits, mass adds up because of its monopole character, whereas spin, being of dipole character, usually averages out in absence of external forces; hence matter in its macro physical regime can be dynamically characterized only by the energy-momentum tensor. If we wish to extend GR to include micro physics, we must take into account, therefore, that matter is dynamically described by mass and spin density, and since mass is related to curvature via EM tensor in framework of GR, spin should be related, through spin density tensor, to some other geometrical property of space time in the spirit of geometric theory of gravity. This requirement is satisfied by EC or ECSK theory. 

EC theory removes the restriction for the affine connection to be symmetric which was considered in GR. The antisymmetric part of the affine connection commonly known as 'torsion'($Q_{\alpha\beta}^{\:\:\:\:\:\:\mu}$), transforms like a third rank tensor and is known as Cartan's torsion tensor. It is seen that torsion couples to the  intrinsic spin angular momentum of particles \cite{hehl_RMP} just as the symmetric part of the connection (which gets expressed completely in terms of metric and its derivatives) couples to the mass. Since torsion is a geometrical quantity, spin modifies space-time and the resultant space-time is known as 'Riemann-Cartan' space-time ($U_4$) The field equations that follow are known as Einstein-Cartan field equation. The ($U_4$) manifold is also metric compatible (See section explained below) and hence can describe physical space-time in agreement with equivalence principle. 

Physically, torsion is related to the translation of vector like curvature is related to rotation, when a vector is displaced along infnitesimal path on $U_4$ manifold. Hence torsion allows for translations to be included and converts the local lorentz symmetry group of GR to the Poincare' group \cite{hehl_RMP}; which is essential because, in microscopic regime, elementary particles are the irreducible representations of Poincare' group, labeled by mass and spin. A detailed account of this motivation to include torsion can be looked in \cite{hehl_RMP}. Another motivation is that in the absence of external forces, the correct conservation law of total angular momentum arises only if torsion, whose origin is spin density, is included into gravitation \cite{hammond_necessity}, 

First we define a connection $\Gamma_{\alpha\beta}^{\:\:\:\:\:\:\:\mu}$ on a general affine manifold $(A_4)$ to allow for the parallel transport of tensorial objects. We define a torsion tensor out of this connection and it is given by,
\begin{equation}\label{6}
	Q_{\alpha\beta}^{\:\:\:\:\:\:\mu} = \Gamma_{[\alpha\beta]}^{\:\:\:\:\:\:\:\mu} = \frac{1}{2}(\Gamma_{\alpha\beta}^{\:\:\:\:\:\:\mu} - \Gamma_{\beta\alpha}^{\:\:\:\:\:\:\mu})
\end{equation}
\\
It is a third rank tensor that is antisymmetric in its first two indices and has 24 independent components. It can be shown that the general connection $\Gamma_{\alpha\beta}^{\:\:\:\:\:\:\:\mu}$ on ($A_4$) can be expressed in terms of metric, torsion tensor, and tensor of non metricity ($N_{\alpha\beta\mu} = \nabla_{\mu} g_{\alpha\beta}$)

\begin{align}\label{7}
	\Gamma_{\alpha\beta}^{\:\:\:\:\:\:\:\mu} = \bfrac{\mu}{\alpha\beta}- K_{\alpha\beta}^{\:\:\:\:\:\:\:\mu}-V_{\alpha\beta}^{\:\:\:\:\:\:\:\mu}
\end{align}
\\
where $\bfrac{\mu}{\alpha\beta}$ is the Christoffel symbol, $K_{\alpha\beta}^{\:\:\:\:\:\:\:\mu} = -Q_{\alpha\beta}^{\:\:\:\:\:\:\:\mu} - Q^{\mu}_{\:\:\:\alpha\beta} + Q_{\beta \:\:\:\alpha}^{\:\:\:\mu}$ is the contorsion tensor and $V_{\alpha\beta}^{\:\:\:\:\:\:\:\mu} = \dfrac{1}{2}[N_{\alpha\beta}^{\:\:\:\:\:\mu} - N^{\mu}_{\:\:\alpha\beta} - N_{\beta\:\:\:\alpha}^{\:\:\:\mu}]$ is the definition of V .

Einstein-Cartan manifold ($U_4$) is a particular case of a general affine manifold in which the metric tensor is covariantly constant. 

\begin{equation}\label{8}
	N_{\alpha\beta\mu} = \nabla_{\mu} g_{\alpha\beta} = 0
\end{equation}
This condition, which preserves scalar products (and then the invariance of lengths and angles) under parallel displacement is called metricity postulate. It secures the local Minkowski structure of space-time in agreement with principle of equivalence. The connection satisfying the condition of eqn (\ref{8}) is called metric compatible connection. The connection of Riemann Cartan manifold ($U_4$) is then written as:
\begin{align}\label{9}
	\Gamma_{\alpha\beta}^{\:\:\:\:\:\:\:\mu} = \bfrac{\mu}{\alpha\beta}- K_{\alpha\beta}^{\:\:\:\:\:\:\:\mu}
\end{align}
\\
Other quantities such as covariant derivative, Riemann tensor, Ricci tensor and Einstein tensor are defined in a similar fashion as in GR, the only difference being that the Christoffel symbols are replaced by the total connection as defined in equation (\ref{9}),
\begin{align}
	A^{\mu}_{\:\:\: ;\beta} &= \partial_{\beta}A^{\mu} + \Gamma_{\beta\alpha}^{\:\:\:\:\:\:\:\mu} A^{\alpha}\\
	R_{\alpha \beta \mu}^{\:\:\:\:\:\:\:\:\:\nu} &= \partial_{\alpha}\Gamma_{\beta\mu}^{\:\:\:\:\:\:\:\nu} - \partial_{\beta}\Gamma_{\alpha\mu}^{\:\:\:\:\:\:\:\nu} + \Gamma_{\alpha\lambda}^{\:\:\:\:\:\:\:\nu}\Gamma_{\beta\mu}^{\:\:\:\:\:\:\:\lambda} - \Gamma_{\beta\lambda}^{\:\:\:\:\:\:\:\nu}\Gamma_{\alpha\mu}^{\:\:\:\:\:\:\:\lambda}\\
	R_{\mu\nu} &= R_{\alpha\mu\nu}^{\:\:\:\:\:\:\:\:\:\alpha}\\
	G_{\mu\nu} &= R_{\mu\nu} - \frac{1}{2} g_{\mu\nu} R
\end{align}
\\
However it must be noted that $R_{\mu\nu}$ and $G_{\mu\nu}$ are no longer symmetric. Riemann Tensor has 36 independent components. The Bianchi identities can be defined in a similar way; following the usual definitions. It is worth investigating the anti-symmetric part of $G_{\mu\nu}$. We can show that 
\begin{equation}
	G_{[\mu\nu]} = R_{[\mu\nu]} = \nabla_{\alpha}T_{\mu\nu}^{~~\alpha} + 2Q_{\alpha} T_{\mu\nu}^{~~\alpha} = \widetilde{\nabla}_{\alpha}T_{\mu\nu}^{~~\alpha}~~~~~~~~~~ where ~~~~~\widetilde{\nabla}_{\alpha} = \nabla_{\alpha} + 2Q_{\alpha}
\end{equation}
where $T_{\mu\nu}^{~~\alpha} = Q_{\mu\nu}^{~~\alpha} + \delta_{\mu}^{\alpha}Q_{\nu} - \delta_{\nu}^{\alpha}Q_{\mu}$ is called as the modified torsion tensor (This is a very important quantity which, as we will see appears in filed equations of EC theory) and the quantity $Q_{\nu}$ is the trace of torsion, given by $Q_{\nu}= Q_{\nu \alpha}^{~~~\alpha}$. $G^{\mu\nu}$ can also be expressed as \cite{hehl_RMP},
\begin{align}\label{eq:for G = G_riemann + extra} 
	G^{\mu\nu}(\Gamma) = &G^{\mu\nu}(\{\})+ \widetilde{\nabla}_{\alpha}[T^{\mu\nu\alpha}+T^{\alpha\mu\nu}-T^{\nu\alpha\mu}]  \\
	& +\Big[4T^{\mu\alpha}_{\:\:\:\:\:\:[\beta}T^{\nu\beta}_{\:\:\:\:\:\:\alpha]} + 2T^{\mu\alpha\beta}T^{\nu}_{\:\:\:\alpha\beta} - T^{\alpha\beta\mu}T_{\alpha\beta}^{\:\:\:\:\:\:\nu} - \frac{1}{2}g^{\mu\nu}(4T_{\alpha\:\:\:[\gamma}^{\beta}T^{\alpha\gamma}_{\:\:\:\:\:\:\beta]} + T^{\alpha\beta\gamma}T_{\alpha\beta\gamma})\Big] \nonumber
\end{align}
We adopt an important convention henceforth:
\begin{itemize}
	\item The symbol $\nabla$ is used to indicate total covariant derivative. The symbol $\{\}$ is used to indicate christofell connection. So, $\nabla^{\{\}}$ would mean covariant derivative w.r.t christofell connection.
	\item Whenever there is a bracket like $(\{\})$ this in front of any object, it indicates the value of object calculated using Christoffel connection. We would also call it ``Riemann part of the object" often. 
\end{itemize}
Hence $G^{\mu\nu}(\{\})$ is the Riemann part of Einstein's tensor (the one occurring in GR). By definition, it is symmetric. However it doesn't capture the full symmetric part of total $G^{\mu\nu}$. Hence all the additional part to $G^{\mu\nu}(\{\})$ is asymmetric. 
\section{Lagrangian and the corresponding Field equations of EC theory}\label{b.2}

The field equations for the Einstein-Cartan theory may be obtained by the usual procedure where the action is constructed and then varied w.r.t. the geometric and matter fields in the Action. Lagrangian of EC theory will have matter lagrangian and a kinetic term for the gravitational field. We apply minimal coupling procedure, where Minkowski metric $\eta _{\mu\nu}$ is replaced by world metric $g_{\mu\nu}$ and partial with covariant derivatives of EC theory (defined later in next section). We keep $\mathcal{L}_{g} = R$ as in normal GR. We justify this by knowing the fact that in the limit of vanishing torsion, the original field equations of GR are obtained. The action is given by:

\begin{equation} \label{14}
	S = \int d^{4}x \sqrt{-g} \Big[\mathcal{L}_{m} (\psi, \nabla\psi, g) - \frac{1}{2\rchi} R(g, \partial g, Q)\Big]
\end{equation}
\\
Here $\rchi = \frac{8\pi G}{c^4}$ and $\mathcal{L}_{m}$ denotes the matter Lagrangian density and describes the distribution of matter field. The second term on the RHS represents the Lagrangian density due to the gravitational field. There are 3 fields in this Lagrangian viz. $\psi$ (matter field), $g_{\mu\nu}$ (metric field) and $K_{\alpha\beta\mu}$ (Contorsion field)  \\
\\
\textbf{varying w.r.t the matter field $\psi$}

\begin{equation}\label{15}
	\frac{\delta(\sqrt{-g}\mathcal{L}_m)}{\delta \psi} = 0  ------ E.O.M ~ for ~ matter ~ field.
\end{equation}
\textbf{Varying w.r.t. the metric field},
\begin{equation}\label{16}
	\frac{1}{\sqrt{-g}\chi}\frac{\delta(\sqrt{-g}R)}{\delta g_{\mu\nu}} = \frac{2}{\sqrt{-g}} \frac{\delta(\sqrt{-g}\mathcal{L}_{m})}{\delta g_{\mu\nu}} = T^{\mu\nu}
\end{equation} 
\textbf{Varying w.r.t. Contorsion field},
\begin{equation}\label{17}
	\frac{1}{\sqrt{-g}\chi}\frac{\delta(\sqrt{-g}R)}{\delta K_{\alpha\beta\mu}} = \frac{2}{\sqrt{-g}} \frac{\delta(\sqrt{-g}\mathcal{L}_{m})}{\delta K_{\alpha\beta\mu}} = S^{\mu\beta\alpha}
\end{equation}
\\
These are the generic field equations of Einstein-Cartan theory. $T_{\mu\nu}$ on the RHS of eqn (\ref{16}) dynamical Energy-Momentum Tensor. Similarly, $S^{\mu\beta\alpha}$ on the RHS of eqn (\ref{17}) is the dynamical spin density tensor defined in equation (\ref{spin density1})

Therefore we notice that, just as mass/energy density of the matter is coupled to the Riemann curvature of space-time via $T_{\mu\nu}$, the spin of matter is coupled to torsion of the space time via $S_{\mu\beta\alpha}$. 
Using the definition of the curvature tensor and torsion tensor defined in the earlier section, we obtain:

\begin{align}
	\frac{1}{\sqrt{-g}}\frac{\delta(\sqrt{-g}R)}{\delta g_{\mu\nu}} &= G^{\mu\nu} - \widetilde{\nabla}_{\alpha}[T^{\mu\nu\alpha}+T^{\alpha\mu\nu}-T^{\nu\alpha\mu}] = \chi T^{\mu\nu} \label{varying R w.r.t g}\\
	\frac{1}{\sqrt{-g}}\frac{\delta(\sqrt{-g}R)}{\delta K_{\alpha\beta\mu}} &= T^{\mu\beta\alpha} = \chi S^{\mu\beta\alpha}\label{varying R w.r.t K}
\end{align}
Equation (\ref{varying R w.r.t g})  and (\ref{varying R w.r.t K}) can be together written as, 

\begin{align}
	G^{\mu\nu} &= \chi T^{\mu\nu}  +  \widetilde{\nabla}_{\alpha}[T^{\mu\nu\alpha}+T^{\alpha\mu\nu}-T^{\nu\alpha\mu}] \\
	&= \chi T^{\mu\nu} + \chi\widetilde{\nabla}_{\alpha} [S^{\mu\nu\alpha}+S^{\alpha\mu\nu}-S^{\nu\alpha\mu}]\\
	G^{\mu\nu} &= \chi \Sigma^{\mu\nu} 
\end{align}
Where, $\Sigma^{\mu\nu}$ is the canonical energy momentum tensor. Field equations of EC theory can be summarized below \cite{hehl_RMP}, \cite{Hehl1971}, \cite{gasperini}. 

\begin{align}
	G^{\mu\nu} &= \rchi \Sigma^{\mu\nu} \label{eq:generic EOM of mass-metric} \\
	T^{\mu\nu\alpha} &= \rchi S^{\mu\nu\alpha} \label{eq:generic EOM of spin-torsion}\\
	\frac{\delta(\sqrt{-g}\mathcal{L}_m)}{\delta \bar{\psi}} &= 0 \label{eq:generic EOM of matter}
\end{align}
We now find th explicit expression for $G^{\mu\nu}(\{\})$ using equations (\ref{eq:for G = G_riemann + extra}), (\ref{eq:generic EOM of mass-metric}), (\ref{eq:generic EOM of spin-torsion}). 
\begin{equation}\label{20}
	G^{\mu\nu}(\{\}) = \rchi T^{\mu\nu} + \rchi^{2}\tau^{\mu\nu} = \chi \widetilde{\sigma}^{\mu\nu}; ~~~~~~~~~~~~~ \widetilde{\sigma}^{\mu\nu} = T^{\mu\nu} + \rchi\tau^{\mu\nu} 
\end{equation}
where 
\begin{equation}\label{21}
	\tau^{\mu\nu} = c^2\Big(4S^{\mu\alpha}_{\:\:\:\:\:\:[\beta}S^{\nu\beta}_{\:\:\:\:\:\:\alpha]} - 2s^{\mu\alpha\beta}S^{\nu}_{\:\:\:\alpha\beta} + S^{\alpha\beta\mu}S_{\alpha\beta}^{\:\:\:\:\:\:\nu} + \frac{1}{2}g^{\mu\nu}(4S_{\alpha\:\:\:[\gamma}^{\beta}S^{\alpha\gamma}_{\:\:\:\:\:\:\beta]} + S^{\alpha\beta\gamma}S_{\alpha\beta\gamma})\Big)
\end{equation}
We again note an important point here though $\widetilde{\sigma}^{\mu\nu}$ defined above is symmetric by definition, it doesn't capture the full symmetry of $\Sigma^{\mu\nu}$. 

This term on RHS of equation (\ref{20}) is completely dependent on the spin of the particle. Some important observations can be made from above field equations. eqn (\ref{eq:generic EOM of spin-torsion}) is an algebric equation; suggesting that torsion can't propogate outside matter field in the EC theory. It is confined to the region of matter fields. However, Spin of the matter fields modifies the Energy momentum tensor as given by eqn (\ref{20}), which in turn modifies the metric, which propogates upto infinity. \textbf{The spin content of the matter can influence the geometry outside the matter, though indirectly (through metric) and very weakly.} 

\section{Coupling of EC theory with Dirac field: Einstein-Cartan-Dirac (ECD) theory}

We will consider particles with spin-1/2, described by the Dirac field. The matter Lagrangian density for Dirac field is given by ,

\begin{equation}\label{Lagrangian for Dirac spin 1/2 field}
	\mathcal{L}_{m} = \frac{i\hbar c}{2}(\overline{\psi}\gamma^{\mu}\nabla_{\mu}\psi - \nabla_{\mu}\overline{\psi}\gamma^{\mu}\psi) - mc^{2}\overline{\psi}\psi
\end{equation}
\\
Here $\psi$ is a spinor. Transformation properties of Spinors are defined in a flat Minkowski space; locally tangent to the $U_4$ manifold. We know that, at each point, we have a coordinate basis vector field $\hat{e}^{\mu} = {\partial_{\mu}}$. This coordinate basis field is covariant under General coordinate transformation. However, a spinor (as defined on flat Minkowski space-time) is associated with the basis vectors which are covariant under local Lorentz transformations. To this aim, we define at each point of our manifold, a set of 4 orthonormal basis field (called tetrad field), Given by ${\hat{e}^i (x)}$. These are 4 vectors (one for each $\mu$) et every point. This tetrad field is governed by a relation ${\hat{e}^i (x)} = e^i_{\mu} (x) \hat{e}^{\mu}$ where trasformation matrix $e^i_{\mu}$ is such that,

\begin{equation}\label{eq:tetrad-metric transformation}
	e^{(i)}_{\mu} e^{(k)}_{\nu} \eta_{{(i)}{(k)}} = g_{\mu\nu}
\end{equation}

The detail account of Tetrad formalism is given in Appendix [\ref{Tetrad_formalism_CD_for_spinors}]. Here we will use some results and definitions from this section. Trasformation matrix $e^{(i)}_{\mu}$ allows us to convert the components of any world tensor (tensor which transforms according to general coordinate transformation) to the corresponding components in local Minkowskian space (These latter components being covariant under local Lorentz transformation). Greek indices are raised or lowered using the metric $g_{\mu\nu}$, while the Latin indices are raised or lowered using $\eta_{(i)(k)}$. parenthesis around indices is just a matter of convention. 
\\
\underline{We adopt an important conventions for the remainder of paper}\\
\begin{itemize}
	\item Greek indices e.g. $\alpha, \zeta, \delta $ refer to world components (which transform according to \textbf{general coordinate transformation}).
	\item Latin indices with parenthesis e.g. (a) or (i) refer to tetrad index. (which transform according to \textbf{local Lorentz transformation} in flat tangent space).
	\item Latin index without parenthesis e.g. i,j,b,c would just mean objects in Minkowski space (which transform according to \textbf{global Lorentz transformation}).
	\item {0,1,2,3} indicate world index and {(0),(1),(2),(3)} indicate tetrad index.
	\item The symbol $\nabla$ is used to indicate total covariant derivative. The symbol $\{\}$ is used to indicate christofell connection. So, $\nabla^{\{\}}$ would mean covariant derivative w.r.t christofell connection.
	\item The symbol comma (,) is used to indicate partial derivatives and  the symbol semicolon (;) is used to indicate Riemannian covariant derivative. So for tensors, (;) and $\nabla^{\{\}}$ are same, but for spinors (;) would have partial derivatives and riemannian part of spin connection ($\gamma$) as described below.
\end{itemize}

Just as we define affine connection $\Gamma$ to facilitate parallel transport of geometrical objects with world (greek) indices, we define Spin connection $\omega$ for anholonomic objects (those having latin index). As affine connection $\Gamma$ has 2 parts- riemannian ($\{\}$) part coming from christofell connection and torsional part (made up of contorsion tensor K), similarly, spin connection $\omega$ also has 2 parts - Riemannian (denoted by $\gamma$) and torsional part (again made up of contorsion tensor K). $\gamma$, $\gamma^o$ and K are related by following equation.  These symbols and notations ae taken from \cite{jogia_Griffiths}. All the mathematics is explained in Appendix [\ref{Tetrad_formalism_CD_for_spinors}].

\begin{equation}\label{eq:total spin connection as sum of riemannian and torsional part}
	\gamma_{\mu}^{\:\:\:(i)(k)} = \gamma_{\mu}^{o\:\:\:(i)(k)} - K_{\mu}^{\:\:\:(k)(i)}
\end{equation}
Here, $\gamma_{\mu}^{o\:\:\:(i)(k)}$ is riemannian part and  $K_{\mu}^{\:\:\:(k)(i)}$ is the contorsion (torsional part)\\
\\
The relation between spin conection and affine connection is as follows

\begin{equation}\label{eq:relation between spin conection and affine connection is as follows}
	\begin{split}
		\gamma_{\mu}^{\:\:\:(i)(k)} &= e^{(i)}_{\alpha}e^{\nu (k)} \Gamma_{\mu\nu}^{\:\:\:\:\:\:\:\alpha} - e^{\nu (k)}\partial_{\mu}e^{(i)}_{\nu}\\ &= e^{(i)}_{\alpha}e^{\nu (k)} \bfrac{\alpha}{\mu \nu} - K_{\mu}^{\:\:\:(k)(i)} - e^{\nu(k)} \partial_{\mu}e^{(i)}_{\nu} 
	\end{split}
\end{equation}
\\
From above two equations, one can obtain the following crucial equation for Riemannian part of spin connection, entirely in terms of Christoffel symbols and tetrads.\cite{gasperini}

\begin{equation}\label{26}
	\bfrac{\alpha}{\mu \nu} = e_{(i)}^{\alpha}e_{\nu (k)} \gamma_{\mu}^{o\:\:\:(k)(i)} +  e_{(i)}^{\alpha}\partial_{\mu}e^{(i)}_{\nu}
\end{equation}
\\
Using all the results mentioned above, we define covariant derivative (CD) for Spinors on $V_4$ and $U_4$

\begin{align}
	\psi_{;\mu} &= \partial_{\mu} \psi + \frac{1}{4}\gamma^o_{\mu (b)(c)}\gamma^{[(b)}\gamma^{(c)]}\psi  ----------------CD~on~ [V_4]\label{CD on V4}\\
	\nabla_{\mu} \psi &= \partial_{\mu} \psi +  \frac{1}{4}\gamma^o_{\mu (c)(b)}\gamma^{[(b)}\gamma^{(c)]}\psi - \frac{1}{4} K_{\mu (c)(b)}\gamma^{[(b)}\gamma^{(c)]}\psi -------CD~on~ [U_4]\label{CD on U4}
\end{align}
\\
Substituting this into eqn (\ref{Lagrangian for Dirac spin 1/2 field}), we obtain the explicit form of Lagrangian density; which we vary w.r.t. $\overline{\psi}$ as in eqn (\ref{eq:generic EOM of matter}) to obtain Dirac equation on $V_4$ and $U_4$.

\begin{align}
	i\gamma^{\mu}\psi_{;\mu}- \frac{mc}{\hbar}\psi &= 0 ------Dirac~ Eqn~ on~ [V_4] \label{eq:DE V4}\\
	i\gamma^{\mu}\psi_{;\mu} + \frac{i}{4}K_{(a)(b)(c)}\gamma^{[(a)}\gamma^{(b)}\gamma^{(c)]}\psi - \frac{mc}{\hbar}\psi &= 0  ------Dirac~ Eqn~ on~ [U_4] \label{eq:DE U4 with K}
\end{align}
We next obtain gravitational field equations on both $V_4$ and $U_4$ using eqn (\ref{eq:generic EOM of mass-metric}) and Lagangian density defined in eqn (\ref{Lagrangian for Dirac spin 1/2 field}).

\begin{align}
	G^{\mu\nu}(\{\}) &= \frac{8\pi G}{c^4} T^{\mu\nu}---------------Gravitation~ Eqn~ on~ [V_4] \label{eq:GE V4}\\
	G^{\mu\nu}(\{\}) &= \frac{8\pi G}{c^4} T^{\mu\nu} - \frac{1}{2} \bigg{(}\frac{8\pi G}{c^4}\bigg{)}^2 g_{\mu\nu}  S^{\alpha\beta\lambda}S_{\alpha\beta\lambda}----Gravitation~ Eqn~ on~ [U_4] \label{eq:GE U4}
\end{align}
Here, $T^{\mu\nu}$ is the dynamical EM tensor which is symmetric and defined below:
\begin{equation}\label{dynamic EM tensor}
	T_{\mu\nu} = \Sigma_{(\mu\nu)}(\{\}) = \frac{i\hbar c}{4}\Big[\bar{\psi}\gamma_{\mu} \psi_{;\nu} + \bar{\psi}\gamma_{\nu} \psi_{;\mu} - \bar{\psi}_{;\mu} \gamma_{\nu}\psi -\bar{\psi}_{;\nu} \gamma_{\mu}\psi  \Big]
\end{equation}
\underline{Equations [\ref{eq:DE V4} and \ref{eq:GE V4}] together form a system of equations of Einstein-Dirac theory.} \\

We now aim to establish the field equations of Einstein-Cartan-Dirac theory. First let's define Spin density tensor using Lagrangian density defined in eqn (\ref{Lagrangian for Dirac spin 1/2 field}) 

\begin{equation}\label{eq:spin density}
	S^{\mu\nu\alpha} = \frac{-i\hbar c}{4}\bar{\psi}\gamma^{[\mu}\gamma^{\nu}\gamma^{\alpha]}\psi
\end{equation}
Using equations (\ref{eq:spin density}) and (\ref{eq:generic EOM of spin-torsion}), eqn (\ref{eq:DE U4 with K}) can be simplified to give us the Hehl-Datta equation \cite{hehl_RMP}, \cite{Hehl1971} ('$L_{Pl}$' being the Planck length). This, Along with equation (\ref{eq:GE U4}) and the equation which couples modified torsion tensor and spin density tensor together define the field equations of: \\ \underline{Einstein-Cartan-Dirac (ECD) theory; as summarized below}

\begin{align}
	i\gamma^{\mu}\psi_{;\mu} &= +\frac{3}{8}L_{Pl}^{2}\overline{\psi}\gamma^{5}\gamma_{(a)}\psi\gamma^{5}\gamma^{(a)}\psi + \frac{mc}{\hbar}\psi   \label{eq:HDFull_std}\\
	G^{\mu\nu}(\{\}) &= \frac{8\pi G}{c^4} T^{\mu\nu} - \frac{1}{2} \bigg{(}\frac{8\pi G}{c^4}\bigg{)}^2 g_{\mu\nu}  S^{\alpha\beta\lambda}S_{\alpha\beta\lambda}\label{eq:ECD_gravity_std}\\
	T^{\mu\nu\alpha} &=  - K^{\mu\nu\alpha} = \frac{8\pi G}{c^4}S^{\mu\nu\alpha} \label{eq:ECD_algebric_std}
\end{align}

\chapter{Introducing unified length scale $L_{cs}$ in quantum gravity}\label{chp:introducing L_cs}

\section{Brief review of quantum theories of gravity}
This section is mainly based upon the review article titled  ``Conceptual Problems in Quantum Gravity and Quantum Cosmology" by Claus Kiefer \cite{kiefer_conceptual_problems_inQG}. According to our present knowledge; strong, weak, electromagnetic and gravitational interactions are the four fundamental interactions in nature. The first three are described by standard model of particle physics (whose framework is of Quantum field theory) and fourth one is described by GR (whose framework is classical). Though no empirical evidence goes against GR; from purely theoretical point of view, the situation is not satisfactory. The main field equation of GR (\ref{GRmaineq}) would no longer have the same form if we consider the quantum nature of fields in $T_{\mu\nu}$. The `semiclassical Einstein equations' with $T_{\mu\nu}$ replaced by its expectation value `$<\psi|\mathbb{T}_{\mu\nu}|\psi>$' leads to problems \cite{SN_with_QG2}. In the 1957 Chapel Hill Conference, Richard Feynman gave the argument suggesting that `It is the superposition principle of QM which strongly points towards the need for quantizing gravity \cite{feynman_chapelhillconf}'. Apart from this, `unavoidable presence of singularities in GR \cite{singularity_in_GR}' and `problem of time in QM \cite{Time_in_QM},\cite{kiefer_conceptual_problems_inQG} ' forms the motivation for quantizing gravity amongst few other motivations. On a side note, it should be noted that the idea of `emergent gravity' by Padmanabhan \cite{paddy} is an alternative to the approach of direct quantization of gravitational fields. In brief, we can divide the approaches to quantum gravity in 4 broad groups \cite{isham2},\cite{isham}: 1) Quantize general relativity [2 methods are used in this approach -covariant and canonical quantum gravity.] 2) ‘General-relativise’ quantum theory [trying to adapt standard quantum theory to the needs of classical general relativity]. 3) General relativity is the low-energy limit of a quantum theory of something quite different [The most notable example of this type is the theory of closed superstrings]. 4) Start ab initio with a radical new theory. [Both classical general relativity and standard quantum theory ‘emerge’ from a deeper theory that involves a fundamental revision of the concepts of space, time and matter.]  We will now introduce the idea of unified length scale ($L_{cs}$) in quantum gravity.

\section{The idea of $L_{cs}$}
Einstein's theory of gravity (GR) and relativistic Quantum mechanics (Ex: Dirac theory for spin-1/2 particles) are the 2 most successful theories of the description of Universe at micro and macro level (in terms of mass `m' which is being described). Given a relativistic particle of mass `m', we can associate 2 length scales to it- characterizing its quantum and relativistic behavior. Quantum nature of the particle is associated with its Compton wavelength; given by $\lambda_C = (\hbar/mc)$ and the relativistic nature is associated to the Schwarzschild radius given by $R_S= (2GM/c^2)$. It is through these length scales, that the mass `m' enters the equation of description of their motion. Example, mass enters Dirac equation through $\lambda_C$ and it enters GR equations through $R_S$. Also, It is important to note that neither $\lambda_C$ (having $\hbar$ and c as fundamental constants) nor $R_S$ (having G and c as fundamental constants) could be used individually to define mass (or units of mass).

Both Dirac theory and general relativity claim to hold for all values of $m$ and it is only through experiments that we find that Dirac equation holds if $m \ll m_p$ or $\lambda_C \gg l_p$ while Einstein equations hold if  $m \gg m_p$ or $R_S \gg l_p$. \textbf{``From the theoretical viewpoint, it is unsatisfactory that the two theories should have to depend on the experiment to establish their domain of validity''} \cite{TP_1}. If we assert the fact that plank length is the smallest physically meaningful length, then it makes no sense to talk of $R_S<L_{pl}$ when $m<m_{pl}$ and to talk of $\lambda_C<L_{pl}$ when $m>m_{pl}$. Instead it is more reasonable to think of universal length scale which remains above $L_{pl}$ for all masses and whose limiting cases give $\lambda_C$ for small mass and $R_S$ for large mass. One such way to define a universal length scale is given in \cite{TP_1} as follows

\begin{equation}
	\frac{L_{CS}}{2l_p} := \frac{1}{2}\left(\frac{2m}{m_p}+\frac{m_p}{2m}\right) := \cosh{z}
\end{equation}.
where $z = \ln{\frac{2m}{m_p}}$. These ideas are discussed in details in the recent works of Tejinder P. Singh \cite{TP_1}, \cite{TP_2}. The dynamical process for mass `m' now involves $L_{cs}$ (mass enters the dynamics through $L_{cs}$). An action principle has been proposed with this new length scale and Dirac equation and Einstein GR equations are shown to be mutually dual limiting cases of this underlying modified action. The proposed action for this underlying gravitation theory, which gives the required limits is as follows 
\begin{figure}\label{fig:{Lcs Vs mass behavior}}
	\begin{center}
		\includegraphics[width=16cm]{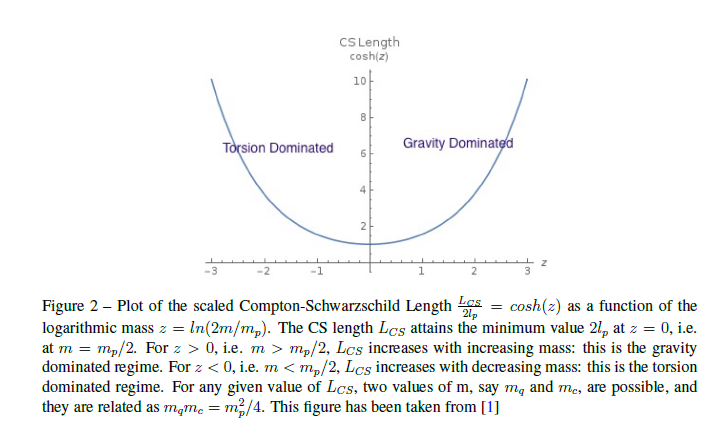}
		\caption{$L_{cs}$ Vs. mass behavior and its description \cite{TP_1}}
	\end{center}
\end{figure}
\begin{equation}
	\dfrac{L_{pl}^2}{\hbar} S = \int d^4x \sqrt{-g} [R-(1/2) L_{CS}\bar{\psi}\psi +  L_{CS}^2 \bar{\psi} i \gamma^{\mu} \partial_{\mu}\psi  ]
\end{equation} 
Generalizing this on a curved space-time, the action is:
\begin{equation}\label{eq:action for Lcs_ curved}
	\dfrac{L_{pl}^2}{\hbar} S = \int d^4x \sqrt{-g} \Bigg[R-\frac{1}{2}L_{CS}\bar{\psi}\psi + \pi \hbar i L_{CS}^2\Big(\bar{\psi}  \gamma^{\mu} \nabla_{\mu}\psi - \nabla_{\mu}\bar{\psi}\gamma^{\mu} \psi \Big) \Bigg]
\end{equation} 
If $\nabla$ and `R' are taken on $V_4$, the system is called 'Einstein-Dirac' system. In such a system , for small mass limit, couping to EM tensor in Einstein's equation is through $\hbar$ and not G. Hence we expect gravity to vanish. This creates an unpleasant situation for Einstein's equations. Because vanishing of gravity makes LHS 0; but RHS is non-zero (it is EM tensor coupled through $\hbar$). This compels us to introduce torsion in the theory. Because it would now add torsion field in the LHS and then it couples to EM tensor via $\hbar$. Further arguments can be looked up in \cite{TP_2}. So the idea of $L_{cs}$ is more coherent with the framework of Einstein-Cartan manifold ($U_4$ manifold). For $m\ll m_{pl}$, the spin density is more important than mass density. Mass density can be neglected and spin density sources the torsion (coupling is through $\hbar$). Whereas, $m\gg m_{pl}$, mass density dominates spin density. spin density can be neglected and as usual, mass density sources the gravity (coupling is through G). \textbf{Spin density and torsion are significant in micro-regime; whereas gravity and mass density are important in macro-regime.} In this manner there exists a symmetry between small mass and large mass in the sense that small mass is the source for torsion and large mass is the source for gravity. The solution for small mass is dual to the 'wave-function collapsed' solution for large mass in the sense that both the solutions have same value for $L_{cs}$  which is the only free parameter in the theory.\cite{TP_2}

\section{ECD equations with $L_{cs}$} \label{sec:ECD_with_Lcs}
The set of ECD field equations with the $L_{CS}$ incorporated in them are obtained by varying the Action (\ref{eq:action for Lcs_ curved}) w.r.t all the 3 fields (Here we have also given gravity equation with riemannian part of Einstein tensor.)\cite{TP_2}

\begin{align}
	G_{\mu\nu} &= \frac{8\pi L_{CS}^{2}}{\hbar c} \Sigma_{\mu\nu} \\
	G_{\mu\nu}(\{\}) &= \frac{8\pi L_{CS}^{2}}{\hbar c} T_{\mu\nu} - \frac{1}{2} \left(\frac{8\pi L_{CS}^{2} }{\hbar c}\right)^2  g_{\mu\nu}  S^{\alpha\beta\lambda}S_{\alpha\beta\lambda} \label{eq:gravity eqn with lcs}\\
	T^{\mu\nu\alpha} &= \frac{8\pi L_{CS}^{2}}{\hbar c} S^{\mu\nu\alpha} \\
	i\gamma^{\mu} \psi_{;\mu} &= \frac{3}{8} L_{CS}^2 \bar{\psi}\gamma^5\gamma_{\nu}\psi\gamma^5\gamma^{\nu}\psi + \frac{1}{2L_{CS}} \psi = 0 \label{eq:HD eqn with lcs}
\end{align}
\textbf{\underline{Important notation to be used henceforth:}} We use the symbol 'l' to denote a 'length scale' in the theory. For standard ECD theory, it is either plank length $l = L_{pl} = \sqrt{\frac{G\hbar}{c^3}}$ or half Compton wavelength $l = \frac{\lambda_C}{2} = \frac{\hbar}{2mc}$ or Schwarzschild radius $l=R_s = \frac{2GM}{c^2}$. For modified theory with new unified length scales$L_{cs}$, the length scale we use is $l = L_{cs}$. Every equation written in terms of generic `l' henceforth is valid for both length scales. We will mention in each case what this `l' refers to in the standard theory.

\chapter{Non-relativistic limit of ECD field equations}\label{chp:NR limit}

\section{Theoretical background and notations/ representations used in this chapter}

\subsection{Notations, conventions, representations and Ansatz's used}

\begin{itemize}
	\item Greek indices e.g. $\alpha, \zeta, \delta $ refer to world components (which transform according to \textbf{general coordinate transformation}).
	\item Latin indices with parenthesis e.g. (a) or (i) refer to tetrad index. (which transform according to \textbf{local Lorentz transformation} in flat tangent space).
	\item Latin index without parenthesis e.g. i,j,b,c would just mean objects in Minkowski space (which transform according to \textbf{global Lorentz transformation}).
	\item {0,1,2,3} indicate world index and {(0),(1),(2),(3)} indicate tetrad index.
	\item The Lorentz Signature used in this report is Diag(+, -, -, -).
	\item We use Dirac basis to represent the gamma matrices. These are basically matrix representation of clifford algebra $Cl_{1,3}[\mathbb{R}]$
	
	\begin{align}
		\gamma^0 = \beta = \begin{pmatrix}
			\mathbb{I}_2 & 0 \\ 0 & -\mathbb{I}_2
		\end{pmatrix}, \gamma^i = \begin{pmatrix}
			0 & \sigma^i\\ -\sigma^i & 0 \end{pmatrix}, \gamma^5 = \frac{i}{4!}\epsilon_{ijkl}\gamma^i\gamma^j\gamma^k\gamma^l = \begin{pmatrix}
			0 & \mathbb{I}_2 \\ \mathbb{I}_2 & 0 \end{pmatrix}, \alpha^i = \beta\gamma^i = \begin{pmatrix}
			0 & \sigma^i\\ \sigma^i & 0 \end{pmatrix}
	\end{align}
	
	\item \underline{Ansatz for Dirac spinor:} We want to choose an appropriate ansatz for spinor so as to fetch non-relativistic limit. We expand $\psi (x,t)$ as  $\psi (x,t) = e^{[iS(x,t)\hbar]}$: (which can be done for any complex function of x and t). Here S is Hamilton's principle function. We can either expand S as a perturbative power series in the parameters $\sqrt{\hbar}$ or $(1/c)$ and obtain the semi-classical and non-relativistic limit respectively at various orders. The scheme for non-relativistic limit has been employed by C.Kiefer and TP Singh \cite{kiefer_Singh_1991}. Guillini and Grobsardt in their works \cite{guilini_grosardt}, combines both these schemes and constructs a new ansatz in the parameter $\frac{\sqrt{\hbar}}{c}$ as follows:
	
	\begin{equation}\label{eq:Spinor_ansatz_NRlimit}
		\psi(\textbf{r},t) = e^{\frac{ic^2}{\hbar}S(\textbf{r},t)}\sum_{n=0}^{\infty}\bigg(\frac{\sqrt{\hbar}}{c}\bigg)^n a_n(\textbf{r},t)
	\end{equation}
	where $S(\textbf{r},t)$ is a scalar function and $a_n(\textbf{r},t)$ is a spinor field. We use this ansatz in our calculations. 
	
	\item \underline{Ansatz for metric:}
	We first express the generic form of the metric in a power series with parameter same as that used to expand spinor viz. $\frac{\sqrt{\hbar}}{c}$.
	
	\begin{equation} \label{eq:metric_ansatz_NRlimit_generic}
		g_{\mu\nu}(\vec{X}, t) = \eta_{\mu\nu} + \sum_{n = 1}^{\infty} \bigg{(}\frac{\sqrt{\hbar}}{c} \bigg{)}^n g_{\mu\nu}^{[n]}(\vec{X}, t)
	\end{equation}
	
	where, $g_{\mu\nu}^{(n)}(x)$ are infinite metric functions indexed by 'n'. In non-relativistic scheme, gravitational potentials can't produce velocities comparable to c. they are weak potentials. $\therefore$ we have assumed that the leading function $g_{\mu\nu}^{[0]}(x) = \eta_{\mu\nu}$. With this, we get the following generic power series for tetrads and spin coefficients and Einstein tensor. 
	
	\begin{align}
		e^{\mu}_{(i)} &=  \delta^{\mu}_{(i)} + \sum_{n = 1}^{\infty} \bigg{(}\frac{\sqrt{\hbar}}{c} \bigg{)}^n e^{\mu[n]}_{(i)} ~~~~~~~~~~~~~~~~~~~~~~ \gamma_{(a)(b)(c)} = \sum_{n = 1}^{\infty} \bigg{(}\frac{\sqrt{\hbar}}{c} \bigg{)}^n \gamma^{[n]}_{(a)(b)(c)} \label{eq:geeric expansions1}\\
		e^{(i)}_{\mu} &=\delta^{(i)}_{\mu} + \sum_{n = 1}^{\infty} \bigg{(}\frac{\sqrt{\hbar}}{c} \bigg{)}^n e_{\mu}^{(i)[n]} ~~~~~~~~~~~~~~~~~~~~~~ G_{\mu\nu} = \sum_{n = 1}^{\infty} \bigg{(}\frac{\sqrt{\hbar}}{c} \bigg{)}^n G^{[n]}_{\mu\nu} \label{eq:geeric expansions2}
	\end{align}
	
	where $e^{\mu(n)}_{(i)} [g_{\mu\nu}^{[n]}]$, $e_{\mu}^{(i)[n]} [g_{\mu\nu}^{[n]}]$, $\gamma^{[n]}_{(a)(b)(c)}[g_{\mu\nu}^{[n]}]$ and $G^{[n]}_{\mu\nu}$  are infinite tetrad, spin coefficient and Einstein tensor functions indexed by 'n'. They are functions of metric functions $g_{\mu\nu}^{[n]}$ and their various derivatives.
	
\end{itemize}

\section{Analysis of Einstein-Dirac system with our Ansatz}
Einstein-Dirac system is the self-gravitating Dirac field on Riemann manifold [$V_4$].
\subsection{Analyzing Dirac equation with our Ansatz}
We will now evaluate Dirac equation on $V_4$ as given in eqn (\ref{eq:DE V4}) with these Ansatz. We also note that $\gamma^{(a)}\psi_{;(a)}= e^{(a)}_{\mu} e^{\nu}_{(a)} \gamma^{\mu}\psi_{;\nu} = \delta^{\mu}_{\nu}\gamma^{\mu}\psi_{;\nu} = \gamma^{\mu}\psi_{;\mu}$.

\begin{align}\label{eq:DE}
	&i\gamma^{\mu}\psi_{;\mu} - \frac{mc}{\hbar} \psi = 0 \\
	\Rightarrow ~~ & i\gamma^{0}\partial_{0}\psi + \frac{i}{4}\gamma^{(0)}\gamma^o_{(0)(b)(c)}\gamma^{[(b)}\gamma^{(c)]}\psi + i\gamma^{\alpha}\partial_{\alpha}\psi + \frac{i}{4}\gamma^{(j)}\gamma^o_{(j) (b)(c)}\gamma^{[(b)}\gamma^{(c)]}\psi - \frac{mc}{\hbar}\psi = 0  \label{Ieq:4.1}
\end{align}
We separate spatial and temporal parts. Substituting appropriate expansions from (\ref{eq:geeric expansions1}), (\ref{eq:geeric expansions2}) into above equations and multiplying by $\gamma^{(0)}c$ on both sides yields:

\begin{align} \label{Ieq:4.2}
	\begin{split}
		\Rightarrow ~~ & \bigg{[} 1+ \sum_{n = 1}^{\infty} \bigg{(}\frac{\sqrt{\hbar}}{c} \bigg{)}^n e^{0[n]}_{(0)}\bigg{]} i\partial_{t}\psi + \frac{ic}{4}\bigg{[}\sum_{n = 1}^{\infty} \bigg{(}\frac{\sqrt{\hbar}}{c} \bigg{)}^n \gamma^{o[n]}_{(0)(b)(c)}\bigg{]}\gamma^{[(b)}\gamma^{(c)]}\psi + \\ 
		& \bigg{[} 1+ \sum_{n = 1}^{\infty} \bigg{(}\frac{\sqrt{\hbar}}{c} \bigg{)}^n e^{\alpha[n]}_{(a)}\bigg{]}ic\vec{\alpha}.\nabla\psi + \frac{ic}{4}\alpha^{(j)} \bigg{[}\sum_{n = 1}^{\infty} \bigg{(}\frac{\sqrt{\hbar}}{c} \bigg{)}^n \gamma^{o[n]}_{(j)(b)(c)}\bigg{]}\gamma^{[(b)}\gamma^{(c)]}\psi - \frac{\beta mc^2}{\hbar}\psi = 0 \\
	\end{split}
\end{align}

Dividing both sides by $\bigg{[} 1+ \sum_{n = 1}^{\infty} \bigg{(}\frac{\sqrt{\hbar}}{c} \bigg{)}^n e^{0[n]}_{(0)}\bigg{]}$, we obtain

\begin{align}\label{Ieq:4.3}
	\begin{split}
		i\partial_{t}\psi &= - \frac{ic}{4}\frac{\bigg{[}\sum_{n = 1}^{\infty} \bigg{(}\frac{\sqrt{\hbar}}{c} \bigg{)}^n \gamma^{o[n]}_{(0)(b)(c)}\bigg{]}}{\bigg{[} 1+ \sum_{n = 1}^{\infty} \bigg{(}\frac{\sqrt{\hbar}}{c} \bigg{)}^n e^{0[n]}_{(0)}\bigg{]}}\gamma^{[(b)}\gamma^{(c)]}\psi -  \frac{\bigg{[} 1+ \sum_{n = 1}^{\infty} \bigg{(}\frac{\sqrt{\hbar}}{c} \bigg{)}^n e^{\alpha[n]}_{(a)}\bigg{]}}{\bigg{[} 1+ \sum_{n = 1}^{\infty} \bigg{(}\frac{\sqrt{\hbar}}{c} \bigg{)}^n e^{0[n]}_{(0)}\bigg{]}}ic\vec{\alpha}.\nabla\psi - \\ 
		& \frac{ic}{4}\alpha^{(j)} \frac{\bigg{[}\sum_{n = 1}^{\infty} \bigg{(}\frac{\sqrt{\hbar}}{c} \bigg{)}^n \gamma^{o[n]}_{(j)(b)(c)}\bigg{]}}{\bigg{[} 1+ \sum_{n = 1}^{\infty} \bigg{(}\frac{\sqrt{\hbar}}{c} \bigg{)}^n e^{0[n]}_{(0)}\bigg{]}}\gamma^{[(b)}\gamma^{(c)]}\psi + \frac{1}{\bigg{[} 1+ \sum_{n = 1}^{\infty} \bigg{(}\frac{\sqrt{\hbar}}{c} \bigg{)}^n e^{0[n]}_{(0)}\bigg{]}}\frac{\beta mc^2}{\hbar}\psi 
	\end{split}
\end{align}
We consider the terms of order $c^2$,c,1 and neglect the terms having order of $O\Big(\frac{1}{c^n}\Big)$; n$\geq$1. This is sufficient to get the behavior of some functions in spinor Ansatz. It will turn out later that this is also sufficient to get the equation which is followed by leading order spinor term $a_0$.  We obtain following equation:

\begin{align} \label{eq:DE with generic metric ansatz}
	\begin{split}
		i\partial_{t}\psi & + \frac{i\sqrt{\hbar}}{4}\gamma^{o[1]}_{(0)(b)(c)}\gamma^{[(b)}\gamma^{(c)]}\psi + ~ ic\vec{\alpha}.\nabla\psi + \frac{i\sqrt{\hbar}}{4}\alpha^{(j)}\gamma^{o[1]}_{(j)(b)(c)}\gamma^{[(b)}\gamma^{(c)]}\psi  \\ - & \beta\frac{ mc^2}{\hbar}\psi  + \beta \frac{mc}{\sqrt{\hbar}}e^{0[1]}_{(0)}\psi - \beta m \bigg{[}\bigg{(}e^{0[1]}_{(0)}\bigg{)}^2 - e^{0[2]}_{(0)} \bigg{]} \psi = 0
	\end{split}
\end{align}

Substituting the Spinor Ansatz i.e. eqn (\ref{eq:Spinor_ansatz_NRlimit})in equation (\ref{eq:DE with generic metric ansatz}), the various terms are evaluated as follows: \\

\textbf{Term 1}
\begin{align}\label{eq:term1}
	i\partial_t\psi &= i\partial_t\Big[e^{\frac{ic^2S}{\hbar}}\sum_{n=0}^{\infty}\bigg(\frac{\sqrt{\hbar}}{c}\bigg)^n a_n\Big]\nonumber \\
	&= ie^{\frac{ic^2S}{\hbar}}\frac{c^2}{\hbar}\sum_{n=0}^{\infty}\bigg(\frac{\sqrt{\hbar}}{c}\bigg)^n\Big[\dot{a}_{n-2} + i\dot{S}a_n\Big]\nonumber \\
	&=e^{\frac{ic^2S}{\hbar}}\frac{c^3}{\hbar^{3/2}}\sum_{n=0}^{\infty}\bigg(\frac{\sqrt{\hbar}}{c}\bigg)^n\Big[-\dot{S}a_{n-1} + i\dot{a}_{n-3}\Big]
\end{align}

\textbf{Term 2}
\begin{align}\label{eq:term2}
	+\frac{i\sqrt{\hbar}}{4}\gamma^{o[1]}_{(0)(b)(c)}\gamma^{[(b)}\gamma^{(c)]}\psi &=
	+\frac{i\sqrt{\hbar}}{4}\gamma^{o[1]}_{(0)(b)(c)}\gamma^{[(b)}\gamma^{(c)]}\Big[e^{\frac{ic^2S}{\hbar}}\sum_{n=0}^{\infty}\bigg(\frac{\sqrt{\hbar}}{c}\bigg)^n a_n\Big]\\
	& =e^{\frac{ic^2S}{\hbar}}\frac{c^3}{\hbar^{3/2}}\sum_{n=0}^{\infty}\bigg(\frac{\sqrt{\hbar}}{c}\bigg)^n\Big[i\sqrt{\hbar} \gamma^{o[1]}_{(0)(b)(c)}\gamma^{[(b)}\gamma^{(c)]} a_{n-3}\Big]
\end{align}

\textbf{Term 3}
\begin{align}\label{eq:term3}
	ic\alpha^j\partial_j\psi &= ic\overrightarrow{\alpha}\cdot\overrightarrow{\nabla}\Big[e^{\frac{ic^2S}{\hbar}}\sum_{n=0}^{\infty}\bigg(\frac{\sqrt{\hbar}}{c}\bigg)^n a_n\Big]\nonumber \\
	&=ic\overrightarrow{\alpha}\cdot\Big[e^{\frac{ic^2S}{\hbar}}\frac{c^2}{\hbar}\sum_{n=0}^{\infty}\bigg(\frac{\sqrt{\hbar}}{c}\bigg)^n\Big(i\overrightarrow{\nabla}S a_n + \overrightarrow{\nabla}a_{n-2}\Big)\Big]\nonumber \\
	&=e^{\frac{ic^2S}{\hbar}}\frac{c^3}{\hbar^{3/2}}\sum_{n=0}^{\infty}\bigg(\frac{\sqrt{\hbar}}{c}\bigg)^n\Big[-\sqrt{\hbar}\overrightarrow{\alpha}\cdot\overrightarrow{\nabla}Sa_n + i\sqrt{\hbar}\overrightarrow{\alpha}\cdot\overrightarrow{\nabla}a_{n-2}\Big]
\end{align}

\textbf{Term 4}

\begin{align}\label{eq:term4}
	+\frac{i\sqrt{\hbar}}{4}\alpha^{(j)}\gamma^{o[1]}_{(j)(b)(c)}\gamma^{[(b)}\gamma^{(c)]}\psi &=
	+\frac{i\sqrt{\hbar}}{4}\alpha^{(j)}\gamma^{o[1]}_{(j)(b)(c)}\gamma^{[(b)}\gamma^{(c)]}\Big[e^{\frac{ic^2S}{\hbar}}\sum_{n=0}^{\infty}\bigg(\frac{\sqrt{\hbar}}{c}\bigg)^n a_n\Big]\\
	& =e^{\frac{ic^2S}{\hbar}}\frac{c^3}{\hbar^{3/2}}\sum_{n=0}^{\infty}\bigg(\frac{\sqrt{\hbar}}{c}\bigg)^n\Big[i\sqrt{\hbar} \alpha^{(j)}\gamma^{o[1]}_{(j)(b)(c)}\gamma^{[(b)}\gamma^{(c)]} a_{n-3}\Big]
\end{align}

\textbf{Term 5}
\begin{align}\label{eq:term5}
	-\beta\frac{mc^2}{\hbar}\psi &= -\beta\frac{mc^2}{\hbar}e^{\frac{ic^2S}{\hbar}}\sum_{n=0}^{\infty}\bigg(\frac{\sqrt{\hbar}}{c}\bigg)^n a_n \nonumber \\
	&= e^{\frac{ic^2S}{\hbar}}\frac{c^3}{\hbar^{3/2}}\sum_{n=0}^{\infty}\bigg(\frac{\sqrt{\hbar}}{c}\bigg)^n (-\beta m a_{n-1})
\end{align}

\textbf{Term 6}
\begin{align}\label{eq:term6}
	+ \beta \frac{mc}{\sqrt{\hbar}}e^{0[1]}_{(0)}\psi &= + \beta \frac{mc}{\sqrt{\hbar}}e^{0[1]}_{(0)}\Big[e^{\frac{ic^2S}{\hbar}}\sum_{n=0}^{\infty}\bigg(\frac{\sqrt{\hbar}}{c}\bigg)^n a_n\Big]\\ &= e^{\frac{ic^2S}{\hbar}}\frac{c^3}{\hbar^{3/2}}\sum_{n=0}^{\infty}\bigg(\frac{\sqrt{\hbar}}{c}\bigg)^n \Big[\beta m ~ e^{0[1]}_{(0)} a_{n-2}\Big]
\end{align}

\textbf{Term 7}

\begin{align}\label{eq:term7}
	- \beta m \bigg{[}\bigg{(}e^{0[1]}_{(0)}\bigg{)}^2 - e^{0[2]}_{(0)} \bigg{]} \psi &=- \beta m \bigg{[}\bigg{(}e^{0[1]}_{(0)}\bigg{)}^2 - e^{0[2]}_{(0)} \bigg{]}  \Big[e^{\frac{ic^2S}{\hbar}}\sum_{n=0}^{\infty}\bigg(\frac{\sqrt{\hbar}}{c}\bigg)^n a_n\Big] \\ &=
	-e^{\frac{ic^2S}{\hbar}}\frac{c^3}{\hbar^{3/2}}\sum_{n=0}^{\infty}\bigg(\frac{\sqrt{\hbar}}{c}\bigg)^n \Bigg[\beta m \Bigg( \big(e^{0[1]}_{(0)}\big)^2 - e^{0[2]}_{(0)}  \Bigg)\Bigg] a_{n-3}
\end{align}

After substituting equations (\ref{eq:term1}), (\ref{eq:term2}), (\ref{eq:term3}),(\ref{eq:term4}),(\ref{eq:term5}),(\ref{eq:term6}) and (\ref{eq:term7}) into (\ref{eq:DE with generic metric ansatz}) and sorting by powers of $n$ we get,

\begin{equation}\label{eq:DE_generic_after substituting spinor, metric ansatz}
	\begin{split}
		e^{\frac{ic^2S}{\hbar}}\frac{c^3}{\hbar^{3/2}}\sum_{n=0}^{\infty}\bigg(\frac{\sqrt{\hbar}}{c}\bigg)^n &
		\Bigg[\Big(-\sqrt{\hbar}\overrightarrow{\alpha}\cdot\overrightarrow{\nabla}S \Big) a_n - \Big(\dot{S} + \beta m  \Big)a_{n-1} + \Big(i\sqrt{\hbar}\overrightarrow{\alpha}\cdot\overrightarrow{\nabla}  + \beta m ~ e^{0[1]}_{(0)} \Big)a_{n-2} \\ + i\dot{a}_{n-3} + \Bigg( i\sqrt{\hbar} &\gamma^{o[1]}_{(0)(b)(c)}\gamma^{[(b)}\gamma^{(c)]} + i\sqrt{\hbar} \alpha^{(j)}\gamma^{o[1]}_{(j)(b)(c)}\gamma^{[(b)}\gamma^{(c)]} - \beta m \Big( \big(e^{0[1]}_{(0)}\big)^2 - e^{0[2]}_{(0)} \Big) \Bigg)a_{n-3} \Bigg] = 0
	\end{split}
\end{equation}
At order $n=0$ the equation reduces to,
\begin{equation}\label{eq:nablaS=0}
	\overrightarrow{\nabla}S = 0
\end{equation}
which implies the scalar 'S'is a function of time only i.e., $S = S(t)$.
Dirac spinor is a 4 component spinor $a_n = (a_{n,1},a_{n,2},a_{n,3},a_{n,4})$. We split it into two two-component spinors $a_n^> = (a_{n,1}, a_{n,2})$ and $a_n^< = (a_{n,3}, a_{n,4})$. For order $n = 1$, the equation is $\Big(\dot{S} + \beta m  \Big) = 0$; which can be written as following 2 equations:
\begin{subequations}
	\begin{align}
		(m + \dot{S})a_0^> &= 0\label{48a} \\
		(m - \dot{S})a_0^< &= 0\label{48b}
	\end{align}
\end{subequations}
This implies that either $S = -mt$ and $a_0^< = 0$ or $S = +mt$ and $a_0^> = 0$.\\
We will consider the former i.e. $S = -mt$ and $a_0^< = 0$, which represents positive energy solutions. We stop at this point and analyze dynamical EM tensor now with the results obtained in equation (\ref{eq:nablaS=0}) and the fact that $a_0^< = 0$. 

\subsection{Analyzing the Energy momentum tensor $T_{ij}$ with our Ansatz}

The dynamical Energy momentum tensor given in equation (\ref{dynamic EM tensor}). Lets consider the "k$T_{00}$" component.\\ 
\underline{\textbf{Analyzing $k T_{00}$}}:
\begin{equation} \label{kT_00_generic}
	kT_{00} = \frac{4i\pi G\hbar}{c^4}\Bigg[\bar{\psi}\gamma^0 \Big(\partial_t\psi + \frac{c}{4}[\gamma^o_{0(i)(j)}\gamma^{[(i)}\gamma^{(j)]}]\psi \Big)- \Big(\partial_t\bar{\psi} + \frac{c}{4}[{\gamma}^o_{0(i)(j)}\gamma^{[(i)}\gamma^{(j)]}]\bar{\psi} \Big)\gamma^0\psi\Bigg]
\end{equation}
\begin{align}
	\begin{split}
		\Rightarrow kT_{00} = \frac{4i\pi G\hbar}{c^4}\Big( 1+ \sum_{n = 1}^{\infty} \big(\frac{\sqrt{\hbar}}{c} \big)^n e^{0[n]}_{(0)}\Big) &\Bigg[\bar{\psi}\gamma^{(0)} \Big(\partial_t\psi + \frac{c}{4}[\gamma^o_{0(i)(j)}\gamma^{[(i)}\gamma^{(j)]}]\psi \Big)\\ &-\Big(\partial_t\bar{\psi} + \frac{c}{4}[{\gamma}^o_{0(i)(j)}\gamma^{[(i)}\gamma^{(j)]}]\bar{\psi} \Big)\gamma^{(0)}\psi\Bigg]
	\end{split}
\end{align}
After putting spinor anstaz eqn (\ref{eq:Spinor_ansatz_NRlimit}) in eqn (\ref{kT_00_generic}), we obtain following power series for k$T_{00}$. We have given expression for the leading order only.

\begin{equation}
	\begin{split}
		kT_{00}  = \frac{4i\pi G}{c^2}\bigg\{\bigg(\sum_{n=0}^{\infty}\bigg(\frac{\sqrt{\hbar}}{c}\bigg)^n a^\dagger_{n}\bigg)\bigg(\sum_{m=0}^{\infty}\bigg(\frac{\sqrt{\hbar}}{c}\bigg)^m\Big[ i\dot{S}a_m + \dot{a}_{m-2} \Big]\bigg)\\ + \bigg(\sum_{n=0}^{\infty}\bigg(\frac{\sqrt{\hbar}}{c}\bigg)^n\Big[ i\dot{S}a^{\dagger}_{n} - \dot{a}^{\dagger}_{n-2} \Big]\bigg)\bigg(\sum_{n=0}^{\infty}\bigg(\frac{\sqrt{\hbar}}{c}\bigg)^m a_{m}\bigg)  \bigg\} + \sum_{n=3}^{\infty} O\Big( \frac{1}{c^n}\Big)
	\end{split}
\end{equation}
Explicit expression for leading order is obtained by considering $(n+m = 0)$ as follows:

\begin{align}
	kT_{00}&= \frac{4\pi G i}{c^2}\bigg\{i(-m)a_{0}^{>\dagger}a_{0}^{>} + i(-m)a_{0}^{>\dagger}a_{0}^{>}\bigg\} + \sum_{n=3}^{\infty} O\Big( \frac{1}{c^n}\Big)\\
	kT_{00}&= \frac{8\pi G m\, |a_{0}^{>}|^2}{c^2} + \sum_{n=3}^{\infty} O\Big( \frac{1}{c^n}\Big) \label{result:T_00}
\end{align}
\\
\underline{\textbf{Analyzing $k T_{0\mu}$}}:
\begin{equation} \label{kT_0a_generic}
	\begin{split}
		kT_{0\mu} = \frac{2i\pi G\hbar}{c^4}&\Bigg[
		c\bar{\psi}\gamma^0\Big(\partial_{\mu}\psi + \frac{1}{4}[\gamma^o_{\mu(i)(j)}\gamma^{[(i)}\gamma^{(j)]} \psi\Big) - c\bar{\psi}\gamma^{\mu} \Big(\partial_0\psi + \frac{1}{4}[\gamma^o_{0(i)(j)}\gamma^{[(i)}\gamma^{(j)]} \psi\Big) \\ &-c\Big(\partial_{\mu}\bar{\psi} + \frac{1}{4}[{\gamma}^o_{\mu(i)(j)}\gamma^{[(i)}\gamma^{(j)]} \bar{\psi}\Big)\gamma^0\psi + c\Big(\partial_0\bar{\psi} + \frac{1}{4}[{\gamma}^o_{0(i)(j)}\gamma^{[(i)}\gamma^{(j)]} \bar{\psi}\Big)\gamma^{\mu}\psi \Bigg]
	\end{split}
\end{equation}
We will first find the coefficient of the term of order $\frac{1}{c^2}$ which is the leading order of $T_{00}$. Now, all the terms containing spin coefficients $\gamma_{\mu(i)(j)}$ have leading order of  $\frac{1}{c^3}$. So it won't contribute at  the order $\frac{1}{c^2}$. So what we get is:

\begin{align} \label{kT_0mu_generic}
	kT_{0\mu} &= \frac{2i\pi G\hbar}{c^4}\Bigg[
	c\bar{\psi}\gamma^0\partial_{\mu}\psi - c\bar{\psi}\gamma^{\mu} \partial_0\psi -c\partial_{\mu}\bar{\psi}\gamma^0\psi + c\partial_0\bar{\psi}\gamma^{\mu}\psi \Bigg] \\
	&= \frac{-2i\pi G\hbar}{c^3}\Big( 1+ \sum_{n = 1}^{\infty} \bigg{(}\frac{\sqrt{\hbar}}{c} \bigg{)}^n e^{0[n]}_{(0)}\Big) \Bigg[\bar{\psi}\gamma^{(0)}\partial_{\mu}\psi  -\partial_{\mu}\bar{\psi}\gamma^{(0)}\psi \Bigg]  \\ 
	& + \frac{2i\pi G\hbar}{c^4}\Big( 1+ \sum_{n = 1}^{\infty} \bigg{(}\frac{\sqrt{\hbar}}{c} \bigg{)}^n e^{\mu[n]}_{(a)}\Big)\Bigg[ \partial_t\bar{\psi}\gamma^{(a)}\psi -\bar{\psi}\gamma^{(a)} \partial_t\psi \Bigg] \nonumber
\end{align}
There are 2 types of terms in equation above. One having coefficient $\frac{2i\pi G\hbar}{c^3}$ and other with coefficient $\frac{2i\pi G\hbar}{c^4}$. We call them term 1 and 2 respectively. We analyze both of them independently. Term 1 gives

\begin{align}
	(term~1) &= \frac{2i\pi G\hbar}{c^3}\sum_{n = 0}^{\infty} \bigg{(}\frac{\sqrt{\hbar}}{c} \bigg{)}^n\Big( a_{n_1}^{\dagger} \partial_{\mu}a_{n_2} - \partial_{\mu}a_{n_1}^{\dagger} a_{n_2}\Big) ; ~~~~ n=n_1+n_2 \nonumber\\
	&= \sum_{n=3}^{\infty} O\Big( \frac{1}{c^n}\Big) \\
	\begin{split}
		(term~2) &= \frac{2i\pi G}{c^2}\bigg\{\bigg(\sum_{n=0}^{\infty}\bigg(\frac{\sqrt{\hbar}}{c}\bigg)^n a^\dagger_{n}\bigg)\alpha^{(a)}\bigg(\sum_{m=0}^{\infty}\bigg(\frac{\sqrt{\hbar}}{c}\bigg)^m\Big[ i\dot{S}a_m + \dot{a}_{m-2} \Big]\bigg)\\ &+ \bigg(\sum_{n=0}^{\infty}\bigg(\frac{\sqrt{\hbar}}{c}\bigg)^n\Big[ i\dot{S}a^{\dagger}_{n} - \dot{a}^{\dagger}_{n-2} \Big]\bigg)\alpha^{(a)}\bigg(\sum_{n=0}^{\infty}\bigg(\frac{\sqrt{\hbar}}{c}\bigg)^m a_{m}\bigg)  \bigg\} + \sum_{n=3}^{\infty} O\Big( \frac{1}{c^n}\Big)
	\end{split} \nonumber \\
	&= \frac{4\pi Gm}{c^2}(a_0^{\dagger}\alpha^{(a)}a_0) + \sum_{n=3}^{\infty} O\Big( \frac{1}{c^n}\Big) \nonumber\\
	&= \frac{4\pi Gm}{c^2}\Bigg[ \begin{pmatrix} a_0^> & 0 \end{pmatrix}^{\dagger} \begin{pmatrix}0&\sigma^{(a)} \\\sigma^{(a)} &0   \end{pmatrix}\begin{pmatrix} a_0^> \\ 0 \end{pmatrix} \bigg]   + \sum_{n=3}^{\infty} O\Big( \frac{1}{c^n}\Big) \nonumber\\
	& =\sum_{n=3}^{\infty} O\Big( \frac{1}{c^n}\Big) 
\end{align}
So we find, in both term 1 and 2, terms of the $O\Big( \frac{1}{c^2}\Big)$ are ZERO. Hence

\begin{equation}\label{result:T_0 mu}
	kT_{0\mu} = \sum_{n=3}^{\infty} O\Big( \frac{1}{c^n}\Big)
\end{equation}
\\
\underline{\textbf{Analyzing $k T_{\mu\nu}$}}
\begin{equation} \label{kT_mu nu_generic1}
	\begin{split}
		kT_{\mu\nu} = \frac{2i\pi G\hbar}{c^3}&\Bigg[
		-\bar{\psi}\gamma^{\mu}\Big(\partial_{\nu}\psi + \frac{1}{4}[\gamma^o_{\nu(i)(j)}\gamma^{[(i)}\gamma^{(j)]} \psi\Big) - \bar{\psi}\gamma^{\nu} \Big(\partial_{\mu}\psi + \frac{1}{4}[\gamma^o_{\mu(i)(j)}\gamma^{[(i)}\gamma^{(j)]} \psi\Big) \\ &+\Big(\partial_{\nu}\bar{\psi} + \frac{1}{4}[{\gamma}^o_{\nu(i)(j)}\gamma^{[(i)}\gamma^{(j)]} \bar{\psi}\Big)\gamma^{\mu}\psi + \Big(\partial_{\mu}\bar{\psi} + \frac{1}{4}[{\gamma}^o_{\mu(i)(j)}\gamma^{[(i)}\gamma^{(j)]} \bar{\psi}\Big)\gamma^{\nu}\psi \Bigg]
	\end{split}
\end{equation}
Here also, we will first find the coefficient of the term of order $\frac{1}{c^2}$ which is the leading order of $kT_{00}$. All the terms containing spin coefficients $\gamma_{\mu(i)(j)}$ have leading order of  $\frac{1}{c^3}$. So it won't contribute at  the order $\frac{1}{c^2}$. So what we get is:

\begin{align} \label{kT_mu nu_generic2}
	kT_{\mu\nu} &= \frac{2i\pi G\hbar}{c^3}\Bigg[
	-\bar{\psi}\gamma^{\mu}\partial_{\nu}\psi - \bar{\psi}\gamma^{\nu} \partial_{\mu}\psi +\partial_{\nu}\bar{\psi}\gamma^{\mu}\psi + \partial_{\mu}\bar{\psi}\gamma^{\nu}\psi \Bigg] \nonumber\\
	&= \frac{2i\pi G\hbar}{c^3}\Big( 1+ \sum_{n = 1}^{\infty} \bigg{(}\frac{\sqrt{\hbar}}{c} \bigg{)}^n e^{\mu[n]}_{(a)}\Big) \Bigg[\psi^{\dagger}\alpha^{(a)}\partial_{\nu}\psi  -\partial_{\nu}\psi^{\dagger}\alpha^{(a)}\psi \Bigg] \nonumber \\ 
	& + \frac{2i\pi G\hbar}{c^3}\Big( 1+ \sum_{n = 1}^{\infty} \bigg{(}\frac{\sqrt{\hbar}}{c} \bigg{)}^n e^{\nu[n]}_{(b)}\Big)\Bigg[ \partial_{\mu}\psi^{\dagger}\alpha^{(b)}\psi -\psi^{\dagger}\alpha^{(b)} \partial_{\mu}\psi \Bigg] \nonumber \\
	&= \frac{2i\pi G\hbar}{c^3}\sum_{n = 0}^{\infty} \bigg{(}\frac{\sqrt{\hbar}}{c} \bigg{)}^n\Big( a_{n_1}^{\dagger}\alpha^{(a)} \partial_{\nu}a_{n_2} - \partial_{\nu}a_{n_1}^{\dagger}\alpha^{(a)} + a_{n_1}^{\dagger}\alpha^{(b)} \partial_{\mu}a_{n_2} - \partial_{\mu}a_{n_1}^{\dagger}\alpha^{(b)} a_{n_2}\Big)\\
	kT_{\mu\nu}& =\sum_{n=3}^{\infty} O\Big( \frac{1}{c^n}\Big) \label{result:T_mu nu}
\end{align}

From order analysis of components of EM tensor, summarized in equations (\ref{result:T_00}) , (\ref{result:T_0 mu}) and (\ref{result:T_mu nu}), we proved a very crucial result here viz.
\begin{align}
	\frac{|T_{00}|}{|T_{0i}|} &\ll 1, ~~~~~~ \frac{|T_{00}|}{|T_{ij}|} \ll 1,~~~~~~ k|T_{00}| \sim O\Big(\frac{1}{c^2}\Big)~~~~; i,j \in (1,2,3) 
\end{align}
Owing to Einstein's equations, the same relation exists amongst the components of Einstein Tensor as well viz.
\begin{align}
	\frac{|G_{00}|}{|G_{0i}|} &\ll 1, ~~~~~~ \frac{|G_{00}|}{|G_{ij}|} \ll 1,~~~~~~ |G_{00}| \sim O\Big(\frac{1}{c^2}\Big)~~~~; i,j \in (1,2,3) 
\end{align}

\subsection{Constraints imposed on metric as an implication of this section}
We proved a very important fact from previous 2 sections viz. $|G_{00}| \sim O\Big(\frac{1}{c^2}\Big)$ and all other components of G are higher orders. For a generic metric ansatz, $G_{\mu\nu}$ has been explicitly calculated in Appendix [\ref{NR:generic G}]. \textbf{At this point, we make an important assumption -- Metric field is asymptotically flat}. Which means we cannot allow for non-trivial solution to the equations like $\square g_{\mu\nu}^{[1]} = 0$ and $\square g_{\mu\nu}^{[2]} = 0$. Because allowing for non-trivial solution to such equations (which are basically gravitational wave solutions) would contradict the assumption of asymptotic flatness of metric. This fact \textbf{suggests the following important constraints on metric components.} \\
1) $G_{\mu\nu}^{[1]} = 0$ ($\forall \mu,\nu$) and non-allowance of solutions which don't respect asymptotic flatness of metric gives:
\begin{align}
	g_{\mu\nu}^{[1]} = 0, ~~~~~~ e^{\mu[1]}_{(i)} = 0, ~~~~~~~ e^{(i)[1]}_{\mu} = 0, ~~~~~~ \gamma_{(i)(j)(k)}^{[1]} = 0 ~~~~~ \forall ~~ij,k,\mu,\nu \in(0,1,2,3)
\end{align}
2) We also have $G_{\mu\nu}^{[2]} = 0$ (\textbf{except for $\mu= 0$ and $\nu = 0$}). This imposes different kind of restrictions on $g_{\mu\nu}^{[2]}$. From Appendix [\ref{NR:generic G}], we see that the form which $g_{\mu\nu}^{[2]}$ can take is $g_{\mu\nu}^{[2]} = F(\vec{x},t) \delta_{\mu\nu}$ for some field $F(\vec{x},t)$. Here also, we respect asymptotic flatness of metric. The full metric is then given by:

\begin{equation}\label{eq:metric_ansatz_NRlimit_special}
	g_{\mu\nu}(\vec{x}, t) = \begin{bmatrix} 1 & 0 & 0 & 0  \\ 0 & -1 & 0 & 0 \\ 0 & 0 &  -1& 0 \\ 0 & 0 & 0 &-1 \end{bmatrix} + \bigg{(}\frac{\hbar}{c^2}\bigg{)}\begin{bmatrix} F  & 0 & 0 & 0  \\ 0 & F  & 0 & 0 \\ 0 & 0 & F & 0 \\ 0 & 0 & 0 & F \end{bmatrix} (\vec{x},t)+ \sum_{n = 3}^{\infty} \bigg{(}\frac{\sqrt{\hbar}}{c} \bigg{)}^n \begin{bmatrix} g^{[n]}_{00}  & g^{[n]}_{01} & g^{[n]}_{02} &g^{[n]}_{03}  \\ g^{[n]}_{10} & g^{[n]}_{11}& g^{[n]}_{12} &g^{[n]}_{13} \\ g^{[n]}_{20} &g^{[n]}_{21} & g^{[n]}_{22}&g^{[n]}_{23} \\ g^{[n]}_{30}& g^{[n]}_{31} &g^{[n]}_{32} & g^{[n]}_{33} \end{bmatrix}(\vec{x},t)
\end{equation}

where $g^{[2]}_{00} = g^{[2]}_{11} = g^{[2]}_{22} = g^{[2]}_{33} = F(\vec{x},t)$

With this form of metric, all the other objects (tetrads, spin coefficients etc.)
have been calculated in Appendix sections [\ref{NR:metric}], [\ref{NR:spin connection}], [\ref{NR:tetrad}] and [\ref{NR:G}]. We have used these results in the next section. 
\section{Non-Relativistic (NR) limit of ECD field equations with standard length scale}

\subsection{NR limit of Einstein-Dirac system}\label{NRlimit_ED}
\underline{\textbf{Dirac equation}}\\
Equation (\ref{eq:DE_generic_after substituting spinor, metric ansatz}) becomes following 

\begin{align}\label{eq:DE_specific_after substituting spinor, metric ansatz}
	e^{\frac{ic^2S}{\hbar}}\frac{c^3}{\hbar^{3/2}}\sum_{n=0}^{\infty}\bigg(\frac{\sqrt{\hbar}}{c}\bigg)^n \Big[&m~a_{n-1} + i\dot{a}_{n-3} + i\sqrt{\hbar}\overrightarrow{\alpha}\cdot\overrightarrow{\nabla}a_{n-2} - \beta m a_{n-1} - \beta\frac{mF(\vec{x},t)}{2}a_{n-3}\Big] = 0
\end{align}
We have already used the results from analysis of this equation for n=0 and n=1. We now analyze it for n=2 and n=3.
At order $n = 2$ the equation (\ref{eq:DE_generic_after substituting spinor, metric ansatz}) results in,
\begin{align}\label{49}
	\left (
	\begin{tabular}{cc}
		$\dot{S}$ + m & 0 \\
		0 & $\dot{S}$ - m
	\end{tabular}
	\right )
	\left (
	\begin{array}{c}
		a_1^> \\
		a_1^<
	\end{array}
	\right )
	- i\sqrt{\hbar} \left (
	\begin{tabular}{cc}
		0 & $\overrightarrow{\sigma}\cdot\overrightarrow{\nabla}$ \\
		$\overrightarrow{\sigma}\cdot\overrightarrow{\nabla}$ & 0 
	\end{tabular}
	\right )
	\left (
	\begin{array}{c}
		a_0^> \\
		a_0^<
	\end{array}
	\right ) = 0
\end{align}
The first of these is trivially satisfied. The second one yields an expression for $a_1^<$ in terms of $a_0^>$,

\begin{equation}\label{50}
	a_1^< = \frac{-i\sqrt{\hbar}\overrightarrow{\sigma}\cdot\overrightarrow{\nabla}}{2m} a_0^>
\end{equation}
At order $n = 3$,
\begin{align}\label{51}
	\left (
	\begin{tabular}{cc}
		$\dot{S}$ + m & 0 \\
		0 & $\dot{S}$ - m
	\end{tabular}
	\right )
	\left (
	\begin{array}{c}
		a_2^> \\
		a_2^<
	\end{array}
	\right )
	- i\sqrt{\hbar} \left (
	\begin{tabular}{cc}
		0 & $\overrightarrow{\sigma}\cdot\overrightarrow{\nabla}$ \\
		$\overrightarrow{\sigma}\cdot\overrightarrow{\nabla}$ & 0 
	\end{tabular}
	\right )
	\left (
	\begin{array}{c}
		a_1^> \\
		a_1^<
	\end{array}
	\right ) \nonumber \\
	-\left (
	\begin{tabular}{cc}
		$i\partial_t -  \frac{m F(\vec{x},t)}{2}$ & 0 \\
		0 & $i\partial_t + \frac{m F(\vec{x},t)}{2}$
	\end{tabular}
	\right )
	\left (
	\begin{array}{c}
		a_0^> \\
		a_0^<
	\end{array}
	\right ) = 0
\end{align}
\\
Upon using equation (\ref{50}), the first branch of (\ref{51}) yields,

\begin{equation}\label{Shrodinger equation with F}
	i\hbar\frac{\partial a_0^>}{\partial t} = -\frac{\hbar^2}{2m} \nabla^2 a_0^> + \frac{m\hbar F(\vec{x},t)}{2} a_0^>
\end{equation}
\\
\underline{\textbf{Einstein's equation}}  \\
Next, we go to Einstein's equation. $G_{00}$ is evaluated in Appendix [\ref{NR:G}]. We equate it with k$T_{00}$ and obtain:

\begin{equation}
	\frac{\hbar \nabla^2 F(\vec{x},t)}{c^2} + \sum_{n=3}^{\infty} O\Big( \frac{1}{c^n}\Big) = \frac{8\pi G m\, |a_{0}^{>}|^2}{c^2} + \sum_{n=3}^{\infty} O\Big( \frac{1}{c^n}\Big) 
\end{equation}
Equating the functions at order $\frac{1}{c^2}$, we obtain:

\begin{equation}\label{poisson equation with F}
	\nabla^2 F(\vec{x},t) = \frac{8\pi G m\, |a_{0}^{>}|^2}{\hbar}
\end{equation}
If we recognize the quantity $\frac{\hbar F(\vec{x},t)}{2}$ as the potential $\phi$,
then we get Schr\"{o}dinger-Newton system of equations with $m\phi$ as the gravitational potential energy and $m\, |a_{0}^{>}|^2$ as mass density $\rho(\vec{x},t)$. The physical picture, which this
system of equations suggest is given in the introduction. 
\begin{align}
	i\hbar\frac{\partial a_0^>}{\partial t} &= -\frac{\hbar^2}{2m} \nabla^2 a_0^> + m \phi(\vec{x},t) a_0^> \\
	\nabla^2\phi(\vec{x},t) &= 4\pi G m\, |a_{0}^{>}|^2 = 4\pi G \rho(\vec{x},t)\\
	i\hbar\frac{\partial a_0^>}{\partial t} &= -\frac{\hbar^2}{2m} \nabla^2 a_0^{>}  - G m^2\,\int \frac{|a_{0}^{>}(\vec{r}^{'}, t)|^2}{|\vec{r} - \vec{r}^{'}|}d^3\vec{r}^{'}a_0^>
\end{align}

\subsection{NR limit of Einstein-Cartan-Dirac system}

Dirac equation on $U_4$ (which is famously called Hehl-Datta equation) is given by equation (\ref{eq:HDFull_std})
\begin{equation}
	i\gamma^{\mu}\psi_{;\mu} - \frac{3}{8}L_{Pl}^{2}\overline{\psi}\gamma^{5}\gamma_{(a)}\psi\gamma^{5}\gamma^{(a)}\psi - \frac{mc}{\hbar}\psi = 0
\end{equation}
We have already evaluated first and the last term after putting Ansatz for spinor (\ref{eq:Spinor_ansatz_NRlimit}) and metric (\ref{eq:metric_ansatz_NRlimit_special}). the second term (arising because of torsion) induces non-linearity into the Dirac equation. We now evaluate this term by following similar procedure as we did for the other 2 terms. First we multiply the mid-term by $\gamma^{0}c$ as done while getting equation (\ref{Ieq:4.2}) from (\ref{Ieq:4.1}) and get the following:
\begin{equation} \gamma^{(0)}\frac{3c}{8}L_{Pl}^{2}\overline{\psi}\gamma^{5}\gamma_{(a)}\psi\gamma^{5}\gamma^{(a)}\psi = -\frac{3c}{8} l_{Pl}^2 e^{\frac{ic^2S}{\hbar}}\bigg(\sum_{n=0}^{\infty}\bigg(\frac{\sqrt{\hbar}}{c}\bigg)^n a^\dagger_{n}\bigg) \gamma_{(a)} \bigg(\sum_{l=0}^{\infty}\bigg(\frac{\sqrt{\hbar}}{c}\bigg)^l a_l\bigg)\gamma_{5} \gamma^{(a)}\bigg(\sum_{m=0}^{\infty}\bigg(\frac{\sqrt{\hbar}}{c}\bigg)^l a_m\bigg)
\end{equation}
Next, we divide it by $\bigg{[} 1+ \sum_{n = 1}^{\infty} \bigg{(}\frac{\sqrt{\hbar}}{c} \bigg{)}^n e^{0[n]}_{(0)}\bigg{]}$ as done while getting equation (\ref{Ieq:4.3}) from (\ref{Ieq:4.2}). This is equivalent to dividing by  $\Big[1 - \frac{\hbar F(\vec{x},t)}{2c^2} +\sum_{n=3}^{\infty} O\Big( \frac{1}{c^n}\Big) \Big]$ or equivalently multiplying by $\Big[1 + \frac{\hbar F(\vec{x},t)}{2c^2} +\sum_{n=3}^{\infty} O\Big( \frac{1}{c^n}\Big) \Big]$ as given in (\ref{NR:tetrad}). We get following:

\underline{\textbf{The non-linear term}}

\begin{equation}\label{HD NL term with all ansatz}
	e^{\frac{ic^2S}{\hbar}}\frac{c^3}{\hbar^{3/2}}\Bigg[1 + \frac{\hbar F(\vec{x},t)}{2c^2} +\sum_{n=3}^{\infty} O\Big( \frac{1}{c^n}\Big)\Bigg]\frac{3G}{8} \bigg(\sum_{n_1,n_2,n_3=0}^{\infty}\bigg(\frac{\sqrt{\hbar}}{c}\bigg)^{n} a^{\dagger}_{n_1-i}\gamma_{a} a_{n_2-j}\gamma^5\gamma^a a_{n_3-k}\bigg)
\end{equation}
where $n = n_1 + n_2 + n_3$. This term modifies 
Equation (\ref{eq:DE_specific_after substituting spinor, metric ansatz}) as follows 

\begin{equation}\label{eq:HD_specific_after substituting spinor, metric ansatz}
	\begin{split}
		e^{\frac{ic^2S}{\hbar}}\frac{c^3}{\hbar^{3/2}}\sum_{n=0}^{\infty}& \bigg(\frac{\sqrt{\hbar}}{c}\bigg)^n \Bigg[m~a_{n-1} + i\dot{a}_{n-3} + i\sqrt{\hbar}\overrightarrow{\alpha}\cdot\overrightarrow{\nabla}a_{n-2} - \beta m a_{n-1} - \beta\frac{mF(\vec{x},t)}{2}a_{n-3} \\
		& + \frac{3G}{8}\Big(\sum_{n_1,n_2,n_3=0}^{\infty}\bigg(\frac{\sqrt{\hbar}}{c}\bigg)^{n} a^{\dagger}_{n_1-i}\gamma_{a} a_{n_2-j}\gamma^5\gamma^a a_{n_3-k}\Big) \Bigg] = 0
	\end{split}
\end{equation}
where $n = n_1 + n_2 + n_3, i+j+k = 5$ and, whatever value of $i,j,k,n_1,n_2,n_3$ is chosen from (0,1,2,3,4,5) the fact that $i\leq n_1$, $j\leq n_2$ and $k\leq n_3$ is to be respected. We find from the above expression that the non-linear term with starts contributing finitely from n = 5 onwards. So, the analysis for n = 0,1,2,3 as given in Appendix [\ref{NRlimit_ED}] remains as it is and we obtain Schr\"{o}dinger equation for $a_0^>$ viz.$ i\hbar\frac{\partial a_0^>}{\partial t} = -\frac{\hbar^2}{2m} \nabla^2 a_0^> + \frac{m\hbar F(\vec{x},t)}{2} a_0^>$.
\\

Next, we go to Einstein's equation (gravitation equation of ECD theory). The equations of interest here are as given by eqn (\ref{eq:ECD_gravity_std}) as $G^{\mu\nu}(\{\}) = \rchi T^{\mu\nu} - \frac{1}{2}\rchi^{2}g_{\mu\nu} S^{\alpha\beta\lambda}S_{\alpha\beta\lambda}$ \\
$G^{\mu\nu}$ and $T^{\mu\nu}$ are already analyzed in above section (\ref{NRlimit_ED}). We will analyze the second term on the RHS, which is ($- \frac{1}{2}\rchi^{2}g_{\mu\nu} S^{\alpha\beta\lambda}S_{\alpha\beta\lambda}$). It contains the products of spin density tensor which is given by eqn (\ref{eq:spin density}). We consider only first term in the expansion of metric because other terms combined with the coupling constant are already higher orders.

\begin{align}
	\frac{-1}{2}\rchi^{2}g_{00} S^{\alpha\beta\lambda}S_{\alpha\beta\lambda} &= -g_{00}\frac{2\pi^2G^2\hbar^2}{c^6} \sum_{N=0}^{\infty}\Big(\sum_{k=0}^{\infty} \sum_{l=0}^{\infty} a_{k}^{\dagger} \gamma^{0} \gamma^{[c} \gamma ^{a} \gamma^{b]}\Big) \Big( \sum_{m=0}^{\infty} \sum_{n=0}^{\infty} a_{m}^{\dagger} \gamma^{0} \gamma_{[c} \gamma_{a} \gamma_{b]} n_m \Big) =\sum_{n=6}^{\infty} O\Big( \frac{1}{c^n}\Big)
\end{align}
We find that this addition doesn't contribute at the order $1/c^2$ on the RHS of equation (\ref{eq:ECD_gravity_std}). Hence we get back Poisson equation. Recognizing the quantity $\frac{\hbar F(\vec{x},t)}{2}$ as the potential $\phi$, at leading order, we find that ECD theory also yields Schr\"{o}dinger-Newton equation. Torsion doesn't contribute at leading order.

\section{Non-relativistic limit of ECD field equations with new length scale $L_{cs}$}

\subsection{Analysis for Higher mass limit of $L_{cs}$} 
Higher mass limit of $L_{cs}$ is $\frac{2Gm}{c^2}$. The Einstein equation in the Riemann-Cartan spacetime with new length scale $L_{CS}$ is given by (\ref{eq:gravity eqn with lcs})

\begin{equation}
	G^{\mu\nu}(\{\}) = \frac{8\pi L_{CS}^{2}}{\hbar c}  T^{\mu\nu} - \frac{1}{2} \Big(\frac{8\pi L_{CS}^{2} }{\hbar c}\Big)^2 g_{\mu\nu}  S^{\alpha\beta\lambda}S_{\alpha\beta\lambda}
\end{equation}
We neglect terms higher order in $L_{CS}$ because it is easy to deduce from the fact that $L_{CS}^2$ in higher mass limit is already 4th order in (1/c). So only first term of RHS is significant. We consider the $"00"$ component of the above equation.

\begin{equation}\label{74}
	G_{00} = \frac{8\pi L_{CS}^{2} }{\hbar c} T_{00}
\end{equation}
\\
The stress tensor is given by (\ref{dynamic EM tensor}). Its $"00"$ component is given by [We neglect orders greater than $1/c^2$].

\begin{equation}\label{75}
	T_{00} = \frac{i\hbar c}{4}\bigg[2 \bar{\psi}\gamma^{0} \psi_{;0} - 2 \bar{\psi}_{;0}\gamma^{0} \psi\bigg]
\end{equation}
The Dirac equation with $L_{cs}$ in its higher mass limit is given as: (\ref{eq:HD eqn with lcs})

\begin{equation}
	i\gamma^{\mu}\psi_{;\mu} = +\frac{3}{8}L_{cs}^{2}\overline{\psi}\gamma^{5}\gamma_{(a)}\psi\gamma^{5}\gamma^{(a)}\psi + \frac{1}{2L_{CS}} \psi 
\end{equation}
\\
Now, for large masses ($m >> m_{Pl}$), amplitude of state $\psi$ is negligible (except in a very narrow region where mass m gets localized). This is possible if we assume localization process, like collapse of wave function \cite{bassi2013models}. In such case, The kinetic energy term can be neglected and we obtain following equations  
\begin{equation}\label{77}
	\begin{split}
		\psi_{;0} &= -\frac{3}{8}i\gamma^0 L_{CS}^2 \bar{\psi}\gamma^5\gamma_{a}\psi\gamma^5\gamma^a\psi - \frac{i\gamma^{0}}{2L_{CS}}\psi \\
		\psi_{;0}^{\dagger} &= \frac{3}{8}i L_{CS}^2(\gamma^0 \bar{\psi}\gamma^5\gamma_{a}\psi\gamma^5\gamma^a\psi)^{\dagger} + \frac{i}{2L_{CS}} \psi^{\dagger}\gamma^0
	\end{split}
\end{equation}
\\
Substituting above equation (\ref{77}) in eqn (\ref{75}) and neglecting higher order terms in $L_{CS}$ we get,

\begin{equation}\label{78}
	\frac{8\pi L_{CS}^{2} }{\hbar c} T_{00} = 4\pi L_{CS}(\psi^{\dagger}\gamma^{0}\psi)
\end{equation}
\\
Substituting for $L_{CS}$ in the large mass limit in eqn (\ref{78}) ,

\begin{equation}\label{79}
	\frac{8\pi L_{CS}^{2} T_{00}}{\hbar c} = 4\pi L_{CS}(\psi^{\dagger}\gamma^{0}\psi) = \frac{8\pi G m \bar{\psi}\psi}{c^2}
\end{equation}
\\
In the localization process we replace $\bar{\psi}\psi$ with a spatial Dirac delta function \cite{TP_2}. Substituting equation (\ref{79}) and $G_{00}$ from Appendix [\ref{A5}] in equation (\ref{74}) and equating at order $\frac{1}{c^2}$, we get the Poisson equation as the non relativistic weak field limit of the modified Einstein equation in the large mass limit,

\begin{equation}
	\nabla^{2}F(\vec{x},t) = \frac{8\pi G m \delta (\vec{x})}{\hbar}
\end{equation}
As earlier, we recognize $\frac{\hbar F}{2}$ as Newtonian potential $\phi$ and hence, we get 
\begin{equation}
	\nabla^{2}\phi = 4\pi G m \delta (\vec{x})
\end{equation}
\subsection{Analysis for lower mass limit of $L_{cs}$}

Lower mass limit of $L_{cs}$ is $\frac{\lambda_C}{2} = \frac{\hbar}{2mc}$. The Dirac equation in the Riemann-Cartan spacetime with new length scale $L_{CS}$ in its lower mass limit is given by  (\ref{eq:ECD_HD_Lcs}):

\begin{equation}
	i\gamma^{\mu}\psi_{;\mu} = \frac{3\hbar^2}{32m^2c^2}\overline{\psi}\gamma^{5}\gamma_{(a)}\psi\gamma^{5}\gamma^{(a)}\psi + \frac{1}{2L_{CS}} \psi 
\end{equation}
We have already evaluated first and the last term after putting Ansatz for spinor (\ref{eq:Spinor_ansatz_NRlimit}) and metric (\ref{eq:metric_ansatz_NRlimit_special}). the second term (arising because of torsion) induces non-linearity into the Dirac equation. We now evaluate this term by following similar procedure as we did for the other 2 terms. First we multiply the mid-term by $\gamma^{0}c$ as done while getting equation (\ref{Ieq:4.2}) from (\ref{Ieq:4.1}) and get the following:
\begin{equation} \gamma^{(0)}\frac{3c}{32}\lambda_{C}^{2}\overline{\psi}\gamma^{5}\gamma_{(a)}\psi\gamma^{5}\gamma^{(a)}\psi = \frac{3c}{32} \lambda_{C}^2 e^{\frac{ic^2S}{\hbar}}\bigg(\sum_{n=0}^{\infty}\bigg(\frac{\sqrt{\hbar}}{c}\bigg)^n a^\dagger_{n}\bigg) \gamma_{(a)} \bigg(\sum_{l=0}^{\infty}\bigg(\frac{\sqrt{\hbar}}{c}\bigg)^l a_l\bigg)\gamma_{5} \gamma^{(a)}\bigg(\sum_{m=0}^{\infty}\bigg(\frac{\sqrt{\hbar}}{c}\bigg)^l a_m\bigg)
\end{equation}
Next, we divide it by $\bigg{[} 1+ \sum_{n = 1}^{\infty} \bigg{(}\frac{\sqrt{\hbar}}{c} \bigg{)}^n e^{0[n]}_{(0)}\bigg{]}$ as done while getting equation (\ref{Ieq:4.3}) from (\ref{Ieq:4.2}). This is equivalent to dividing by  $1 - \frac{\hbar F(\vec{x},t)}{2c^2} +\sum_{n=3}^{\infty} O\Big( \frac{1}{c^n}\Big)$ as given in (\ref{NR:tetrad}) or multiplying by $1 + \frac{\hbar F(\vec{x},t)}{2c^2} +\sum_{n=3}^{\infty} O\Big( \frac{1}{c^n}\Big)$ . We get:

\underline{\textbf{The non-linear term}}

\begin{equation}\label{HD NL term with all ansatz_Lcs}
	e^{\frac{ic^2S}{\hbar}}\frac{c^3}{\hbar^{3/2}}\Bigg[1 + \frac{\hbar F(\vec{x},t)}{2c^2} +\sum_{n=3}^{\infty} O\Big( \frac{1}{c^n}\Big)\Bigg]\frac{3\hbar^{3/2}}{32m^2} \bigg(\sum_{n_1,n_2,n_3=0}^{\infty}\bigg(\frac{\sqrt{\hbar}}{c}\bigg)^{n} a^{\dagger}_{n_1-i}\gamma_{a} a_{n_2-j}\gamma^5\gamma^a a_{n_3-k}\bigg)
\end{equation}
where $n = n_1 + n_2 + n_3, i+j+k = 4$ and, whatever value of $i,j,k,n_1,n_2,n_3$ is chosen from (0,1,2,3,4) the fact that $i\leq n_1$, $j\leq n_2$ and $k\leq n_3$ is to be respected. We find from the above expression that the non-linear term with $L_{CS}$ starts contributing finitely from n = 4 onwards. So, the analysis for n = 0,1,2,3 as given in section (\ref{NRlimit_ED}) remains as it is and we obtain Schr\"{o}dinger equation for $a_0^>$ viz.$ i\hbar\frac{\partial a_0^>}{\partial t} = -\frac{\hbar^2}{2m} \nabla^2 a_0^> + \frac{m\hbar F(\vec{x},t)}{2} a_0^>$.
\\
\\
Now, the gravitational equation of ECD with $L_{cs}$ in its lower mas limit is given by (\ref{eq:gravity eqn with lcs}). We will consider terms only up till second order in (1/c). So we stick to equation for 00 component. We neglect the 2nd term om the RHS of (\ref{eq:gravity eqn with lcs}) because it is already much higher in order. The equation for 00 component is:
\begin{align}
	G_{00} &= \Big(\frac{2\pi\hbar}{m^2 c^3}\Big)\Big(\frac{i\hbar c}{4}\Big)\bigg[2 \bar{\psi}\gamma_{0} \psi_{;0} - 2 \bar{\psi}_{;0}\gamma_{0} \psi  \bigg] \\
	G_{00} &= e^0_{(0)}\Big(\frac{i\pi\hbar^2}{m^2 c^3}\Big)\bigg[\psi^{\dagger} (\partial_t\psi)-(\partial_t\psi^{\dagger}) \psi  \bigg]
\end{align}
After substituting spinor ansatz (\ref{eq:Spinor_ansatz_NRlimit}), we obtain following equation for RHS

\begin{align}
	G_{00} = \Big(\frac{i\pi\hbar}{m^2 c}\Big) \Bigg[&\Bigg(\sum_{m=0}^{\infty}\bigg(\frac{\sqrt{\hbar}}{c}\bigg)^m a^\dagger_{m}\Bigg) \Bigg( \sum_{n=0}^{\infty}\bigg(\frac{\sqrt{\hbar}}{c}\bigg)^n [\dot{a}_{n-2} + i\dot{S}a_n]\Bigg) \\
	-& \Bigg(  \sum_{m=0}^{\infty}\bigg(\frac{\sqrt{\hbar}}{c}\bigg)^m [\dot{a}_{m-2}^{\dagger} - i\dot{S}a_m^{\dagger}]\Bigg) \Bigg( \sum_{n=0}^{\infty}\bigg(\frac{\sqrt{\hbar}}{c}\bigg)^n a_{n}\Bigg) \Bigg] \nonumber 
\end{align}

The equation which we get after 

\begin{equation}
	\frac{\hbar\nabla^2 F}{c^2} + \sum_{n=3}^{\infty} O\Big( \frac{1}{c^n}\Big)= \frac{1}{c} \Bigg(\frac{2\pi\hbar}{m} |a_0^>|^2\Bigg) + \frac{1}{c^2} \Bigg(\frac{2\pi\hbar^{3/2}}{m}\Big[a_1^{>\dagger}a_0^> + a_0^{>\dagger}a_1^>\Big]\Bigg)+ \sum_{n=3}^{\infty} O\Big( \frac{1}{c^n}\Big)
\end{equation}
This leads us to conclude that $a_0^> =0$ and Hence 
\begin{equation}
	\nabla^2 F = 0 \Longrightarrow \ \nabla^2 \phi = 0
\end{equation}
\subsection{Some comments on analysis for intermediate mass}
For an intermediate mass $L_{cs}$ is given by equation (\ref{eq:explicit form of Lcs}). With this, the ECD equations become as follows:

\begin{align}
	i\gamma^{\mu}\psi_{;\mu} &= \frac{3}{8}\Big(\frac{2Gm}{c^2} + \frac{\hbar}{2mc} \Big)^2\overline{\psi}\gamma^{5}\gamma_{(a)}\psi\gamma^{5}\gamma^{(a)}\psi + \frac{1}{\Big(\frac{4Gm}{c^2} + \frac{\hbar}{mc}\Big)} \psi  \label{eq:ECD_HD_Lcs_explicit}\\
	G^{\mu\nu}(\{\}) &= \frac{8\pi }{\hbar c} \Big(\frac{2Gm}{c^2} + \frac{\hbar}{2mc} \Big)^2 T^{\mu\nu} -  \frac{32\pi^2}{\hbar^2 c^2}\Big(\frac{2Gm}{c^2} + \frac{\hbar}{2mc} \Big)^4 g_{\mu\nu}  S^{\alpha\beta\lambda}S_{\alpha\beta\lambda}\label{eq:ECD_gravity_Lcs_explicit}
\end{align}
First we will analyze HD equation. The 3 non-linear terms appear in this equation with coefficients $\frac{3G^2m^2}{2c^4}$, $\frac{3L_{pl}^2}{4}$ and $\frac{3\hbar^2}{32m^2c^2}$. We have already done the order analysis of all these terms and shown to be higher order; not contributing to the equation at leading order. So we neglect them. What we get is: 
\begin{align}
	&i\gamma^{\mu}\psi_{;\mu} = \frac{1}{\Big(\frac{4Gm}{c^2} + \frac{\hbar}{mc}\Big)} \psi = \frac{mc\psi}{\hbar} \Bigg(\frac{1}{1+\frac{4m^2}{m_{pl}^2}}\Bigg) \label{eq:ECD_HD_Lcs_explicit2} \\
	\Longrightarrow& \Bigg[1+ \frac{4m^2}{m_{pl}^2}\Bigg] i\gamma^{\mu}\psi_{;\mu} = \frac{mc\psi}{\hbar} 
\end{align}
This is a very interesting equation. If mass m is too small compared to $m_{pl}$, we can neglect 2nd term on LHS and this basically gives Schrodinger's equation. On the other hand, if mass is too large, we neglect the first term on LHS. and then the equation becomes such that we can safely assume localization process. [basically it justifies eq. (\ref{77})]. We will investigate the intermediate mass case more rigorously in future (see future plans)

\section{Summary of important results}\label{sec:summary_NR}

\begin{itemize}
	\item At leading order, non-relativistic limit of self-gravitating Dirac field on $V_4$ (commonly called as Einstein-Dirac system) is  Schr\"{o}dinger-Newton equation with no assumption of symmetry on metric. 
	\item Non-relativistic limit of self-gravitating Dirac field on $U_4$ (commonly called as Einstein-Cartan-Dirac system) is also  Schr\"{o}dinger-Newton equation at leading order. 
	\item Non-relativistic limit of ECD theory with $L_{cs}$ in its low mass limit produces a source-free Poisson equation. This will be interpreted in chapter (\ref{chp:discussions}). 
	\item Non-relativistic limit of ECD theory with $L_{cs}$ in its higher mass limit produces Poisson equation with delta function source. This will be interpreted in chapter (\ref{chp:discussions}).
\end{itemize}
\textbf{The work in this chapter is based on the paper titled ``The non-relativistic limit of the Einstein-Cartan-Dirac equations" which is \texttt{"under preparation"}} \cite{in_preparation_NR}


\chapter{Brief review of Newmann-Penrose (NP) formalism and formulation of ECD equations in NP formalism}\label{Np/ECDinNP}

There has been a variety of different (physically and mathematically equivalent) ways of writing the field equations of General theory of relativity. Initially, it was formulated in standard coordinate-basis version using the metric tensor components as the basic variable and the Christoffel symbols as connection. Later various methods like that of differential forms developed by Cartan (Lovelock and Rund, 1975), the space-time (orthonormal) tetrad version of Ricci (Levy, 1925) and the spin coefficient version of Newman and Penrose (Newman and Penrose, 1962; Geroch et al, 1973; Penrose, 1968; Penrose and Rindler, 1984; Penrose and Rindler, 1986; Newman and Tod, 1980; Newman and Unti, 1962) are developed. All references in parenthesis are taken from Scholarpedia article titled ``Spin-Coefficient formalism".

Dirac equation on $V_4$ has been studied extensively in NP formalism. It's detail account can be seen in \cite{Chandru}. From this chapter onwards, we follows the notations/ representations/ conventions and symbols of this celebrated book ``The mathematical theory of black holes" By S. Chandrasekhar \cite{Chandru}. Our aim in this chapter is as follows
\begin{itemize}
	\item We know that Contorsion tensor is completely expressible in terms of components of Dirac spinor. We want to find an explicit expression for Contorsion spin coefficients (in Newman-Penrose) in terms of Dirac spinor components. 
	\item Dirac equation on $V_4$ is presented in equation (108) of \cite{Chandru}. We aim to modify these equations on $U_4$. 
\end{itemize}
We will first present a brief review of NP formalism and then formulate ECD equations in NP formalism.

\section{Newman-Penrose formalism}
NP formalism was formulated by Neuman and Penrose in their work \cite{NP_original_paper}. It is a special case of tetrad formalism (introduced in Appendix [\ref{Tetrad_formalism_CD_for_spinors}]); where we choose our tetrad as a set of four null vectors viz. 
\begin{equation}
	e_{(0)}^{\mu} = l^{\mu},~~~ e_{(1)}^{\mu} = n^{\mu},~~~ e_{(2)}^{\mu} = m^{\mu}, ~~~e_{(3)}^{\mu} = \bar{m}^{\mu}
\end{equation}
$l^{\mu}, n^{\mu}$ are real and $m^{\mu}, \bar{m}^{\mu}$ are complex. The tetrad indices are raised and lowered by flat space-time metric 
\begin{equation}
	\eta_{(i)(j)} = \eta^{(i)(j)} = \begin{pmatrix}
		0 & 1& 0 & 0 \\
		1 & 0& 0 & 0 \\
		0 & 0& 0 & -1 \\
		0 & 0& -1 & 0 \\
	\end{pmatrix}
\end{equation}
and the tetrad vectors satisfy the equation $g_{\mu\nu} = e_{\mu}^{(i)}e_{\nu}^{(j)}\eta_{(i)(j)} $.
In the formalism, we replace tensors by their tetrad components and represent these components with distinctive symbols. These symbols are quite standard and used everywhere in literature. It was Hermann Bondi who first suggested the use of null-tetrads for the analysis of electromagnetic and gravitational radiation since they propogate along these null directions. Some important features of NP formalism are can be jot down as follows: (these are partially also the reasons why we adopted this formalism to represent our equations)
\begin{itemize}
	\item With NP formalism, equations can be partially grouped together into sets of linear equations (Newman and Unti, 1962)
	\item All are complex equations; thereby reducing the total number of equations by half
	\item It allows one to concentrate on individual 'scalar' equations with particular physical or geometric significance.
	\item  It allows one to search for solutions with specific special features, such as the presence of one or two null directions that might be singled out by physical or geometric considerations. Ex. it turns out to be a very useful tool in solving problems involving massless fields etc.\item Newman and Penrose also showed that their formalism is completely equivalent to the SL(2,$\mathbb{C}$) spinor approach. [We are gonna follow SL(2,$\mathbb{C}$) spinor approach]
	\item In NP formalism, equations are written out explicitly without the use of the index and summation conventions.
	\item While dealing with Spinors on curved space-times, it becomes very easy to establish the knowledge of physical/ geometric properties of complicated space-times (e.g. space-time around Kerr black hole etc.) and the knowledge of various properties of Spinors simultaneously in a common vocabulary of NP formalism. Various commonly occurring space-times have been formulated in NP formalism in \cite{Chandru}. This point is the main reason why we adopt this formalism. 
\end{itemize}

\begin{figure}[h!]\label{fig:12 spin coefficients their `torsional' parts}
	\begin{center}
		\includegraphics[width=10cm]{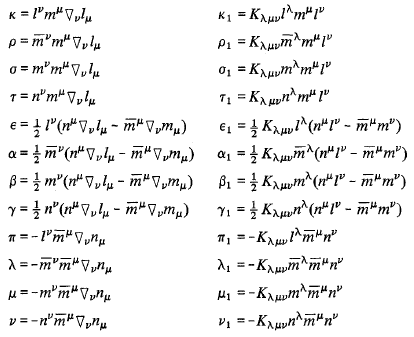}
		\caption{12 complex spin coefficients their `torsional' parts}
	\end{center}
\end{figure}

The orthonormality condition on null tetrads imply 
$l.m = l.\bar{m} = n.m = n.\bar{m} = 0$, $l.l = n.n = m.m = \bar{m}.\bar{m} = 0$ and $l.n = 1$ and $ m.\bar{m} = -1$.  The Ricci rotation coefficients (defined in appendix [\ref{Tetrad_formalism_CD_for_spinors}]) for null tetrads are called spin coefficients and are defined as follows 

\begin{equation}\label{def:spin coefficients1}
	\gamma_{(l) (m) (n)} = e^{\nu}_{(n)} e^{\mu}_{(m)} \nabla_{\nu} e_{(l)\mu}
\end{equation}
The covariant derivative defined in the above equation can be taken w.r.t both \textbf{$V_4$ and $U_4$ manifold.} We are here interested in $U_4$. Spin coefficients are denoted by following symbols
\begin{equation}\label{symbols for spin coefficients}
	\begin{split}
		\kappa = \gamma_{(2)(0)(0)}~~~~~~~ \rho=\gamma_{(2)(0)(3)} ~~~~~~~ \epsilon=\frac{1}{2}(\gamma_{(1)(0)(0)}+\gamma_{(2)(3)(0)})  \\
		\sigma =\gamma_{(2)(0)(2)}  ~~~~~~~ \mu =\gamma_{(1)(3)(2)}  ~~~~~~~ \gamma=\frac{1}{2}(\gamma_{(1)(0)(1)}+\gamma_{(2)(3)(1)})  \\
		\lambda =\gamma_{(1)(3)(3)} ~~~~~~~ \tau=\gamma_{(2)(0)(1)}  ~~~~~~~ \alpha=\frac{1}{2}(\gamma_{(1)(0)(3)} + \gamma_{(2)(3)(3)})\\
		\nu=\gamma_{(1)(3)(1)}  ~~~~~~~ \pi=\gamma_{(1)(3)(0)} ~~~~~~~ \beta=\frac{1}{2}(\gamma_{(1)(0)(2)}+\gamma_{(2)(3)(2)}) \\
	\end{split}
\end{equation}
These are $12$ complex spin coefficients, corresponding to $24$ real components of $\gamma$. We separate the Riemann part and the torsional part from the covariant derivative of equation (\ref{def:spin coefficients1}). The result is 

\begin{align}\label{eq:spin coefficients in terms of CD of tetrads}
	\gamma_{(l) (m) (n)} &= e^{\nu}_{(n)} e^{\mu}_{(m)} \nabla_{\nu} e_{(l)\mu} \\ 
	&= e^{\nu}_{(n)} e^{\mu}_{(m)} \bigg{[}\delta^{\alpha}_{\mu} \partial_{\nu} - \bfrac{\alpha}{\mu\nu} + K_{\nu\mu}^{\:\:\:\:\:\:\alpha} \bigg{]}e_{(l)\alpha} \nonumber\\
	&= \gamma^o_{(l) (m) (n)} + K_{(n) (m) (l)} \nonumber 
\end{align}
In terms of the symbols (defined in equation (\ref{symbols for spin coefficients})), we adopt notation of \cite{jogia_Griffiths} where $\kappa =\kappa^o + \kappa_1$ and so on for all the 12 spin coefficients. $\kappa^o$ denotes Riemann part and and $\kappa_1$ denote torsional part. The torsional part of spin coefficients (which distinguishes it from $V_4$) is called Contorsion spin coefficients. The spin coefficients and contorsion spin coefficients are given in the figure (\ref{fig:12 spin coefficients their `torsional' parts}). 

The directional derivatives w.r.t these null tetrads are given by 

\begin{equation}
	\begin{split}
		D = l^\mu\frac{\partial}{\partial x^\mu} = e_0 ~~~~\Delta = n^\mu\frac{\partial}{\partial x^\mu} = e_1~~~~
		\delta = m^\mu\frac{\partial}{\partial x^\mu} = e_2~~~~ \delta^*= \bar{m}^\mu\frac{\partial}{\partial x^\mu} = e_3 
	\end{split}
\end{equation}
\section{ECD equations in NP formalism} 
\subsection{Notations/ representations and spinor analysis}
\begin{itemize}
	\item The Lorentz Signature used in this chapter is Diag (+ - - -)
	\item The 4 component Dirac-spinor is 
	\begin{equation}\label{notation:Dirac Spinor} 
		\psi = \begin{bmatrix} P^A \\ \bar{Q}_{B'} \end{bmatrix}  
	\end{equation}
	
	where $P^A$ and $\bar{Q}_{B'}$ are 2-dim complex vectors in $\mathbb{C}^2$ (also called spinors) Please see section for details. We use following notations for Dirac spinor components (consistent with the notations of Chandra's book \cite{Chandru}) $P^0 = F_1$, $P^1 = F_2$, $\bar{Q}^{1'} = G_1$ and $\bar{Q}^{0'} = -G_2$. \\
	\item We define 4 null vectors (and their corresponding co-vectors) on Minkowski space
	\begin{align}
		l^a &= \frac{1}{\sqrt{2}} (1,0,0,1), ~ m^a =  \frac{1}{\sqrt{2}} (0,1,-i,0),~ \bar{m}^a = \frac{1}{\sqrt{2}} (0,1,i,0), ~ n^a = \frac{1}{\sqrt{2}} (1,0,0,-1)  \\
		l_a &= \frac{1}{\sqrt{2}} (1,0,0,-1),~ m_a =  \frac{1}{\sqrt{2}} (0,-1,i,0),~ \bar{m}_a = \frac{1}{\sqrt{2}} (0,-1,-i,0),~ n_a = \frac{1}{\sqrt{2}} (1,0,0,1)
	\end{align} 
	We also define, what is called as \textbf{Van der Waarden} symbols as follows:
	
	\begin{equation}\label{def:vanderwrden symbols}
		\sigma^a = \sqrt{2}\begin{bmatrix} l^a & m^a \\ \bar{m}^a & n^a \end{bmatrix}~~~~~~~~~~~~~~~~~~~~
		\tilde{\sigma}^a = \sqrt{2}\begin{bmatrix} n^a & -{m}^a \\ -\bar{m}^a & l^a \end{bmatrix}  
	\end{equation}

	\item We use following representation of gamma matrices [its the \textbf{complex version of Weyl or chiral representation}]
	\begin{equation}\label{notation:Gamma matrix_ECDinNP}
		\gamma^a = \begin{bmatrix} 0 & (\tilde{\sigma}^a)^* \\ (\sigma^a)^* & 0 \end{bmatrix} (a =0,1,2,3) ~~~where~~~ \gamma^0 = \begin{bmatrix} 0 & \mathbb{I} \\ \mathbb{I} & 0 \end{bmatrix},~~ \gamma^i = \begin{bmatrix} 0 & (-\sigma^{i})^* \\ (\sigma^{i})^* & 0 \end{bmatrix}
	\end{equation}
\end{itemize}
The reason for choosing complex Weyl representation is the fact that the spinor and gamma matrix defined in equation (\ref{notation:Dirac Spinor}) and (\ref{notation:Gamma matrix_ECDinNP}) gives us equation (97) and (98) of section (103) given in Chandra's book \cite{Chandru}. \textbf{We want to keep everything in accordance with \cite{Chandru} as a standard reference}. (Equation (99) is the complex version of what we will get). For representing equations or physical objects having spinors and gamma matrices on a curved space time, we adopt Tetrad formalism. Using tetrads, we follow the prescription described briefly in \cite{SVD_geometry_fields_cosmology}. We summarize and comment on it as follows:-

Given a curved manifold $\mathcal{M}$ with all conditions necessary for the existence of spin structure. Let U be a chart on $\mathcal{M}$  with coordinate functions ($x^{\alpha}$), then the prescription for representing spinorial objects (objects with spinors and gamma matrices) is as follows:-\\
1) choose an Orthonormal tetrad field $e^{\mu}_{(a)}(x^{\alpha})$ on U\\
2) Define the Van der Waarden symbols (the $\sigma^{(a)}$ and $\tilde{\sigma}^{(a)}$) in this tetrad basis exactly as defined on Minkowski space in equation (\ref{def:vanderwrden symbols}). Choose a representation of gamma matrix (we will stick to the one chosen above in equation (\ref{notation:Gamma matrix_ECDinNP}))\\
3) The $\sigma$'s in local coordinate frame are obtained through following equation:-

\begin{equation}
	\sigma^{\mu}(x^{\alpha}) = e^{\mu}_{(a)}(x^{\alpha}) \sigma^{(a)} = \sqrt{2}\begin{bmatrix} l^{\mu} & m^{\mu} \\ \bar{m}^{\mu} & n^{\mu} \end{bmatrix} ~~~~~~~~~~~~~~~\tilde{\sigma}^{\mu} =e^{\mu}_{(a)}\tilde{\sigma}^{(a)} \sqrt{2}\begin{bmatrix} n^{\mu} & -m^{\mu} \\ -\bar{m}^{\mu} & l^{\mu} \end{bmatrix} 
\end{equation}
and similar transformation for and gamma matrix. \textbf{So components of any world object which is indexed by the components of gamma matrices or Spinors is now a function of chosen orthonormal tetrad}. It is defined a-priori in a local tetrad basis [whose components are exactly the same as defined on a flat Minkowski space] and then carried to curved space via tetrad. (This is unlike a normal world objects which are first defined naturally at a point on a manifold and then carried to local tangent space via tetrad).

Dirac equation on $V_4$ has been studied extensively in NP formalism. It's detail account can be seen in \cite{Chandru}(Dirac equation on $V_4$ is presented in equation (108)). We aim to modify these equations on $U_4$. To this aim, we want to modify section 102(d) of Chandra's book \cite{Chandru} \textbf{to include torsion} in the theory and modify Dirac equation accordingly on $U_4$. For calculating covariant derivative of spinor, we require the spinor affine connection coefficients. They are defined through the requirement that $\epsilon_{AB}$ and $\sigma$'s are covariantly constant. The whole analysis remains as it is up till eqn (91) of Chandra's book except, everywhere, the covariant derivative would now be evaluated on $U_4$. The covariant derivatives are defined as:

\begin{align}
	\nabla_{\mu}P^A &= \partial_{\mu} P^A + \Gamma^{A}_{\mu B} P^B  \label{eq:CD-Spinor}\\
	\nabla_{\mu} \bar{Q}^{A'} &= \partial_{\mu}\bar{Q}^{A'} + \bar{\Gamma}^{A'}_{\mu B'} \bar{Q}^{B'} \label{eq:CD-CoSpinor}
\end{align}

Here $\Gamma$ terms are the terms that add to the partial derivative while calculating the full derivative of spinorial objects on $U_4$. Their values can be determined completely in terms of Spin coefficients and we now evaluate its tetrad components. \textbf{Using Friedman's lemma} (proved on page 542 of Chandra's book \cite{Chandru}), we can express various spin coefficients $\Gamma_{(a)(b)(c)(d')}$ in terms of covariant derivative of basis null vectors (which we had defined earlier viz. l,n,m,$\bar{m}$). The covariant derivative here is exactly the same as defined in equation equation 3.3 (and explicitly written in eqn 3.5) of \cite{jogia_Griffiths}. We have also defined in (\ref{eq:spin coefficients in terms of CD of tetrads}). Using this covariant derivative, it can be easily seen how equations (95) and (96) will get modified. For instance, Chandra's equations (95) and (96) gets modified as $\Gamma_{0000'} = \kappa^o + \kappa_1$ and $\Gamma_{1101'} = \mu^o + \mu_1$. Here the subscript 0 in $\kappa^o$ and  $\mu^o$ is just used to indicate the original $\kappa$ and $\mu$ defined on $V_4$ as in, those original equations of Chandra's book. Likewise, 12 independent spin coefficients are calculated in terms of covariant derivatives of null vectors and defined in tabular equation (\ref{tab:generic spin coefficients U4}).

\begin{equation}\label{tab:generic spin coefficients U4}
	\Gamma_{(a)(b)(c)(d')} = \begin{tabular}{|l||*{3}{c|}}\hline
		\backslashbox{(c)(d')}{(a)(b)}
		&\makebox[3em]{00}&\makebox[3em]{01 or 10}&\makebox[3em]{11}
		\\\hline\hline
		00' &$\kappa^o + \kappa_1 $&$\epsilon^o+ \epsilon_1$&$\pi^o + \pi_1$\\\hline
		10' &$\rho^o + \rho_1$& $\alpha^o + \alpha_1$& $\lambda^o + \lambda_1$\\\hline
		01' &$\sigma^o + \sigma_1$&$\beta^o+\beta_1$&$\mu^o +\mu_1$\\\hline
		11' &$\tau^o+\tau_1$&$\gamma^o+\gamma_1$&$\nu^o + \nu_1$\\\hline\hline
	\end{tabular}
\end{equation}	
We note that, for generic case, all the 12 terms will have Contorsion spin coefficients.

\subsection{Contorsion Spin Coefficients in terms of Dirac spinor components}\label{sec:Contorsion interms of spinor}

Spin Density tensor of matter ($S^{ijk}$) which is made up of Dirac spinor and gamma matrices, when expressed as a world tensor on $U_4$ manifold, is given by 

\begin{equation}\label{eq:Spin density of Dirac particles}
	S^{\mu\nu\alpha} = \frac{-i\hbar c}{4}\bar{\psi}\gamma^{[\mu} \gamma^{\nu} \gamma^{\alpha]} \psi
\end{equation}

The ECD field equations suggest that $T^{\mu\nu\alpha} = kS^{\mu\nu\alpha} $ where $T^{\mu\nu\alpha}$ is modified torsion tensor defined in eqn (2.3) of \cite{hehl_RMP}. It can be shown that, for Dirac field,  $T^{\mu\nu\alpha} =  -K^{\mu\nu\alpha} = kS^{\mu\nu\alpha}$ as in eqn (5.6) of \cite{Hehl1971}. Here k is a gravitational coupling constant with the length scale $ l$ viz. $\frac{8\pi l ^2}{\hbar c}$. \textbf{For the standard theory, $l=L_{pl}$ and for modified theory, its $l = L_{cs}$.} Substituting eqn (\ref{eq:Spin density of Dirac particles}) in field equations, we obtain following
\begin{equation}
	K^{\mu\nu\alpha} = - kS^{\mu\nu\alpha} = 2i\pi l ^2 \bar{\psi}\gamma^{[\mu} \gamma^{\nu} \gamma^{\alpha]} \psi
\end{equation} 

where gamma matrix $\gamma^{\mu}$ is just the gamma matrix as given in eqn (\ref{notation:Gamma matrix_ECDinNP}) being generalized to world index using orthonormal tetrads. Only 4 \textbf{independent} components of $K^{\mu\nu\alpha}$ are excited by Dirac field. Next, we replace this tensor by its null tetrad components (The Newman-Penrose formalism) as follows
\begin{equation}
	K_{(i)(j)(k)} = e_{(i)\mu}e_{(j)\nu}e_{(k)\alpha} K ^{\mu\nu\alpha}
\end{equation} 
where $e_{(i)\mu} = (l_{\mu},n_{\mu},m_{\mu},\bar{m}_{\mu})$; i = 1,2,3,4. After evaluating everything, the eight \textbf{non-zero} spin coefficients excited by Dirac Spinor given in eqn (\ref{notation:Dirac Spinor}) are as follows (Out of total 12  Contorsion spin coefficient, 8 are non-zero and 4 of them are independent). We have shown the explicit calculation of $\rho_1$ in appendix section (\ref{app:Contorsion_spin interms of spinor}). The calculations of other Contorsion spin coefficients are similar.

\begin{align} 
	\tau_1 &= k_{123} = 2\sqrt{2}i\pi  l ^2 (P^1\bar{P}^0 -  Q^1\bar{Q}^0) = 2\sqrt{2}i\pi  l ^2 (F_2\bar{F}_1+ G_2\bar{G}_1) \label{eq:tau1} \\
	\pi_1 &= k_{124} = 2\sqrt{2}i\pi  l ^2 (Q^0\bar{Q}^1 -  P^0\bar{P}^1) = 2\sqrt{2}i\pi  l ^2 (-F_1\bar{F}_2 - G_1\bar{G}_2) \\
	\mu_1 &= -k_{234} = 2\sqrt{2}i\pi  l ^2 (P^0\bar{P}^0 - Q^0\bar{Q}^0 ) = 2\sqrt{2}i\pi  l ^2 (F_1\bar{F}_1 - G_2\bar{G}_2) \\
	\rho_1 &= -k_{134} = 2\sqrt{2}i\pi  l ^2 (Q^1\bar{Q}^1 - P^1\bar{P}^1) = 2\sqrt{2}i\pi  l ^2 (G_1\bar{G}_1 - F_2\bar{F}_2) \\
	\epsilon_1 &= \frac{-1}{2}\rho_1 = -\sqrt{2}i\pi  l ^2 (G_1\bar{G}_1 - F_2\bar{F}_2)\\
	\alpha_1 &= \frac{-1}{2}\pi_1 = \sqrt{2}i\pi  l ^2 (F_1\bar{F}_2 + G_1\bar{G}_2) \\
	\beta_1 &= \frac{-1}{2}\tau_1 = -\sqrt{2}i\pi  l ^2 (F_2\bar{F}_1+ G_2\bar{G}_1)  \\
	\gamma_1 &= \frac{-1}{2}\mu_1 = -\sqrt{2}i\pi  l ^2 (F_1\bar{F}_1 - G_2\bar{G}_2) \label{eq:gamma1}
\end{align}

From above relations we can deduce that

\begin{align}
	\mu_1 &= -\mu_1^*\\
	\rho_1 &= -\rho_1^*\\
	\pi_1 &= + \tau_1^*
\end{align}

As we saw, Eight Contorsion spin coefficients (viz. $\mu,\tau,\rho,\pi, \alpha,\gamma,\epsilon,\beta$) are excited by Dirac particle, we use this to modify table (\ref{tab:generic spin coefficients U4}) as follows. Out of these 8, four are independent. and other four can be expressed in terms of others. 
\begin{equation}\label{tab:Dirac spin coefficients U4}
	\Gamma_{(a)(b)(c)(d')} = \begin{tabular}{|l||*{3}{c|}}\hline
		\backslashbox{(c)(d')}{(a)(b)}
		&\makebox[3em]{00}&\makebox[3em]{01 or 10}&\makebox[3em]{11}
		\\\hline\hline
		00' &$\kappa_0$ & $\epsilon_0 - \frac{\rho_1}{2}$&$\pi_0 + \pi_1$\\\hline
		10' &$\rho_0 + \rho_1$& $\alpha_0 - \frac{\pi_1}{2}$& $\lambda_0$\\\hline
		01' &$\sigma_0$&$\beta_0- \frac{\tau_1}{2}$&$\mu_0 +\mu_1$\\\hline
		11' &$\tau_0+\tau_1$&$\gamma_0- \frac{\mu_1}{2}$&$\nu_0$\\\hline\hline
	\end{tabular}
\end{equation}

\subsection{Dirac equation on $U_4$ (Hehl-Datta equation) in NP formalism} 

The Dirac equation on $U_4$ is 
\begin{equation}
	i\gamma^{\mu}\nabla_{\mu}\psi = \frac{mc}{\hbar}\psi = \frac{\psi}{l}
\end{equation}
Where $\nabla$ here denotes covariant derivative on $U_4$. l = $\lambda_c$ for standard theory and $l = 2L_{cs}$ for modified theory. It can be written in the following matrix form:

\begin{align}
	i\begin{pmatrix} 0 & (\tilde{\sigma}^{\mu})^* \\ (\sigma^{\mu})^* & 0 \end{pmatrix}\nabla_{\mu}\begin{pmatrix} P^A \\ \bar{Q}_{B'} \end{pmatrix} = m\begin{pmatrix} P^A \\ \bar{Q}_{B'} \end{pmatrix}
\end{align}

This leads to the following 2 matrix equations:

\begin{align}
	\begin{pmatrix} \sigma^{\mu}_{00'}  & \sigma^{\mu}_{10'} \\ \sigma^{\mu}_{01'} & \sigma^{\mu}_{11'} \end{pmatrix}\nabla_{\mu}\begin{pmatrix} P^0 \\ P^1 \end{pmatrix} + im\begin{pmatrix} -\bar{Q}^{1'}\\ \bar{Q}^{0'} \end{pmatrix} &= 0 \\
	\begin{pmatrix} \sigma^{\mu}_{11'}  & -\sigma^{\mu}_{10'} \\ -\sigma^{\mu}_{01'} & \sigma^{\mu}_{00'} \end{pmatrix}\nabla_{\mu}\begin{pmatrix} -\bar{Q}^{1'}\\ \bar{Q}^{0'} \end{pmatrix} + im\begin{pmatrix} P^0\\ P^1 \end{pmatrix} &= 0
\end{align}
These are the four Dirac equations as follows:\\
\underline{\textbf{Equation 1:}}\\
\begin{align}
	\sigma^{\mu}_{00'}\nabla_{\mu} P^0 + \sigma^{\mu}_{10'}\nabla_{\mu}P^1 &=\frac{i}{2\sqrt{2} l}\bar{Q}^{1'} \\
	(\partial_{00'}P^0 + \Gamma^0_{~i00'}P^i) + (\partial_{10'}P^1 + \Gamma^1_{~i10'}P^i) &=\frac{i}{2\sqrt{2} l} \bar{Q}^{1'} \\
	[D+\Gamma^0_{~000'}P^0 + \Gamma^0_{~100'}P^1] + [\delta^* + \Gamma^1_{~010'}P^0+ \Gamma^1_{~110'}P^1] &=  \frac{i}{2\sqrt{2} l} \bar{Q}^{1'}  \\
	[D+\Gamma_{1000'} - \Gamma_{0010'}]P^0 + [\delta^* + \Gamma_{1100'} - \Gamma_{0110'}]P^1 &=  \frac{i}{2\sqrt{2} l} \bar{Q}^{1'} \\
	[D + \epsilon^o + \epsilon_1 -\rho^o -\rho_1 ]P^0 + [\delta^* + \pi^o + \pi_1 -\alpha^o -\alpha_1]P^1 &=  \frac{i}{2\sqrt{2} l} \bar{Q}^{1'} \\
	[D + \epsilon^o + \epsilon_1 -\rho^o -\rho_1 ]F_1 + [\delta^* + \pi^o + \pi_1 -\alpha^o -\alpha_1]F_2 &=  \frac{i}{2\sqrt{2} l}G_1\\
	(D+\epsilon_0 - \rho_0)F_1 + (\delta^{*} + \pi_0-\alpha_0)F_2 + \frac{3}{2}(\pi_1F_2 - \rho_1F_1) &= \frac{i}{2\sqrt{2} \lambda_c}G_1
\end{align}
\underline{\textbf{Equation 3:}}\\
\begin{align}
	-\sigma^{\mu}_{11'}\nabla_{\mu} \bar{Q}^{1'} - \sigma^{\mu}_{10'}\nabla_{\mu}\bar{Q}^{0'} + \frac{i}{2\sqrt{2} l} P^0 &= 0\\
	-\bar{\sigma}^{\mu}_{11'}\nabla_{\mu} \bar{Q}^{1'} - \bar{\sigma}^{\mu}_{0'1}\nabla_{\mu}\bar{Q}^{0'} + \frac{i}{2\sqrt{2} l} P^0 &= 0 \\
	(\partial_{11'}\bar{Q}^{1'} + \bar{\Gamma}^{1'}_{~i'1'1}\bar{Q}^{i'}) + (\partial_{10'}\bar{Q}^{0'} + \bar{\Gamma}^{0'}_{~i'0'1}\bar{Q}^{i'}) &=  \frac{i}{2\sqrt{2} l}P^0 
\end{align}
\begin{align}
	[\Delta\bar{Q}^{1'} + \bar{\Gamma}^{1'}_{~0'1'1}\bar{Q}^{0'} + \bar{\Gamma}^{1'}_{~1'1'1}\bar{Q}^{1'}] + [\delta^*\bar{Q}^{0'} + \bar{\Gamma}^{0'}_{~0'0'1}\bar{Q}^{0'} + \bar{\Gamma}^{0'}_{~1'0'1}\bar{Q}^{1'}] &=  \frac{i}{2\sqrt{2} l}P^0  \\
	[\Delta + \bar{\Gamma}_{1'1'0'1} - \bar{\Gamma}_{0'1'1'1}] \bar{Q}^{1'}+ [\delta^* + \bar{\Gamma}_{1'0'0'1} - \bar{\Gamma}_{0'0'1'1}]\bar{Q}^{0'}&=  \frac{i}{2\sqrt{2} l}P^0   \\
	[\Delta + \mu^o+\mu_1- \gamma^o -\gamma_1] \bar{Q}^{1'}+ [\delta^* + \beta^o+\beta_1 - \tau^o - \tau_1]\bar{Q}^{0'}&=  \frac{i}{2\sqrt{2} l}P^0   \\
	[\Delta + \mu^o+\mu_1- \gamma^o -\gamma_1] G_1 -  [\delta^* + \beta^o+\beta_1 - \tau^o - \tau_1]G_2 = im^* F_1\\
	(\Delta + \mu_0^* - \gamma_0^*)G_1 - (\delta^* +\beta_0^* -\tau_0^*)G_2 -
	\frac{3}{2}(\mu_1 G_1 - \pi_1 G_2) &= \frac{i}{2\sqrt{2} \lambda_c}F_1 
\end{align}
Here, We used gamma matrices as defined in (\ref{notation:Gamma matrix_ECDinNP}), compute covariant derivatives using (\ref{eq:CD-Spinor}, \ref{eq:CD-CoSpinor}) and Spin connection in terms of Contorsion spin coefficients as given in (\ref{tab:Dirac spin coefficients U4}). Second Dirac equation can be derived as analogous to equation 1 and fourth equation as analogous to third. The 4 Dirac equations can be summarized below:
\begin{align}
	(D+\epsilon_0 - \rho_0)F_1 + (\delta^{*} + \pi_0-\alpha_0)F_2 + \frac{3}{2}(\pi_1F_2 - \rho_1F_1) &= i b(l)G_1 \label{eq:Dirac eq U4 NP1.1}\\
	(\Delta + \mu_0 - \gamma_0)F_2 + (\delta +\beta_0 -\tau_0)F_1 +
	\frac{3}{2}(\mu_1 F_2 - \tau_1 F_1) &= i b(l)G_2 \\
	(D+\epsilon_0^* - \rho_0^*)G_2 - (\delta + \pi_0^*-\alpha_0^*)G_1 - \frac{3}{2}(\tau_1G_1 - \rho_1 G_2 ) &= i b(l)F_2\\
	(\Delta + \mu_0^* - \gamma_0^*)G_1 - (\delta^* +\beta_0^* -\tau_0^*)G_2 -
	\frac{3}{2}(\mu_1 G_1 - \pi_1 G_2) &=i b(l)F_1 \label{eq:Dirac eq U4 NP1.4}
\end{align}
Substituting the values of Contorsion spin coefficients from equations [\ref{eq:tau1} - \ref{eq:gamma1}] into equations  [\ref{eq:Dirac eq U4 NP1.1} - \ref{eq:Dirac eq U4 NP1.4}], we obtain 

\begin{align}
	(D+\epsilon_0 - \rho_0)F_1 + (\delta^{*} + \pi_0-\alpha_0)F_2 + 3\sqrt{2}i\pi  l ^2((-F_1\bar{F}_2 - G_1\bar{G}_2)F_2 + (F_2\bar{F}_2 - G_1\bar{G}_1)F_1) &= i b(l)G_1 \label{eq:Dirac eq U4 NP2.1}\\
	(\Delta + \mu_0 - \gamma_0)F_2 + (\delta +\beta_0 -\tau_0)F_1+ 3\sqrt{2}i\pi  l ^2((F_1\bar{F}_1 - G_2\bar{G}_2) F_2 -(F_2\bar{F}_1+ G_2\bar{G}_1)F_1) &= i b(l)G_2 \\
	(D+\epsilon_0^* - \rho_0^*)G_2 - (\delta + \pi_0^*-\alpha_0^*)G_1 -  3\sqrt{2}i\pi  l ^2((F_2\bar{F}_2 - G_1\bar{G}_1) G_2 + (F_2\bar{F}_1+ G_2\bar{G}_1)G_1)  &= i b(l)F_2\\
	(\Delta + \mu_0^* - \gamma_0^*)G_1 - (\delta^* +\beta_0^* -\tau_0^*)G_2 - 3\sqrt{2}i\pi  l ^2((F_1\bar{F}_1 - G_2\bar{G}_2) G_1 -(-F_1\bar{F}_2 - G_1\bar{G}_2) G_2)&= i b(l) F_1 \label{eq:Dirac eq U4 NP2.4}
\end{align}
These equations can be condensed into following form:

\begin{align}
	(D+\epsilon_0 - \rho_0)F_1 + (\delta^{*} + \pi_0-\alpha_0)F_2 &= i[b(l)+a(l)\xi]G_1\label{1-DEonU_4final}\\
	(\Delta + \mu_0 - \gamma_0)F_2 + (\delta +\beta_0 -\tau_0)F_1 &= i[b(l)+a(l)\xi]G_2 \\
	(D+\epsilon_0^* - \rho_0^*)G_2 - (\delta + \pi_0^*-\alpha_0^*)G_1 &= i[b(l)+a(l)\xi^*]F_2\\
	(\Delta + \mu_0^* - \gamma_0^*)G_1 - (\delta^* +\beta_0^* -\tau_0^*)G_2  &= i[b(l)+a(l)\xi^*]F_1\label{4-DEonU_4final}
\end{align}

Where $a(l) = 3\sqrt{2}\pi l^2$,  $b(l) = \frac{1}{2\sqrt{2}l}$, $\xi = F_1\bar{G}_1 + F_2\bar{G}_2$ and $
\xi^* = \bar{F}_1 G_1 + \bar{F}_2 G_2$.

\section{Summary of important results}\label{sec:summary_ECDNP}

\begin{itemize}
	\item Dirac equation has been modified on $U_4$ [\ref{1-DEonU_4final} - \ref{4-DEonU_4final}]
	\item Contorsion spin coefficients are expressed completely in terms of Dirac Spinor in section (\ref{sec:Contorsion interms of spinor}). 
	\item Prescription for formulating dynamic EM tensor and Spin density tensor in NP formalism has been presented.
\end{itemize}
\textbf{This work is based on the paper titled ``The non-relativistic limit of the Einstein-Cartan-Dirac equations" which is \texttt{"under preparation"}} \cite{in_preparation_ECDNP}

\chapter{Conjecture: Curvature-Torsion Duality}\label{chp:duality conjecture}

\section{Curvature-Torsion duality}
In chapter (\ref{chp:introducing L_cs}), the idea of $L_{cs}$ is introduced. It asserts a symmetry between small mass (m) and large mass (M), which give the same value of $L_{cs}$. Both the masses enter the ECD equations through the same $L_{cs}$. The solution to the large mass M (for which mass density and correspondingly the `curvature' is dominant) is dual to the solution of small mass m (for which spin density and correspondingly the `torsion' is dominant). Both the solutions are labeled by $L_{cs}$; since it is the only coupling constant in the theory. Qualitatively, we call this the \textbf{`Curvature-Torsion' duality}. We want to establish this duality in the context of ECD system of equations with $L_{cs}$ and make this duality, mathematically more evident. 

\begin{equation}\label{1}
	G_{\mu\nu}(\{\}) = \frac{8\pi L_{CS}^{2}}{\hbar c} T_{\mu\nu} - \frac{1}{2}g_{\mu\nu}\left(\frac{8\pi L_{CS}^{2} }{\hbar c}\right)^2 S_{\alpha\beta\epsilon}S^{\alpha\beta\epsilon}
\end{equation}

\begin{equation}
	i\gamma^{\mu} \psi_{;\mu} = +\frac{3}{8} L_{CS}^2 \bar{\psi}\gamma^5\gamma_{\nu}\psi\gamma^5\gamma^{\nu}\psi + \frac{1}{2L_{CS}} \psi = 0
\end{equation}
This is the system of equations which we have to understand in details, find possible solutions, put bounds etc. By 'a solution', we mean 3 quantities - ($\psi$, g, K) where g and K are metric tensor and Contorsion tensor respectively. These quantities are the 3 independent fields in our theory. 

We know that affine connection is made up of Christoffel symbols and Contorsion tensor. With this affine connection, we construct The total curvature tensor 'R'. It is composed of two terms $R^0$ and Q. This notation, we adopt from \cite{teleparallel_gravity}. It can be written as $R = R^0 + Q$. $R^0$ is the usual Riemann curvature tensor expressible completely in terms of Christoffel symbols and their derivatives and Q is expressible completely in terms of Contorsion tensor K. The full equation is:

\begin{align}
	R^{\alpha}_{~\beta\mu\nu}(\Gamma) &= R^{\alpha}_{~\beta\mu\nu}(\{\}) +\nabla_{\mu}^{\{\}}K^{\alpha}_{~\nu\beta}-  \nabla_{\nu}^{\{\}}K^{\alpha}_{~\mu\beta} + K^{\alpha}_{~\mu\rho}K^{\rho}_{~\nu\beta} - K^{\alpha}_{~\nu\rho}K^{\rho}_{~\mu\beta} \\
	R &= R^0 + Q ------------- Symbolic~ equation
\end{align}
Note that, in the symbolic equation, we have dropped the Indices here. The symbols shouldn't be confused with curvature scalar. Also, Curvature and torsion should be thought of as independent here. Q has information about `torsion' and R has information about `curvature'. In a completely torsion dominated theory (e.g. teleparallel gravity), R = 0; $R^0=-Q$ and in curvature dominated theory (e.g. Einstein's GR), Q=0; $R=R^0$. We know from chapter (\ref{chp:introducing L_cs}), large masses contribute to gravity which is described by the curvature; determined by levi-civita connection (that is $R^0$). Torsion is negligible for large masses. Whereas, for small masses, the total curvature is zero.

\section{Establishing this duality through a conjecture}

We know that, for a given $L_{cs}$, a solution (if it exists) is valid for both  LM (large mass M) and its dual SM (small mass, $m_q$). This leads to an apparent contradiction because 'one solution' which fixes ($\psi$,g,K) can't physically describe both, SM and LM. It will be physically valid either for LM or SM as we expect the large mass solution to be gravity dominated, and the small mass solution to be torsion dominated. This is possible only if for a given $L_{cs}$, there are two solutions, one that is curvature dominated, and another that is torsion dominated. To account this, we propose the following conjecture: Assuming that a solution exists for a given $L_{cs}$, we call it solution (1) [S1]; characterized by three curvature parameters [R(1),$R^0(1)$,Q(1)]. It is governed by equation $R_{(1)}=R^0_{(1)}+Q_{(1)}$. Without loss of generality, we assume it to be curvature dominated. Conjecture is that, given a solution(1), there exists a solution(2) [S2] by construction; characterized by curvature parameters [R(2),$R^0(2)$,Q(2)] and governed by $R_{(2)}=R^0_{(2)}+Q_{(2)}$; such that  

\begin{equation}
	R_{(2)}-Q_{(2)} = -[R_{(1)}-Q_{(1)}] \Rightarrow R^0_{(2)} = -R^0_{(1)}
\end{equation}

This conjecture forces solution(2) to be torsion dominated. The properties of solution(1) and solution(2) are summarized in the table below. In the large mass limit, Q(1) is zero and we have the pure curvature solution R(1)=
$R^0(1)$ (This is general relativity). In the small mass limit, R(2) is zero, and we have the solution Q(2) = - $R^0(2)$ (This is teleparallel gravity). Duality map implies that R(1) = Q(2). These ideas are discussed in details in \cite{GRFessay2018}.

\begin{figure}[h!]\label{fig:curvature-torsion duality}
	\begin{center}
		\includegraphics[width=13cm]{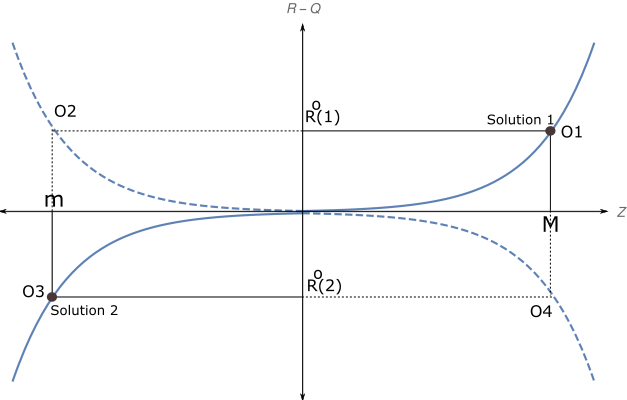}
		\caption{Curvature-Torsion duality}
	\end{center}
\end{figure}

\begin{tabular}{|| p{2.5cm} || c | p{2cm} | p{2.7cm} | c ||}\hline
	\textbf{Solution} & \textbf{Governing eqn} & \textbf{Valid for} &\textbf{Dominated by} & \textbf{Physical for}\\ \hline\hline
	Solution(1) $\rightarrow$& $R_{(1)}=R^0_{(1)}+Q_{(1)}$ & M and m  & 1) \textbf{Curvature}. 2) $R_{(1)}=R^0_{(1)} $& Large mass (M) \\
	\hline
	Solution(2) $\rightarrow$& $R_{(2)}=R^0_{(2)}+Q_{(2)}$ & M and m  & 1) \textbf{Torsion}.  2) $R^0_{(2)}=-Q_{(2)} $& Small mass (m) \\ \hline\hline
\end{tabular} \\
\\
\textbf{This conjecture automatically provides a natural duality between curvature and torsion for Large mass and small mass respectively}. In terms of above vocabulary, we summarize the curvature-torsion duality in figure below (\ref{fig:curvature-torsion duality}).

Here, we have plotted ``R-Q" Vs. $z= ln\Big[\frac{m}{m_{pl}}\Big]$. `M' and `m' have same $L_{cs}$. For M, $R^0$ (or equivalently `R-Q') is positive and dominates as mass goes high. It is shown as ``solution 1" [S1] in the first quadrant. In the limit of very high masses, curvature is fully given by Riemann curvature tensor $R^0$. For small mass m, $R^0$ (or equivalently `R-Q') is negative and goes on becoming more negative as as mass goes further low. As mass tends to zero, the total curvature also tends to zero and torsion balances Reimann curvature tensor $R^0$. Its solution is solution-2 [S2] in third quadrant. At $m = m_{pl}$, we have R-Q = 0 or $R^0$ = 0; where the total curvature is sourced only by torsion. There also exists a unphysical ``mirror universe" in which torsion is sourced by torsion and curvature by large masses. It is shown by dotted graph which rolls down from second quadrant to fourth quadrant.

\section{Attempting Solution(s) for this conjecture to support the curvature-torsion duality}
We proposed the Curvature-torsion duality conjecture in the previous section. At $m = m_{pl}$, R-Q = 0 or $R^0$ = 0. One of the allowed solution to this is on Minkowski space with torsion. So, we next attempt to find the solutions to ECD equations on Minkowski space with torsion and test the duality conjecture. We also propose a `test' which can make our claims falsifiable. First we establish the ingredients of ECD equations on Minkowski space with torsion in this section
\subsection{Dirac equation (Hehl-Datta Equation) on Minkowski Space with Torsion}
Dirac equation on $U_4$; called as Hehl datta (HD) equations are written explicitly in equations [\ref{1-DEonU_4final} - \ref{4-DEonU_4final}]. On Minkowski space with torsion, they are as follows (In NP formalism):
\begin{align}
	D F_1 + \delta^{*}F_2 &= i(b+a\xi)G_1 \label{eq:HD_in_NP_minkowski1}\\
	\Delta F_2 + \delta F_1 &= i(b+a\xi)G_2  \label{eq:HD_in_NP_minkowski2}\\
	DG_2 - \delta G_1 &= i(b+a\xi^*)F_2 \label{eq:HD_in_NP_minkowski3}\\
	\Delta G_1 - \delta^* G_2 &= i(b+a\xi^*)F_1 \label{eq:HD_in_NP_minkowski4}
\end{align}
The HD equations on Minkowski space with torsion in \textbf{Cartesian system}(ct,x,y,z) are as follows:
\begin{align}
	(\partial_0+\partial_3)F_1 + (\partial_1+i\partial_2) F_2 = i\sqrt{2}(b+a\xi)G_1 \label{eq:HD_in_cartesian_minkowski1}\\
	(\partial_0-\partial_3)F_2 + (\partial_1-i\partial_2) F_1 = i\sqrt{2}(b+a\xi)G_2\label{eq:HD_in_cartesian_minkowski2}\\
	(\partial_0+\partial_3)G_2 - (\partial_1-i\partial_2) G_1 = i\sqrt{2}(b+a\xi^*)F_2 \label{eq:HD_in_cartesian_minkowski3}\\
	(\partial_0-\partial_3)G_1 - (\partial_1+i\partial_2) G_2 = i\sqrt{2}(b+a\xi^*)F_1 \label{eq:HD_in_cartesian_minkowski4}
\end{align}
The HD equations on Minkowski space with torsion in \textbf{Cylindrical coordinate system}(ct,r,$\phi$,z) are as follows (we put c= 1):

\begin{align}
	r\partial_t F_1 + e^{i\phi} r\partial_r F_2  + ie^{i\phi}\partial_{\phi} F_2 + r\partial_z F_1 &= ir\sqrt{2} (b+a\xi)G_1 \label{eq:HD_in_cylindrical_minkowski1} \\
	r\partial_t F_2 + e^{-i\phi} r\partial_r F_1  - ie^{-i\phi}\partial_{\phi} F_1 - r\partial_z F_2&= ir\sqrt{2} (b+a\xi)G_2 \label{eq:HD_in_cylindrical_minkowski2}\\
	r\partial_t G_2 - e^{-i\phi}r\partial_r G_1  + ie^{-i\phi}\partial_{\phi} G_1 + cr\partial_z G_2&= ir\sqrt{2} (b+a\xi^*)F_2 \label{eq:HD_in_cylindrical_minkowski3}\\
	r\partial_t G_1 - e^{i\phi} r\partial_r G_2 - ie^{i\phi} \partial_{\phi} G_2 - r\partial_z G_1&= ir\sqrt{2} (b+a\xi^*)F_1 \label{eq:HD_in_cylindrical_minkowski4}
\end{align}
The HD equations on Minkowski space with torsion in \textbf{Spherical polar coordinate system}(ct,r,$\theta$,$\phi$) are as follows (we put c= 1):

\begin{align}
	\partial_t F_1+\cos{\theta} \partial_r F_1 - \frac{\sin{\theta}}{r}\partial_{\theta}F_1 + \frac{ie^{i\phi}}{r\sin{\theta}}\partial_{\phi}F_2 + e^{i\phi}\sin{\theta} \partial_rF_2  +\frac{ e^{i\phi}\cos{\theta}}{r}\partial_{\theta}F_2&= i\sqrt{2}(b+a\xi)G_1 \label{eq:HD_in_SPC_minkowski1}\\
	\partial_t F_2-\cos{\theta} \partial_r F_2 + \frac{\sin{\theta}}{r}\partial_{\theta}F_2 - \frac{ie^{-i\phi}}{r\sin{\theta}}\partial_{\phi}F_1 + e^{-i\phi}\sin{\theta} \partial_rF_1 +\frac{ e^{-i\phi}\cos{\theta}}{r}\partial_{\theta}F_1 &= i\sqrt{2}(b+a\xi)G_2\\
	\partial_t G_2+\cos{\theta} \partial_r G_2 - \frac{\sin{\theta}}{r}\partial_{\theta}G_2  + \frac{ie^{-i\phi}}{r\sin{\theta}}\partial_{\phi}G_1 - e^{-i\phi}\sin{\theta} \partial_rG_1 -\frac{ e^{-i\phi}\cos{\theta}}{r}\partial_{\theta}G_1  &= i\sqrt{2}(b+a\xi^*)F_2 \\
	\partial_t G_1-\cos{\theta} \partial_r G_1 + \frac{\sin{\theta}}{r}\partial_{\theta}G_1 -\frac{ie^{i\phi}}{r\sin{\theta}}\partial_{\phi}G_2 - e^{i\phi}\sin{\theta} \partial_rG_2  -\frac{ e^{i\phi}\cos{\theta}}{r}\partial_{\theta}G_2&= i\sqrt{2}(b+a\xi^*)F_1 \label{eq:HD_in_SPC_minkowski4}
\end{align}

\subsection{The Dynamical EM tensor ($T_{\mu\nu}$) on Minkowski space with torsion}

The dynamical EM tensor given in equation (\ref{dynamic EM tensor}). On Minkowski space, it assumes the following form:

\begin{equation}\label{dynamic EM tensor_minkowski}
	T_{\mu\nu} = \Sigma_{(\mu\nu)}(\{\}) = \frac{i\hbar c}{4}\Big[\bar{\psi}\gamma_{\mu} \partial_{\nu}\psi + \bar{\psi}\gamma_{\nu} \partial_{\mu} \psi - \partial_{\mu}\bar{\psi} \gamma_{\nu}\psi -\partial_{\nu}\bar{\psi} \gamma_{\mu}\psi  \Big]
\end{equation}
Its 10 components are given by following 10 equations:

\begin{equation}
	\begin{split}
		T_{21} =  \frac{i\hbar c}{4} \bigg{(}&
		\bar{F}_1\partial_1 F_1 + \bar{F}_2 \partial_1 F_2 + \bar{G}_1\partial_1 G_1 + \bar{G}_2\partial_1 G_2 -\bar{F}_2 \partial_0 F_1 - \bar{F}_1 \partial_0 F_2 + \bar{G}_2 \partial_0G_1 + \bar{G}_1\partial_0 G_2 \\& -\partial_1\bar{F}_1 F_1 - \partial_1\bar{F}_2 F_2 -\partial_1\bar{G}_1 G_1 - \partial_1\bar{G}_2 G_2 
		+\partial_0\bar{F}_2  F_1 + \partial_0\bar{F}_1  F_2 - \partial_0\bar{G}_2 G_1 - \partial_0\bar{G}_1 G_2 \bigg{)}
	\end{split}
\end{equation}

\begin{equation}
	\begin{split}
		T_{31} = \frac{i\hbar c}{4 } \bigg{(}& \bar{F}_1\partial_2 F_1 + \bar{F}_2 \partial_2 F_2 + \bar{G}_1\partial_2 G_1 + \bar{G}_2\partial_2 G_2  + i\bar{F}_2\partial_0 F_1 - i \bar{F}_1\partial_0 F_2 - i\bar{G}_2 \partial_0G_1 + i \bar{G}_1\partial_0 G_2 \\& -\partial_2\bar{F}_1 F_1 -\partial_2 \bar{F}_2 F_2 - G_1\partial_2 \bar{G}_1 -\partial_2 \bar{G}_2 G_2  -i\partial_0\bar{F}_2 F_1 + i\partial_0 \bar{F}_1 F_2 +i\partial_0 \bar{G}_2 G_1 -i\partial_0  \bar{G}_1 G_2  \bigg{)}
	\end{split}
\end{equation}

\begin{equation}
	\begin{split}
		T_{41} = \frac{i\hbar c}{4 } \bigg{(}& \bar{F}_1\partial_3 F_1 + \bar{F}_2 \partial_3F_2 + \bar{G}_1\partial_3 G_1 + \bar{G}_2\partial_3G_2  -\bar{F}_1 \partial_0 F_1 + \bar{F}_2\partial_0 F_2 + \bar{G}_1 \partial_0G_1 - \bar{G}_2\partial_0 G_2 \\& -\partial_3\bar{F}_1 F_1 - \partial_3\bar{F}_2 F_2 -\partial_3 \bar{G}_1 G_1 -\partial_3\bar{G}_2 G_2 +\partial_0\bar{F}_1 F_1 -\partial_0\bar{F}_2 F_2 -\partial_0 \bar{G}_1 G_1 +\partial_0 \bar{G}_2 G_2 \bigg{)}
	\end{split}
\end{equation}

\begin{equation}
	\begin{split}
		T_{32} = \frac{i\hbar c}{4} \bigg{(}& i\bar{F}_2 \partial_1 F_1 - i \bar{F}_1\partial_1 F_2 - i \bar{G}_2\partial_1 G_1 + i \bar{G}_1\partial_1 G_2  -\bar{F}_2 \partial_2 F_1 - \bar{F}_1\partial_2 F_2 + \bar{G}_2 \partial_2 G_1 + \bar{G}_1\partial_2 G_2 \\ & -i\partial_1\bar{F}_2 F_1 + i\partial_1 \bar{F}_1 F_2 + i\partial_1  \bar{G}_2 G_1 -i\partial_1  \bar{G}_1 G_2 + \partial_2\bar{F}_2F_1 + \partial_2 \bar{F}_1 F_2 -\partial_2\bar{G}_2 G_1 -\partial_2\bar{G}_1 G_2 \bigg{)}
	\end{split}
\end{equation}

\begin{equation}
	\begin{split}
		T_{42}= \frac{i\hbar c}{4} \bigg{(}& -\bar{F}_1 \partial_1 F_1 + \bar{F}_2\partial_1 F_2 + \bar{G}_1\partial_1 G_1 - \bar{G}_2 \partial_1 G_2\  -\bar{F}_2\partial_3 F_1 -\bar{F}_1 \partial_3F_2 +\bar{G}_2\partial_3 G_1 + \bar{G}_1\partial_3 G_2 \\& +\partial_1\bar{F}_1F_1 -\partial_1 \bar{F}_2F_2 -\partial_1 \bar{G}_1G_1 +\partial_1\bar{G}_2 G_2 +\partial_3 \bar{F}_2F_1 + \partial_3 \bar{F}_1 F_2 -\partial_3 \bar{G}_2 G_1 -\partial_3\bar{G}_1G_2	\bigg{)}
	\end{split}
\end{equation}

\begin{equation}
	\begin{split}
		T_{43} = \frac{i\hbar c}{4} \bigg{(}& -\bar{F}_1 \partial_2  F_1 + \bar{F}_2 \partial_2 F_2 + \bar{G}_1 \partial_2 G_1 - \bar{G}_2\partial_2  G_2  + i \bar{F}_2\partial_3 F_1 - i\bar{F}_1\partial_3 F_2 - i \bar{G}_2\partial_3G_1 + i \bar{G}_1\partial_3 G_2  \\&  +\partial_2 \bar{F}_1 F_1 -\partial_2 \bar{F}_2 F_2 -\partial_2 \bar{G}_1 G_1 +\partial_2  \bar{G}_2 G_2  -i\partial_3\bar{F}_2 F_1 + i\partial_3\bar{F}_1F_2 + i\partial_3 \bar{G}_2G_1 - i\partial_3\bar{G}_1 G_2  \bigg{)}
	\end{split}
\end{equation}

\begin{equation}
	\begin{split}
		T_{11} &= \frac{i\hbar c}{2} \bigg{(} \bar{G}_1\partial_0G_1 + \bar{G}_2\partial_0G_2 - \partial_0\bar{G}_1G_1 - \partial_0\bar{G}_2 G_2
		+ \bar{F}_1\partial_0F_1 + \bar{F}_2\partial_0F_2 - \partial_0\bar{F}_1 F_1 - \partial_0\bar{F}_2F_2\bigg{)}
	\end{split}
\end{equation}

\begin{equation}
	\begin{split}
		T_{22}&= \frac{i\hbar c}{2} \bigg{(}-\bar{F}_2\partial_1 F_1 - \bar{F}_1\partial_1 F_2 + \bar{G}_2\partial_1 G_1 + \bar{G}_1\partial_1 G_2  +\partial_1\bar{F}_2F_1 +\partial_1 \bar{F}_1 F_2 -\partial_1\bar{G}_2 G_1 - \partial_1\bar{G}_1 G_2 \bigg{)}
	\end{split}
\end{equation}

\begin{equation}
	\begin{split}
		T_{33}&= \frac{i\hbar c}{2} \bigg{(}  i\bar{F}_2\partial_2 F_1 - i\bar{F}_1 \partial_2 F_2 - i\bar{G}_2\partial_2  G_1 + i\bar{G}_1 \partial_2 G_2
		- i\partial_2 \bar{F}_2 F_1 + i\partial_2  \bar{F}_1 F_2 +i\partial_2  \bar{G}_2 G_1 - i\partial_2 \bar{G}_1 G_2	\bigg{)}
	\end{split}
\end{equation}

\begin{equation}
	\begin{split}
		T_{44} &= \frac{i\hbar c}{2 } \bigg{(} -\bar{F}_1\partial_3 F_1 + \bar{F}_2 \partial_3F_2 + \bar{G}_1 \partial_3G_1 - \bar{G}_2 \partial_3G_2 +\partial_3\bar{F}_1 F_1 - \partial_3\bar{F}_2F_2 -\partial_3\bar{G}_1 G_1 +\partial_3 \bar{G}_2 G_2\bigg{)}
	\end{split}
\end{equation}

\subsection{Calculation of the Spin density part which acts as a correction to $T_{ij}$}
The second term on RHS of equation (\ref{1}) on Minkowski space is given as $\frac{4\pi (L_{cs})^2}{\hbar c}\eta_{ij} S^{abc}S_{abc}$ which can be written as 
\begin{equation}
	\frac{4\pi l^2}{\hbar c}\eta_{\mu\nu} S^{\alpha\beta\gamma}S_{\alpha\beta\gamma} = 6\pi \hbar c l^2\eta_{\mu\nu} (F_1\bar{G}_1 + F_2\bar{G}_2)(\bar{F}_1 G_1 + \bar{F}_2 G_2) = 6\pi \hbar c l^2g_{\mu\nu} \xi\xi^*
\end{equation}

\section{Solutions to HD equation on $M_4$ with torsion and testing duality conjecture}

\subsection{Non static solution to HD equations by reducing it to 1+1 dimension (t,z)}\label{sec:1+1}

Assuming Dirac state to be a function of only t and z, and further assuming the ansatz of the form $F_1 = G_2 $ and $F_2 = G_1 $, the four equations in Cartesian coordinates [\ref{eq:HD_in_cartesian_minkowski1} - \ref{eq:HD_in_cartesian_minkowski4}] as well as four equations in cylindrical coordinates [\ref{eq:HD_in_cylindrical_minkowski1} - \ref{eq:HD_in_cylindrical_minkowski4}] reduce to following 2 independent equations. We note that $\xi = 2 Re(F_1\bar{F}_2)$. hence $\xi = \xi^*$. Also, we set c=1. From henceforth a and b would mean a(l) and b(l). 

\begin{align}
	\partial_t \psi_1 + \partial_z \psi_2 - i\sqrt{2} b \psi_1 +\frac{ia}{\sqrt{2}}(|\psi_2|^2-|\psi_1|^2)\psi_1 &= 0\\
	\partial_t \psi_2 + \partial_z \psi_2 +  i\sqrt{2}b \psi_2 +\frac{ia}{\sqrt{2}}(|\psi_1|^2-|\psi_2|^2)\psi_2 &= 0
\end{align}

where, $\psi_1 = F_1 + F_2$ and $\psi_2 = F_1 - F_2$. We put, $\sqrt{2}b = -m$ and $a = 2\sqrt{2}\lambda$ and obtain following:
\begin{align}
	\partial_t \psi_1 + \partial_z \psi_2 +im\psi_1 +2i\lambda(|\psi_2|^2-|\psi_1|^2)\psi_1 &= 0\\
	\partial_t \psi_2 + \partial_z \psi_2 -im\psi_2 +2i\lambda(|\psi_1|^2-|\psi_2|^2)\psi_2 &= 0
\end{align}

This is exactly same as equation (1) of \cite{alvarez1983numerical} This work by Alvarez finds the solutions to the above set of equations for the following solitary wave as ansatz.

\begin{equation}
	\psi = \begin{pmatrix} \psi_1\\ \psi_2 \end{pmatrix} = \begin{pmatrix} A(z)\\ iB(z)\end{pmatrix} e^{-i\Lambda t}
\end{equation}
Here A and B are real functions of z. Substituting this into above equations we obtain,

\begin{align}
	B' + (m - \Lambda)A - 2\lambda(A^2-B^2)A &= 0 \\
	A' + (m + \Lambda)B - 2\lambda(A^2-B^2)B &= 0
\end{align}

Solving these differential equations gives following solutions for A and B, we obtain folloing solution for A(z) and B(z). 

\begin{align} 
	A(z) &= \frac{-i2^{3/4} (\sqrt{2} b-\Lambda)}{\sqrt{a}} \frac{\sqrt{(\sqrt{2}b+\Lambda)} \cosh(z \sqrt{2 b^2-\Lambda ^2})}{[\Lambda  \cosh (2z\sqrt{2 b^2-\Lambda ^2}) - \sqrt{2} b]} \label {eq:1+1_A(z)}\\
	B(z) &= \frac{-i2^{3/4} (\sqrt{2} b+\Lambda)}{\sqrt{a}}\frac{\sqrt{(\sqrt{2}b-\Lambda)} \sinh(z \sqrt{2 b^2-\Lambda ^2})}{[\Lambda  \cosh (2z\sqrt{2 b^2-\Lambda ^2}) - \sqrt{2} b]} \label {eq:1+1_B(z)}
\end{align}

This is the generalization of the equations for A(z) and B(z) in section III of \cite{alvarez1983numerical}. Putting $\lambda = 0.5$ (equivalently $a = \sqrt{2}$) and $m = 1$ (equivalently $m_0 = -1$) in equations (\ref{eq:1+1_A(z)}), (\ref{eq:1+1_B(z)}), reduces to Alvarez's equations. This solution is also found by \cite{zecca} with a(l) = a($L_{pl}$) and b(l) = b($L_{pl}$). 

\subsubsection{Non static solution in 1+1 dimension (t,z)}

\begin{align}
	F_1&= \frac{\sqrt{(2b^2-\Lambda^2)}}{2}\bigg{[}\frac{-i2^{3/4}}{\sqrt{a}} \frac{\sqrt{(\sqrt{2}b-\Lambda)} \cosh(z \sqrt{2 b^2-\Lambda ^2})}{[\Lambda  \cosh (2z\sqrt{2 b^2-\Lambda ^2}) - \sqrt{2} b]} + \frac{2^{3/4}}{\sqrt{a}}\frac{\sqrt{(\sqrt{2}b+\Lambda)} \sinh(z \sqrt{2 b^2-\Lambda ^2})}{[\Lambda  \cosh (2z\sqrt{2 b^2-\Lambda ^2}) - \sqrt{2} b]} \bigg{]} e^{-i\Lambda t} \\
	F_2&= \frac{\sqrt{(2b^2-\Lambda^2)}}{2}\bigg{[}\frac{-i2^{3/4} }{\sqrt{a}} \frac{\sqrt{(\sqrt{2}b-\Lambda)} \cosh(z \sqrt{2 b^2-\Lambda ^2})}{[\Lambda  \cosh (2z\sqrt{2 b^2-\Lambda ^2}) - \sqrt{2} b]} - \frac{2^{3/4} }{\sqrt{a}}\frac{\sqrt{(\sqrt{2}b+\Lambda)} \sinh(z \sqrt{2 b^2-\Lambda ^2})}{[\Lambda  \cosh (2z\sqrt{2 b^2-\Lambda ^2}) - \sqrt{2} b]} \bigg{]} e^{-i\Lambda t} \\
	\xi &=  \dfrac{-2\sqrt{2}(2b^2-\Lambda^2)(\sqrt{2}b-\Lambda\cosh (2z\sqrt{2 b^2-\Lambda ^2} )}{a[\Lambda  \cosh (2z\sqrt{2 b^2-\Lambda ^2}) - \sqrt{2}b]^2}
\end{align}

\subsubsection{$(T-S)_{ij}$ for non-static Solution 1+1 dimension (t,z)}
\begin{equation}\label{eq:T-S_1+1_f(A,B)}
	(T-S)_{ij} = \hbar c\begin{pmatrix}
		\bigg{(}\Lambda[A^2+B^2] - \frac{a[A^2-B^2]^2}{2\sqrt{2}} \bigg{)}&  0  & -\Lambda AB  & 0 \\
		0 & \bigg{(}\frac{a[A^2-B^2]^2}{2\sqrt{2}} \bigg{)} & 0  & 0  \\
		-\Lambda AB & 0 & \bigg{(}\frac{a[A^2-B^2]^2}{2\sqrt{2}} \bigg{)}  & 0 \\
		0 & 0 & 0 & \bigg{(}[AB' - BA'] + \frac{a[A^2-B^2]^2}{2\sqrt{2}} \bigg{)} 
	\end{pmatrix}
\end{equation}

$\Lambda$ is a free parameter in the solution. We will analyze this tensor "T-S" for various types of values of $\Lambda$.

\underline{\textbf{Case 1: $\Lambda = 0$}} 

A and B reduce to following:

\begin{align} 
	A(z) &= 2i\sqrt{\frac{b}{a}}\cosh(\sqrt{2}zb) = \frac{i}{\sqrt{3\pi  l^3}}\cosh \bigg{(}\frac{z}{2 l}\bigg{)} \label {eq:1+1_A(z)Lambda=0}\\
	B(z) &= 2i\sqrt{\frac{b}{a}}\sinh(\sqrt{2}zb) = \frac{i}{\sqrt{3\pi  l^3}}\sinh\bigg{(}\frac{z}{2 l}\bigg{)} \label {eq:1+1_B(z)Lambda=0}\\
\end{align}

Dirac spinor in this case is:

\begin{equation}
	\begin{pmatrix}
		F_1\\ F_2\\ G_1\\ G_2\\
	\end{pmatrix} = \frac{1}{2} \begin{pmatrix}
		\frac{i}{\sqrt{3\pi  l^3}}\cosh \bigg{(}\frac{z}{2 l}\bigg{)} - \frac{1}{\sqrt{3\pi  l^3}}\sinh\bigg{(}\frac{z}{2 l}\bigg{)} \\ \frac{i}{\sqrt{3\pi  l^3}}\cosh \bigg{(}\frac{z}{2 l}\bigg{)} + \frac{1}{\sqrt{3\pi  l^3}}\sinh\bigg{(}\frac{z}{2 l}\bigg{)} \\ \frac{i}{\sqrt{3\pi  l^3}}\cosh \bigg{(}\frac{z}{2 l}\bigg{)} + \frac{1}{\sqrt{3\pi  l^3}}\sinh\bigg{(}\frac{z}{2 l}\bigg{)} \\ \frac{i}{\sqrt{3\pi  l^3}}\cosh \bigg{(}\frac{z}{2 l}\bigg{)} - \frac{1}{\sqrt{3\pi  l^3}}\sinh\bigg{(}\frac{z}{2 l}\bigg{)}
	\end{pmatrix}
\end{equation}

With this Dirac state, the quantity $\xi$ is a real positive constant and has following value:
\begin{equation}
	\xi = \frac{1}{6\pi  l^3}
\end{equation}

The tensor T-S reduces to following $\forall$ z

\begin{equation}
	(T-S)_{ij}|_{\Lambda=0} = \hbar c\begin{pmatrix}
		\bigg{(}\frac{-1}{6\pi  l^4}\bigg{)}&  0  & 0  & 0 \\
		0 & \bigg{(}\frac{1}{6\pi  l^4}\bigg{)} & 0  & 0  \\
		0 & 0 & \bigg{(}\frac{1}{6\pi  l^4}\bigg{)}  & 0 \\
		0 & 0 & 0 & 0
	\end{pmatrix}
\end{equation}
\underline{\textbf{Case II: $\Lambda = \sqrt{2}b$}} \\
This case makes everything reduce to zero and is z trivial solution. We don't want this type of solution.
\linebreak
\linebreak
\underline{\textbf{Case III: $\sqrt{2}b = 0 \Longrightarrow  l \longrightarrow \infty$}} 
\linebreak
\linebreak
We are not much interested in these kinds of solutions where $ l \longrightarrow \infty$. Because $L_{pl}$ can't go to infinity and $L_{cs}$ will go to infinity only for infinitely large or infinitesimally small masses.
\subsection{Plane wave solutions to HD equations}\label{sec:planewave solutions}

We begin with by substituting following plane wave ansatz in equations [\ref{eq:HD_in_cartesian_minkowski1} - \ref{eq:HD_in_cartesian_minkowski4}] as follows:
\begin{equation}
	\begin{bmatrix}
		F_1 \\ F_2 \\ G_1 \\ G_2 
	\end{bmatrix} = 
	\begin{bmatrix}
		u^0 \\ u^1 \\ \bar{v}_{0'} \\ \bar{v}_{1'}
	\end{bmatrix} e^{ik.x}
\end{equation}

With this ansatz, $\xi$ and $\xi^*$ are as follows

\begin{align}
	\xi &= u^A \bar{v}_{A'} \\
	\xi^* &= \bar{u}^{A'} v_A
\end{align}

We assume $\xi$ to be a real constant such that $\xi = u^A \bar{v}_{A'} = \bar{u}^{A'} v_A = \xi^*$ = (Real constant $\xi$). Putting The above ansatz in the equations [\ref{eq:HD_in_cartesian_minkowski1} - \ref{eq:HD_in_cartesian_minkowski4}], we obtain:

\begin{align}
	(k_0+k_3) u^0 + (k_1+ik_2) u^1 - \mu(\xi) \bar{v}_{0'}&= 0\\
	(k_0-k_3) u^1 + (k_1-ik_2)u^0  -\mu(\xi) \bar{v}_{1'}&= 0\\
	(k_0+k_3) \bar{v}_{1'} - (k_1-ik_2) \bar{v}_{0'} - \mu(\xi) u^1 &= 0\\
	(k_0-k_3) \bar{v}_{0'} - (k_1+ik_2) \bar{v}_{1'} - \mu(\xi) u^0 &= 0
\end{align}

Where $\mu(\xi) = \sqrt{2}(b+a\xi)$. $\mu$ is a function of $\xi$ which remains a undetermined before finding a complete solution. Only we have to make sure the fact that $\xi$ is a real constant. 

\begin{equation}
	\begin{pmatrix}
		(k_0+k_3) & (k_1+ik_2) & -\mu(\xi) &  0 \\
		(k_1-ik_2) & (k_0-k_3) & 0 & -\mu(\xi) \\
		0 & -\mu(\xi) &  - (k_1-ik_2) & (k_0+k_3) \\
		-\mu(\xi) & 0 & (k_0-k_3) & - (k_1+ik_2)
	\end{pmatrix} 
	\begin{pmatrix}
		u^0 \\ u^1 \\  \bar{v}_{0'} \\ \bar{v}_{1'}
	\end{pmatrix} = \begin{pmatrix}
		0 \\ 0 \\ 0 \\ 0
	\end{pmatrix}
\end{equation}

We fisrt assume $k_1 = k_2 = k_3 = 0 $ (This is like attempting a solution in a rest frame). The above equation reduces to 

\begin{equation}
	\begin{pmatrix}
		k_0& 0 & -\mu(\xi) &  0 \\
		0 & k_0 & 0 & -\mu(\xi) \\
		0 & -\mu(\xi) & 0 & k_0 \\
		-\mu(\xi) & 0 & k_0 & 0
	\end{pmatrix} 
	\begin{pmatrix}
		u^0 \\ u^1 \\  \bar{v}_{0'} \\ \bar{v}_{1'}
	\end{pmatrix} = \begin{pmatrix}
		0 \\ 0 \\ 0 \\ 0
	\end{pmatrix}
\end{equation}

For above system to have solution, we must have Det(coefficient matrix in 24) = 0. This gives 

\begin{align*}
	\Rightarrow & [k_0^2 - \mu(\xi)^2]^2 = 0 \\
	\Rightarrow & k_0 = \pm \mu(\xi)
\end{align*}

\subsubsection{The plane wave solution(s) for 2 cases}
\underline{\textbf{Case I: $k_0 = +\mu(\xi)$}}, general solution is of the form:

\begin{equation}
	\begin{pmatrix}
		F_1\\ F_2\\ G_1\\ G_2\\
	\end{pmatrix}=\frac{\alpha_1}{\sqrt{V}}\begin{pmatrix}
		0\\ 1\\ 0\\1\\
	\end{pmatrix}e^{i\mu(\xi) x_0} +  \frac{\beta_1}{\sqrt{V}}\begin{pmatrix}
		1 \\ 0 \\ 1 \\ 0\\
	\end{pmatrix}e^{i\mu(\xi) x_0}
\end{equation}

where, $|\alpha_1|^2+|\beta_1|^2 = 1$ is the normalization condition\\
Here $\xi$ and $\mu$ are as follows:
\begin{align*}
	\xi &= \frac{|\alpha_2|^2+|\beta_2|^2}{V} = \frac{1}{V}\\
	\mu &= \sqrt{2}\bigg{(}b + \frac{a}{V}\bigg{)}
\end{align*}

\underline{\textbf{Case II: $k_0 = -\mu(\xi)$}}, general solution is of the form:

\begin{equation}
	\begin{pmatrix}
		F_1\\ F_2\\ G_1\\ G_2\\
	\end{pmatrix}=\frac{\alpha_2}{\sqrt{V}}\begin{pmatrix}
		0\\ -1\\ 0\\1\\
	\end{pmatrix}e^{-i\mu(\xi) x_0} +  \frac{\beta_2}{\sqrt{V}}\begin{pmatrix}
		-1 \\ 0 \\ 1 \\ 0\\
	\end{pmatrix}e^{-i\mu(\xi) x_0}
\end{equation}

where, $|\alpha_2|^2+|\beta_2|^2 = 1$ is the normalization condition\\
Here $\xi$ and $\mu$ are as follows:
\begin{align*}
	\xi &= \frac{-|\alpha_2|^2-|\beta_2|^2}{V} = \frac{-1}{V}\\
	\mu &= \sqrt{2}\bigg{(}b - \frac{a}{V}\bigg{)}
\end{align*}

\subsubsection{$(T-S)_{ij}$ for Plane wave solutions}
For case I:

\begin{equation}
	(T-S)_{ij} = \hbar c
	\begin{pmatrix}
		-\bigg{(}\frac{V+18\pi  l^3}{V^2  l} \bigg{)}& 0 & 0 & 0\\
		0 & \bigg{(}\frac{6\pi  l^2}{V^2}\bigg{)}& 0 & 0\\
		0 & 0 & \bigg{(} \frac{6\pi  l^2}{V^2}\bigg{)} & 0\\
		0 & 0 & 0 & \bigg{(} \frac{6\pi  l^2}{V^2}\bigg{)}\\
	\end{pmatrix}
\end{equation}

For case II

\begin{equation}
	(T-S)_{ij} = \hbar c
	\begin{pmatrix}
		-\bigg{(}\frac{V-18\pi  l^3}{V^2  l} \bigg{)}& 0 & 0 & 0\\
		0 & \bigg{(}\frac{6\pi  l^2}{V^2}\bigg{)}& 0 & 0\\
		0 & 0 & \bigg{(} \frac{6\pi  l^2}{V^2}\bigg{)} & 0\\
		0 & 0 & 0 & \bigg{(} \frac{6\pi  l^2}{V^2}\bigg{)}\\
	\end{pmatrix}
\end{equation}

Comments:\\

We observe that for both cases that $(T-S)_{ij}$ goes to zero only when $V\longrightarrow \infty$. But V going to $\infty$ implies $\xi$ going to zero. So in case of vanishing torsion only, T-S has any hopes of becoming zero.

\subsection{Solution by reduction to (2+1) Dim in cylindrical coordinates (t,r,$\phi$)}\label{sec:2+1}
\noindent We put z-dependence to zero in the equations [\ref{eq:HD_in_cylindrical_minkowski1} - \ref{eq:HD_in_cylindrical_minkowski4}] and get the following equations:

\begin{align}
	r\partial_t F_1 + cr\partial_r F_2 e^{i\phi} + ic\partial_{\phi} F_2 e^{i\phi}  F_1 &= icr\sqrt{2} (b+a\xi)G_1 \\
	r\partial_t F_2 + cr\partial_r F_1 e^{-i\phi} - ic\partial_{\phi} F_1 e^{-i\phi}&= icr\sqrt{2} (b+a\xi)G_2 \\
	r\partial_t G_2 - cr\partial_r G_1 e^{-i\phi} + ic\partial_{\phi} G_1 e^{-i\phi} &= icr\sqrt{2} (b+a\xi^*)F_2 \\
	r\partial_t G_1 - cr\partial_r G_2 e^{i\phi} - ic\partial_{\phi} G_2 e^{i\phi} &= icr\sqrt{2} (b+a\xi^*)F_1 
\end{align}
We take the ansatz, $F_2 = G_2$ and $F_1 = -G_1$

\begin{align}
	r\partial_t F_1 + r\partial_r F_2 e^{i\phi} + i\partial_{\phi} F_2 e^{i\phi} &= -ir\sqrt{2} (b+a\xi)F_1\\
	r\partial_t F_2 + r\partial_r F_1 e^{-i\phi} - i\partial_{\phi} F_1 e^{-i\phi} &= ir\sqrt{2} (b+a\xi)F_2
\end{align}
We choose following ansatz in the above equation

\begin{align}
	\begin{bmatrix}
		F_1 \\ F_2
	\end{bmatrix} =
	\begin{bmatrix}
		iA(r)e^{\frac{i\phi}{2}}\\ B(r)e^{\frac{-i\phi}{2}}
	\end{bmatrix}e^{-i\omega t}
\end{align}
Putting this ansatz in above equations, we obtain the 2 differential equations as follows:

\begin{align}
	-rB\omega + r\partial_rA + \frac{A}{2} &= r\sqrt{2}[b+a(B^2-A^2)]B\\
	rA\omega + r\partial_rB + \frac{B}{2} &= r\sqrt{2}[b+a(B^2-A^2)]A 
\end{align}
We add and subtract above 2 equations and put following in it:

\begin{align}
	\psi_1 &= B(r) + A(r)\\
	\psi_2 &= B(r) - A(r)
\end{align}
And we obtain:

\begin{align}
	-r\omega \psi_2 + r\psi_1' + \frac{\psi_1}{2} - r\sqrt{2}(b+a\psi_1\psi_2)\psi_1 &= 0\\
	r\omega \psi_1 + r\psi_2' + \frac{\psi_2}{2} + r\sqrt{2}(b+a\psi_1\psi_2)\psi_2 &= 0
\end{align}
We aim to solve this system of equations. With $\omega = 0$, We get 
\begin{align}
	\psi_1 = \bigg{[}\frac{c_2 e^{\sqrt{2}br}}{r^{\big{(}\frac{1-2\sqrt{2}ac_1}{2}\big{)}}} \bigg{]}~~~~~~~~~~~~~~
	\psi_2 = \bigg{[} \frac{c_1 e^{-\sqrt{2}br} r^{\big{(}\frac{-1-2\sqrt{2}ac_1}{2}\big{)}}  }{c_2} \bigg{]}
\end{align}
This is clearly unphysical because $\psi_1$ blows up $\forall$ non-zero $c_2$; and making $c_2$ zero blows up $\psi_2$. So, we conclude that, \textbf{static solution to the above system of equation is unphysical}. So $\omega$ can't be zero. Some further attempts to solve it numerically are given in Appendix. 

\subsection{Solution by reduction to (3+1) Dim in spherical coordinates (t,r,$\theta$,$\phi$)}\label{sec:3+1}

We begin by putting following ansatz in HD equations with spherical coordinates:
\begin{align}
	\begin{bmatrix}
		F_1 \\ F_2 \\G_1 \\ G_2
	\end{bmatrix} =
	\begin{bmatrix}
		R_{-\frac{1}{2}}(r) S_{-\frac{1}{2}}(\theta) e^{+i\phi/2}\\ R_{+\frac{1}{2}}(r)  S_{+\frac{1}{2}}(\theta)e^{-i\phi/2} \\ R_{+\frac{1}{2}}(r) S_{-\frac{1}{2}}(\theta) e^{+i\phi/2} \\ R_{-\frac{1}{2}}(r) S_{+\frac{1}{2}}(\theta)  e^{-i\phi/2}
	\end{bmatrix}e^{-i\omega t}
\end{align}

With this ansatz, equations [\ref{eq:HD_in_SPC_minkowski1} - \ref{eq:HD_in_SPC_minkowski4}] take the following form:

\begin{equation}
	\begin{split}
		\bigg{(}&-i\omega R_{-\frac{1}{2}} S_{-\frac{1}{2}} + \cos{\theta}R_{-\frac{1}{2}}' S_{-\frac{1}{2}}  - \frac{\sin{\theta}}{r}R_{-\frac{1}{2}} S_{-\frac{1}{2}}' + \frac{1}{2r\sin{\theta}}R_{+\frac{1}{2}} S_{+\frac{1}{2}} + \sin{\theta} R_{+\frac{1}{2}}'  S_{+\frac{1}{2}}   +\frac{\cos{\theta}}{r}R_{+\frac{1}{2}}  S_{+\frac{1}{2}}'\bigg{)} \\ &= i\sqrt{2}(b+a\xi)R_{+\frac{1}{2}} S_{-\frac{1}{2}}
	\end{split}
\end{equation}

\begin{equation}
	\begin{split}
		\bigg{(}&-i\omega  R_{+\frac{1}{2}}  S_{+\frac{1}{2}} -\cos{\theta} R_{+\frac{1}{2}}'S_{+\frac{1}{2}} + \frac{\sin{\theta}}{r}R_{+\frac{1}{2}} S_{+\frac{1}{2}}' - \frac{1}{2r\sin{\theta}}R_{-\frac{1}{2}}  S_{-\frac{1}{2}}+ \sin{\theta} R_{-\frac{1}{2}}'  S_{-\frac{1}{2}} +\frac{\cos{\theta}}{r}R_{-\frac{1}{2}}  S_{-\frac{1}{2}}\bigg{)}' \\&= i\sqrt{2}(b+a\xi)R_{-\frac{1}{2}}(r) S_{+\frac{1}{2}}(\theta)
	\end{split}
\end{equation}

\begin{equation}
	\begin{split}
		\bigg{(}&-i\omega R_{-\frac{1}{2}} S_{+\frac{1}{2}}+ \cos{\theta} R_{-\frac{1}{2}}' S_{+\frac{1}{2}} - \frac{\sin{\theta}}{r}R_{-\frac{1}{2}} S_{+\frac{1}{2}}'  + \frac{1}{2r\sin{\theta}}R_{+\frac{1}{2}} S_{-\frac{1}{2}} - \sin{\theta} R_{+\frac{1}{2}}' S_{-\frac{1}{2}} -\frac{\cos{\theta}}{r}R_{+\frac{1}{2}} S_{-\frac{1}{2}}' \bigg{)} \\ &= i\sqrt{2}(b+a\xi^*)R_{+\frac{1}{2}}(r) S_{+\frac{1}{2}}(\theta)
	\end{split}
\end{equation}

\begin{equation}
	\begin{split}
		\bigg{(}&-i\omega R_{+\frac{1}{2}}(r) S_{-\frac{1}{2}}(\theta)-\cos{\theta} R_{+\frac{1}{2}}' S_{-\frac{1}{2}} + \frac{\sin{\theta}}{r}R_{+\frac{1}{2}} S_{-\frac{1}{2}}' -\frac{1}{2r\sin{\theta}}R_{-\frac{1}{2}} S_{+\frac{1}{2}} - \sin{\theta} R_{-\frac{1}{2}}' S_{+\frac{1}{2}} -\frac{ \cos{\theta}}{r}R_{-\frac{1}{2}}S_{+\frac{1}{2}}' \bigg{)} \\&= i\sqrt{2}(b+a\xi^*)R_{-\frac{1}{2}}S_{-\frac{1}{2}}
	\end{split}
\end{equation}

Where 

\begin{align}
	\xi &= R_{-\frac{1}{2}} S_{-\frac{1}{2}} \bar{R}_{+\frac{1}{2}} \bar{S}_{-\frac{1}{2}} + 
	R_{+\frac{1}{2}}S_{+\frac{1}{2}} \bar{R}_{-\frac{1}{2}} \bar{S}_{-\frac{1}{2}} \\
	\xi^* &= \bar{R}_{-\frac{1}{2}} \bar{S}_{-\frac{1}{2}} R_{+\frac{1}{2}} S_{-\frac{1}{2}} + 
	\bar{R}_{+\frac{1}{2}} \bar{S}_{+\frac{1}{2}} R_{-\frac{1}{2}} S_{-\frac{1}{2}} 
\end{align}

\section{Summary}

\begin{itemize}
	\item Curvature-Torsion duality conjecture presented.
	\item Formulated the ECD theory on Minkowski space with torsion.
	\item Solution to Dirac equation on $M_4$ with torsion by reducing the problem to (1+1)- Dim found. However, it cannot make T-S vanish for any values of free parameters. 
	\item  Plane wave solutions to Dirac equation on $M_4$ with torsion exist. Explicit expression of plane wave solutions with only time dependence found. However, it cannot make T-S vanish for any values of free parameters. 
	\item Solution by reducing the problem to (2+1)-Dim attempted. Equations are presented. However solution is not found yet. More has been discussed in chapter (\ref{chp:discussions}) 
	\item  Solution by reducing the problem to (3+1)-Dim attempted. Equations are presented. However solution is not found yet. More has been discussed in chapter (\ref{chp:discussions})
\end{itemize}

\chapter{Discussion}\label{chp:discussions}

\section{Conclusions and outlook}

\# As discussed in (\ref{sec:intro_NR}), we found the non-relativistic limit of Einstein-Dirac system [That is  self-gravitating Dirac field on $V_4$] with \textbf{generic metric} and found that it indeed reproduces the results of \cite{guilini_grosardt} viz. at leading order, the NR limit is Schr\"{o}dinger-Newton equation. In short, We generalized their work (by considering generic metric). Next, we found the NR limit for Einatein-Cartan-Dirac system [That is self-gravitating Dirac field on $U_4$]. At leading order, it also turns out to be Schr\"{o}dinger-Newton equation. This suggests that, at leading order, there is NO effect of torsion in the non-relativistic limit. So in order to experimentally probe the effects of torsion, we will have to go higher orders. Our method of finding NR limit also provides a prescription for finding the correction terms due to torsion. When we compare it with Einstein-Dirac system, we can analyze the orders of the coupled equations which are altered due to torsion through this prescription. This has huge implications for anyone who would like to design experiments to detect torsion in future. This was all w.r.t standard ECD theory (as in, ECD with standard length scales as couplings). We also have some interesting results after we take the NR limit of ECD equations modified with $L_{cs}$. In high mass limit, we obtain Poisson equation with delta function source. We showed that, this result is valid for all energy levels; not only in Non-relativistic limit. This has interesting implications. We know from \cite{bassi2013models} that very large masses are highly localized (In terms of their wave-function, it is already in a collapsed state). So it behaves classically. Hence, we obtain Poisson equation with delta function source (localized source for point particles) even for relativistic case. This is consistent with ordinary GR and Newton's law. It proves that, the modification of theory with $L_{cs}$ is consistent with the known theories in large mass limit. In the small mass limit however, since $L_{cs}$ goes as 1/c as opposed to $1/c^2$ (which was the case with large mass limit), we find that Poisson equation is $\nabla^2\phi = 0$. So for $m \ll m_{pl}$, we find that, gravitational field as well as quantum state vanishes at $1/c^2$. This gives a falsifiable test for the idea of $L_{cs}$. Gravity between very small masses would be weaker than the predictions of GR if one does an experiment to test the inverse square law between the pair of very small masses.

\# In chapter (\ref{Np/ECDinNP}), we formulated ECD theory in NP formalism. Dirac equation is modified on $U_4$ and presented in NP formalism in equations [\ref{1-DEonU_4final} - \ref{4-DEonU_4final}]. We also provided the prescription for finding the expression of EM tensor in NP formalism and also calculated the spin density term which acts as a correction to the dynamic (and symmetrical) EM tensor; together which contribute to the Einstein's tensor made up from Christoffel connection. Contorsion spin coefficients in NP formalism are also expressed in terms of Dirac state.   

\# Chapter (\ref{chp:duality conjecture}) discusses the curvature-torsion duality. As we have mentioned in chapter (\ref{chp:introducing L_cs}), the idea of $L_{cs}$ naturally hints towards a symmetry between higher and lower masses. In this chapter, we have made this duality mathematically more evident through a conjecture. One way to test the conjecture is to find the solutions on Minkowski space with torsion and test the components of tensor ``T-S". This tensor doesn't vanish for the 2 solutions which are presented in section (\ref{sec:1+1}) and section (\ref{sec:planewave solutions}). So these solutions do not support our conjecture. Solutions by reducing the problem to (2+1)-Dim and (3+1)-Dim are under investigation. The big picture which Curvature-torsion duality presents, has been discussed in details in an essay submitted to Gravity research foundation. It can be looked up in \cite{GRFessay2018}.

\section{Future plans}

\begin{itemize}
	\item Continue the self-study of gravitational theories with torsion from both theoretical and experimental perspectives.    
	\item To find the non-relativistic limit of ECD equations with new length scale $L_{cs}$ for the masses which are comparable to plank mass. We speculate that it will be something different from Schr\"{o}dinger-Newton equation. 
	\item To understand the implications of the idea of $L_{cs}$ (in its low mass limit) in the known theories of particle physics. In its low mass limit, Dirac equation has cubic non-linear term with $\lambda_c$ as coupling constant. It can be tested against known experimental data and also to make quantitative predictions for the new experiments. Another plan is to work on the falsifiable test for the idea of $L_{cs}$ presented in the first paragraph of discussions. 
	\item To find a solution to Hehl-Datta equation on Minkowski space with torsion (either by continuing the study of reducing the equations to 2+1 Dim and 3+1 Dim as mentioned in sections (\ref{sec:2+1}) and (\ref{sec:3+1}) or by some other method) such that the tensor ``T-S" becomes zero. The aim is testing the hypothesis of curvature-torsion duality. 
\end{itemize}

\appendix
\chapter{Results of Long calculations used in Chapter 4 - Non-relativistic limit of ECD equations}
\label{app:NR}
\section{Form of Einstein's tensor evaluated from the generic metric upto second order}\label{NR:generic G}

\begin{equation*}
	g_{\mu\nu}(x) = \eta_{\mu\nu} + \sum_{n = 1}^{\infty} \bigg{(}\frac{\sqrt{\hbar}}{c} \bigg{)}^n g_{\mu\nu}^{[n]}(x)
\end{equation*}
The metric and its inverse, up to second order, can be written as following:

\begin{align}
	g_{\mu\nu} &= \eta_{\mu\nu}+ \Big(\frac{\sqrt{\hbar}}{c}\Big) g_{\mu\nu}^{[1]} + \Big(\frac{\hbar}{c^2}\Big) g_{\mu\nu}^{[2]} \\
	g^{\mu\nu} &= \eta^{\mu\nu} - \Big(\frac{\sqrt{\hbar}}{c}\Big) g^{\mu\nu[1]} - \Big(\frac{\hbar}{c^2}\Big)[g^{\mu\nu[1]} + g^{\mu\nu[2]}]
\end{align}
We evaluate Christoffel symbols, Riemann curvature tensor, Ricci tensor and scalar curvature up to second order using above 2 equations and obtain Einstein tensor at the end. Einstein's tensor $G_{\mu\nu}$ is then given by 

\begin{equation}
	G_{\mu\nu} = \Big(\frac{\sqrt{\hbar}}{c}\Big) G_{\mu\nu}^{[1]} + \Big(\frac{\hbar}{c^2}\Big) G_{\mu\nu}^{[2]}
\end{equation}
Where
\begin{align}
	G_{\mu\nu}^{[1]} &= -\frac{1}{2}\square \overline{g}_{\mu\nu}^{[1]}; ~~~~~~~~~~~~~~~~~ where~~\overline{g}_{ij}^{[1]} = g_{\mu\nu}^{[1]} - \frac{1}{2}\eta_{\mu\nu} g^{[1]}; ~~~~~g^{[1]} =(\eta^{\mu\nu}g_{\mu\nu}^{[1]})\\
	G_{\mu\nu}^{[2]} &= -\frac{1}{2}\square \overline{g}_{\mu\nu}^{(2)} + f(g_{\mu\nu}^{[1]})~~~~~~ where~~\overline{g}_{ij}^{[2]} = g_{\mu\nu}^{[2]} - \frac{1}{2}\eta_{\mu\nu} g^{[2]}; ~~~~~~g^{[2]} =(\eta^{\mu\nu}g_{\mu\nu}^{[2]})
\end{align}
f is a function of $g_{\mu\nu}^{[1]}$ and is given by following equation:

\begin{align*}
	f(g^{[1]}_{\mu\nu}) = - \frac{1}{4}\Big[2\partial^{\lambda} g^{[1]}\partial_{\nu} g_{\lambda\mu}^{[1]} - 2\partial ^{\lambda} g^{[1]}\partial_{\lambda} g _{\mu\nu}^{[1]} - \partial_{\rho} g _{\nu}^{\lambda[1]} \partial_{\mu} g _{\lambda}^{\rho[1]} - \partial_{\rho} g _{\nu}^{\lambda[1]} \partial_{\lambda} g _{\mu}^{\rho[1]} +\\
	\partial_{\rho} g _{\nu}^{\lambda[1]} \partial^{\rho} g _{\lambda\mu}^{[1]} + \partial_{\nu} g_{\rho}^{\lambda[1]} \partial_{\mu} g _{\lambda}^{\rho[1]} + \partial_{\nu} g_{\rho}^{\lambda[1]} \partial_{\lambda} g _{\mu}^{\rho[1]} - \partial_{\nu} g_{\rho}^{\lambda[1]} \partial^{\rho} g_{\lambda\mu}^{[1]}\Big] \\
	-\frac{1}{8}\Big[2\partial^{\lambda} g^{[1]}\partial_{\nu} g_{\lambda\mu}^{[1]} - 2\eta_{\mu\nu}\partial^{\lambda} g^{[1]}\partial_{\lambda} g^{[1]} - \partial_{\rho} g _{\nu}^{\lambda[1]} \partial_{\mu} g _{\lambda}^{\rho[1]} - \partial_{\rho} g_{\mu}^{\lambda[1]} \partial_{\lambda} g _{\nu}^{\rho[1]} \\
	+ \partial_{\rho} g_{\mu}^{\lambda[1]} \partial^{\rho} g _{\lambda\nu}^{[1]} + \partial_{\mu} g_{\rho}^{\lambda[1]} \partial_{\nu} g_{\lambda}^{\rho[1]} + \partial_{\mu} g_{\rho}^{\lambda[1]} \partial_{\lambda} g _{\nu}^{\rho[1]} - \partial_{\nu} g_{\rho}^{\lambda[1]} \partial^{\rho} g_{\lambda\mu[1]}\Big]
\end{align*}

\section{Metric and Christoffel symbol components}\label{NR:metric}
\underline{Metric form}:
\begin{align}
	g_{\mu\nu} &= 
	\begin{pmatrix}
		1 + \frac{\hbar F(\vec{x},t)}{c^2} & 0 & 0 & 0 \\
		0 & -1 + \frac{\hbar F(\vec{x},t)}{c^2} & 0 & 0 \\
		0 & 0 & -1 + \frac{\hbar F(\vec{x},t)}{c^2} & 0 \\
		0 & 0 & 0 & -1 + \frac{\hbar F(\vec{x},t)}{c^2}
	\end{pmatrix} +  \sum_{n=3}^{\infty} O\Big( \frac{1}{c^n}\Big) \\
	g^{\mu\nu} &= 
	\begin{pmatrix}
		1 - \frac{\hbar F(\vec{x},t)}{c^2} & 0 & 0 & 0 \\
		0 & -1 - \frac{\hbar F(\vec{x},t)}{c^2} & 0 & 0 \\
		0 & 0 & -1 - \frac{\hbar F(\vec{x},t)}{c^2} & 0 \\
		0 & 0 & 0 & -1 - \frac{\hbar F(\vec{x},t)}{c^2}
	\end{pmatrix} +  \sum_{n=3}^{\infty} O\Big( \frac{1}{c^n}\Big) 
\end{align}
\\
\underline{ Christoffel Connection}:\\
The non-zero Christoffel connection components (up to 2nd order in 1/c) corresponding to metric $g_{\mu\nu}$ defined above are as follows
\begin{align}
	\begin{aligned}
		\Gamma^{0}_{0\mu} &= \frac{-\hbar \partial_{\mu}F(\vec{x},t)}{2c^2} +  \sum_{n=3}^{\infty} O\Big( \frac{1}{c^n}\Big)\\
		\Gamma^{\mu}_{0 0} &= \frac{-\hbar \partial_{\mu}F(\vec{x},t)}{2c^2} + \sum_{n=3}^{\infty} O\Big( \frac{1}{c^n}\Big)\\
		\Gamma^{\mu}_{\mu \mu} &= \frac{+\hbar \partial_{\mu}F(\vec{x},t)}{2c^2} +  \sum_{n=3}^{\infty} O\Big( \frac{1}{c^n}\Big)\\
	\end{aligned}
\end{align}

[Here $\mu = {1,2,3}$ i.e., it refers to the spatial coordinates.]

Other non zero Christoffel connection components have all orders of terms from order 3 viz. $\sum_{n=3}^{\infty} O\Big( \frac{1}{c^n}\Big)$

\section{Tetrad components}\label{NR:tetrad}

\begin{align}
	e_{\mu}^{(i)} &= 
	\begin{pmatrix}
		1 + \frac{\hbar F(\vec{x},t)}{2c^2} & 0 & 0 & 0 \\
		0 & 1 - \frac{\hbar F(\vec{x},t)}{2c^2} & 0 & 0 \\
		0 & 0 & 1 - \frac{\hbar F(\vec{x},t)}{2c^2} & 0 \\
		0 & 0 & 0 & 1 - \frac{\hbar F(\vec{x},t)}{2c^2} 
	\end{pmatrix}+ \sum_{n=3}^{\infty} O\Big( \frac{1}{c^n}\Big) \\
	e^{\mu}_{(i)} &= 
	\begin{pmatrix}
		1 - \frac{\hbar F(\vec{x},t)}{2c^2} & 0 & 0 & 0 \\
		0 & 1 + \frac{\hbar F(\vec{x},t)}{2c^2} & 0 & 0 \\
		0 & 0 & 1 + \frac{\hbar F(\vec{x},t)}{2c^2} & 0 \\
		0 & 0 & 0 & 1 + \frac{\hbar F(\vec{x},t)}{2c^2}   
	\end{pmatrix}+ \sum_{n=3}^{\infty} O\Big( \frac{1}{c^n}\Big) \\
	e_{\nu(k)} &= 
	\begin{pmatrix}
		1 + \frac{\hbar F(\vec{x},t)}{2c^2} & 0 & 0 & 0 \\
		0 & -1 + \frac{\hbar F(\vec{x},t)}{2c^2} & 0 & 0 \\
		0 & 0 & -1 + \frac{\hbar F(\vec{x},t)}{2c^2} & 0 \\
		0 & 0 & 0 & -1 + \frac{\hbar F(\vec{x},t)}{2c^2} 
	\end{pmatrix}+ \sum_{n=3}^{\infty} O\Big( \frac{1}{c^n}\Big)  \\
	e^{\nu(k)} &= 
	\begin{pmatrix}
		1 - \frac{\hbar F(\vec{x},t)}{2c^2} & 0 & 0 & 0 \\
		0 & -1 - \frac{\hbar F(\vec{x},t)}{2c^2} & 0 & 0 \\
		0 & 0 & -1 - \frac{\hbar F(\vec{x},t)}{2c^2} & 0 \\
		0 & 0 & 0 & -1 - \frac{\hbar F(\vec{x},t)}{2c^2}   
	\end{pmatrix}+ \sum_{n=3}^{\infty} O\Big( \frac{1}{c^n}\Big)
\end{align}

\section{Components of the Riemann part of Spin Connection $\gamma^o_{(a)(b)(c)}$}\label{NR:spin connection}
\begin{align}
	\begin{aligned}
		\gamma^o_{(0)(0)(0)} &= \frac{-\hbar\partial_0 F}{2c^2} \frac{\Big(1+\frac{\hbar F}{ 2c^2}\Big)}{\Big(1-\frac{\hbar F}{ 2c^2}\Big)}+ \sum_{n=3}^{\infty} O\Big( \frac{1}{c^n}\Big) ~~~~~~~~~
		\gamma^o_{(i)(0)(0)} = \Big(\frac{-\hbar\partial_i F}{2c^2}\Big) \frac{\hbar F/ 2c^2}{\Big(1+\frac{\hbar F}{ 2c^2}\Big)}+ \sum_{n=5}^{\infty} O\Big( \frac{1}{c^n}\Big)\\
		\gamma^o_{(0)(i)(0)} &= \frac{-\hbar\partial_i F}{2c^2} \frac{\Big(1+\frac{\hbar F}{ 2c^2}\Big)}{\Big(1-\frac{\hbar F}{ 2c^2}\Big)}+ \sum_{n=3}^{\infty} O\Big( \frac{1}{c^n}\Big) ~~~~~~~~~
		\gamma^o_{(0)(0)(i)} =  \frac{\hbar\partial_i F}{2c^2} \frac{1}{\Big(1+\frac{\hbar F}{ 2c^2}\Big)} \\
		\gamma^o_{(i)(i)(i)} &= \frac{\hbar\partial_i F}{2c^2} \frac{\hbar F/ 2c^2}{\Big(1+\frac{\hbar F}{ 2c^2}\Big)}+ \sum_{n=5}^{\infty} O\Big( \frac{1}{c^n}\Big) ~~~~~~~~~~~~
		\gamma^o_{(i)(i)(0)} = \gamma^o_{(i)(0)(i)} = + \sum_{n=3}^{\infty} O\Big( \frac{1}{c^n}\Big)\\
		\gamma^o_{(0)(i)(i)} &= \frac{-\hbar\partial_0F }{2c^2}+ \sum_{n=3}^{\infty} O\Big( \frac{1}{c^n}\Big) ~~~~~~~~~~~~~~~~~~~~~~
		\gamma^o_{(0)(i)(j)} = \gamma^o_{i0j} = \gamma^o_{ij0} = + \sum_{n=3}^{\infty} O\Big( \frac{1}{c^n}\Big) \\
		\gamma^o_{(i)(j)(j)} &=  \frac{-\hbar\partial_0 F}{2c^2} \frac{\Big(1-\frac{\hbar F}{ 2c^2}\Big)}{\Big(1+\frac{\hbar F}{ 2c^2}\Big)}+ \sum_{n=3}^{\infty} O\Big( \frac{1}{c^n}\Big) ~~~~~~~~~ \gamma^o_{(i)(j)(k)} = \gamma^o_{(i)(j)(i)} = \gamma^o_{(j)(j)(i)} = + \sum_{n=3}^{\infty} O\Big( \frac{1}{c^n}\Big)
	\end{aligned}
\end{align}

\section{Components for Einstein's tensor}\label{NR:G}

In this weak field limit we get, 

\begin{equation}
	G_{\mu\nu}^{[2]} = -\frac{1}{2}\square \overline{g}_{\mu\nu}^{[2]}; where ~~~~~~\overline{g}_{\mu\nu}^{[2]} = g_{\mu\nu}^{[2]} - \frac{1}{2}\eta_{\mu\nu} (\eta^{\alpha\beta}h_{\alpha\beta}) 
\end{equation}

\begin{equation}
	\eta^{\mu\nu}h_{\mu\nu}= \left(
	\begin{tabular}{cccc}
		1 & 0 & 0 & 0 \\
		0 & -1 & 0 & 0 \\
		0 & 0 & -1 & 0 \\
		0 & 0 & 0 & -1 
	\end{tabular}
	\right) \left(
	\begin{tabular}{cccc}
		$\frac{\hbar F(\vec{x},t)}{c^2}$ & 0 & 0 & 0 \\
		0 &$\frac{\hbar F(\vec{x},t)}{c^2}$ & 0 & 0 \\
		0 & 0 & $\frac{\hbar F(\vec{x},t)}{c^2}$ & 0 \\
		0 & 0 & 0 & $\frac{\hbar F(\vec{x},t)}{c^2}$ 
	\end{tabular}
	\right) = \frac{-2\hbar F(\vec{x},t)}{c^2}
\end{equation}

It can easily be seen that $G_{ij}$ for $i\neq j$ is equal to $0$.

\bigskip

We now calculate the diagonal components,

\begin{align}\label{A5}
	G_{00} &= -\frac{1}{2}\square \overline{g}_{00}^{[2]}
	= -\frac{\hbar}{c^2} \square F(\vec{x},t) 
	= \Bigg[ -\frac{\hbar\partial_{t}^{2}F(\vec{x},t)}{c^4} + \frac{\hbar\nabla^2 F(\vec{x},t)}{c^2}\Bigg] \\
	G_{\alpha\alpha} &= 0; ~~~~~because~ \overline{g}_{\alpha\alpha}^{[2]} = 0; ~~~~~\alpha \in(1,2,3) 
\end{align}

\section{generic components of $T_{\mu\nu}$}\label{NR:T}
\def\baselinestretch{1}\selectfont
\begin{landscape}
	\begin{equation}\label{62}
		T_{ij} = \frac{i\hbar c}{4}\left(
		\centering
		\resizebox{1.3\textwidth}{!}{%
			\begin{tabular}{cccc}
				$\begin{split}2\bar{\psi}\gamma_0(\partial_0\psi\\ + \frac{1}{4}[\gamma_{00i}\gamma^0\gamma^i + \gamma_{0i0}\gamma^i\gamma^0]\psi)\\-(\partial_0\bar{\psi} + \frac{1}{4}[\gamma_{00i}\gamma^0\gamma^i\\ + \gamma_{0i0}\gamma^i\gamma^0]\bar{\psi})2\gamma_0\psi\end{split}$ & $\begin{split}\bar{\psi}\gamma_0\partial_1\psi + \bar{\psi}\gamma_1(\partial_0\psi\\ + \frac{1}{4}[\gamma_{00i}\gamma^0\gamma^i + \gamma_{0i0}\gamma^i\gamma^0]\psi)\\ - \partial_1\bar{\psi}\gamma_0\psi - (\partial_0\bar{\psi} + \frac{1}{4}[\gamma_{00i}\gamma^0\gamma^i\\ + \gamma_{0i0}\gamma^i\gamma^0]\bar{\psi})\gamma_1\psi ) \end{split}$ & $\begin{split}\bar{\psi}\gamma_0\partial_2\psi + \bar{\psi}\gamma_2(\partial_0\psi\\ + \frac{1}{4}[\gamma_{00i}\gamma^0\gamma^i + \gamma_{0i0}\gamma^i\gamma^0]\psi)\\ - \partial_2\bar{\psi}\gamma_0\psi - (\partial_0\bar{\psi} + \frac{1}{4}[\gamma_{00i}\gamma^0\gamma^i\\ + \gamma_{0i0}\gamma^i\gamma^0]\bar{\psi})\gamma_2\psi ) \end{split}$ & $\begin{split}\bar{\psi}\gamma_0\partial_3\psi + \bar{\psi}\gamma_3(\partial_0\psi\\ + \frac{1}{4}[\gamma_{00i}\gamma^0\gamma^i + \gamma_{0i0}\gamma^i\gamma^0]\psi)\\ - \partial_3\bar{\psi}\gamma_0\psi - (\partial_0\bar{\psi} + \frac{1}{4}[\gamma_{00i}\gamma^0\gamma^i\\ + \gamma_{0i0}\gamma^i\gamma^0]\bar{\psi})\gamma_3\psi ) \end{split}$ \\ \\
				$\begin{split}\bar{\psi}\gamma_1(\partial_0\psi\\ + \frac{1}{4}[\gamma_{00i}\gamma^0\gamma^i + \gamma_{0i0}\gamma^i\gamma^0]\psi)\\
				+ \bar{\psi}\gamma_0\partial_1\psi - (\partial_0\bar{\psi} + \frac{1}{4}[\gamma_{00i}\gamma^{0}\gamma^{i}\\ + \gamma_{0i0}\gamma^{i}\gamma^{0}]\bar{\psi})\gamma_1\psi - \partial_1\bar{\psi}\gamma_0\psi\end{split}$ & $2(\bar{\psi}\gamma_1\partial_1\psi - \partial_1\bar{\psi\gamma_1\psi})$ & $\begin{split}\bar{\psi}\gamma_1\partial_2\psi + \bar{\psi}\gamma_2\partial_1\psi\\ - \partial_2\bar{\psi}\gamma_1\psi - \partial_1\bar{\psi}\gamma_2\psi\end{split}$ & $\begin{split}\bar{\psi}\gamma_1\partial_3\psi + \bar{\psi}\gamma_3\partial_1\psi\\ - \partial_3\bar{\psi}\gamma_1\psi - \partial_1\bar{\psi}\gamma_3\psi\end{split}$ \\ \\
				$\begin{split}\bar{\psi}\gamma_2(\partial_0\psi\\ + \frac{1}{4}[\gamma_{00i}\gamma^0\gamma^i + \gamma_{0i0}\gamma^i\gamma^0]\psi)\\ + \bar{\psi}\gamma_0\partial_2\psi - (\partial_0\bar{\psi} + \frac{1}{4}[\gamma_{00i}\gamma^{0}\gamma^{i}\\ + \gamma_{0i0}\gamma^{i}\gamma^{0}]\bar{\psi})\gamma_2\psi - \partial_2\bar{\psi}\gamma_0\psi\end{split}$ & $\begin{split}\bar{\psi}\gamma_2\partial_1\psi + \bar{\psi}\gamma_1\partial_2\psi\\ - \partial_1\bar{\psi}\gamma_2\psi - \partial_2\bar{\psi}\gamma_1\psi\end{split}$ & $2(\bar{\psi}\gamma_2\partial_2\psi - \partial_2\bar{\psi\gamma_2\psi})$ &  $\begin{split}\bar{\psi}\gamma_2\partial_3\psi + \bar{\psi}\gamma_3\partial_2\psi\\ - \partial_3\bar{\psi}\gamma_2\psi - \partial_2\bar{\psi}\gamma_3\psi\end{split}$ \\ \\
				$\begin{split}\bar{\psi}\gamma_3(\partial_0\psi\\ + \frac{1}{4}[\gamma_{00i}\gamma^0\gamma^i + \gamma_{0i0}\gamma^i\gamma^0]\psi)\\ + \bar{\psi}\gamma_0\partial_3\psi - (\partial_0\bar{\psi} + \frac{1}{4}[\gamma_{00i}\gamma^{0}\gamma^{i}\\ + \gamma_{0i0}\gamma^{i}\gamma^{0}]\bar{\psi})\gamma_3\psi - \partial_3\bar{\psi}\gamma_0\psi\end{split} $ & $\begin{split}\bar{\psi}\gamma_3\partial_1\psi + \bar{\psi}\gamma_1\partial_3\psi\\ - \partial_1\bar{\psi}\gamma_3\psi - \partial_3\bar{\psi}\gamma_1\psi\end{split} $ & $\begin{split}\bar{\psi}\gamma_3\partial_2\psi + \bar{\psi}\gamma_2\partial_3\psi\\ - \partial_2\bar{\psi}\gamma_3\psi - \partial_3 \bar{\psi} \gamma_2\psi\end{split}$ & $2(\bar{\psi}\gamma_3\partial_3\psi - \partial_3\bar{\psi\gamma_3\psi})$ \\ \\
			\end{tabular}%
		} 
		\right)
	\end{equation}\\
\end{landscape}

\def\baselinestretch{1.3}\selectfont
\chapter{Tetrad formalism/ NP formalism and formulating ECD equations in NP formalism}
\label{app:ECDinNP}

\section{Tetrad formalism and formulating Covariant derivative for Spinors}\label{Tetrad_formalism_CD_for_spinors}
The usual method in approaching the solution to the problems in General Relativity was to use a \textbf{local coordinate basis} $\hat{e}^{\mu}$ such that $\hat{e}^{\mu} = {\partial_{\mu}}$. This coordinate basis field is covariant under General coordinate transformation. However, it has been found useful to employ non-coordinate basis techniques in problems involving Spinors. This is the tetrad formalism which consists of setting up four linearly independent basis vectors called a `tetrad basis' at each point of a region of spacetime; which are covariant under local Lorentz transformations. [One of the reason of using tetrad formalism for spinors is essentially this fact that transformation properties of spinors can be easily defined in flat space-time]. The tetrad basis is given by  ${\hat{e}^{(i)} (x)}$. These are 4 vectors (one for each $\mu$) et every point. This tetrad field is governed by a relation ${\hat{e}^i (x)} = e^i_{\mu} (x) \hat{e}^{\mu}$ where trasformation matrix $e^i_{\mu}$ is such that,

\begin{equation}\label{eq:tetrad-metric transformation2}
	e^{(i)}_{\mu} e^{(k)}_{\nu} \eta_{{(i)}{(k)}} = g_{\mu\nu}; 
\end{equation}
Any `object' now can be expressed  in coordinate or tetrad basis as follows:
\begin{align}
	V &= V^{(a)}\hat{e}_{(a)} ------- Tetrad~~ basis \\
	V &= V^{\mu} \partial_{\mu} --------Coordinate~~ basis
\end{align}

Trasformation matrix $e^{(i)}_{\mu}$ allows us to convert the components of any world tensor (tensor which transforms according to general coordinate transformation) to the corresponding components in local Minkowskian space (These latter components being covariant under local Lorentz transformation). [Ex. $T_{\mu\nu} =e^{(i)}_{\mu} e^{(k)}_{\nu} T_{(i)(k)}] $. Greek indices are raised or lowered using the metric $g_{\mu\nu}$, while the Latin indices are raised or lowered using $\eta_{(i)(k)}$. parenthesis around indices is just a matter of convention already defined. A

\begin{align}
	A_{(a),(b)} &= e^{\mu}_{(b)}\frac{\partial}{\partial x^{\mu}}A_{(a)} = e^{\mu}_{(b)}\frac{\partial}{\partial x^{\mu}} [e^{\nu}_{(a)}A_\nu] \\
	&= e^{\mu}_{(b)}[A_\nu\partial_{\mu} e^{\nu}_{(a)} + e^{\nu}_{(a)}\partial_{\mu}A_\nu]\\
	&= e^{\mu}_{(b)}[A^{\rho}\nabla_{\mu} e_{(a)\rho} + e^{\nu}_{(a)}\nabla_{\mu}A_\nu - \Gamma^{\nu}_{\mu\rho}\not e^{\rho}_{(a)} A_{\nu} + \Gamma^{\nu}_{\mu\rho}\not e^{\rho}_{(a)} A_{\nu}] \\
\end{align}
From this, we get the expression for Covariant derivative of object with tetrad index

\begin{align}
	\nabla_{(b)}A_{(a)} &= \partial_{(b)}A_{(a)} - e^{\rho}_{(c)}\nabla_{\mu} e_{(a)\rho}  e^{\mu}_{(b)}A^{(c)}\\
	&= \partial_{(b)}A_{(a)} - \gamma_{(c)(a)(b)}A^{(c)}
\end{align}
where $\gamma_{(c)(a)(b)}$ are called \textbf{Ricci rotation coefficients} which are anti-symmetric in first pair of indices and are defined as
\begin{align}
	\gamma_{(c) (a) (b)} &= e^{\mu}_{(c)} \nabla_{\nu} e_{(a)\mu}e^{\nu}_{(b)} \\ 
	&= e^{\nu}_{(b)} e^{\mu}_{(c)} \bigg{[}\delta^{\alpha}_{\mu} \partial_{\nu} - \bfrac{\alpha}{\mu\nu} + K_{\nu\mu}^{\:\:\:\:\:\:\alpha} \bigg{]}e_{(a)\alpha}\\
	&= \gamma^o_{(c) (a) (b)} - K_{(b) (a) (c)}
\end{align}

\section{Natural connection  between $SL(2,\mathbb{C})$ Spinor formalism and NP formalism}\label{app:NP-SL2C}

4-vector on a Minkowski space can be represented by a hermitian matrix by some transformation law. Unimodular transformations on complex 2-Dim space induces a Lorentz transformation in Minkowski space. Unimodular matrices form a group under multiplication and is  denoted by$SL(2,\mathbb{C})$ - special linear group of 2 x 2 matrices over complex numbers. By a simple counting argument, it has six free real parameters corresponding to those of the Lorentz group. For a Lorentz transformation acting on Minkowski space, there are strictly speaking two transformations $\pm L \in SL(2,\mathbb{C})$. But this sign ambiguity may be resolved by choosing a path connected to the identity transformation. The levi-civita symbol $\epsilon_{AB'}$ acts as metric tensor in this space $\mathbb{C}^2$ which preserves the scalar product under Unimodular transformations. Spinor $P^{A}$ of rank 1 is defined as vector in complex 2-Dim space subject to transformations $\in SL(2, \mathbb{C})$. Similarly higher rank spinor are defined. There are various prescriptions wherein we associate objects in 4-dim Minkowski space with those in 2-dim complex space $\mathbb{C}^2$. The Van der Waarden symbols $\sigma$'s (for whose representation, we have chosen pauli matrices; such a representation is NOT unique) are used to associate tensorial objects with spinorial objects. Few examples are:

\begin{align}
	v^{\mu} &= \sigma^{\mu}_{AA'} V^{AA'}\\
	v^{\mu\nu} &= \sigma^{\mu}_{AA'} \sigma^{\nu}_{BB'} V^{AA'BB'}
\end{align}
(higher rank associations can be defined similarly). Now, analogous to a tetrad in Minkowski space, here we have a spin dyad (a pair of 2 spinors $\zeta_{(0)A}$ and $\zeta_{(1)A}$) such that $\zeta_{(0)A}\zeta_{(1)}^A =1$. A natural connection between dyad formalism and Null tetrad formalism (NP formalism) is evident by observing following association. More details can be looked up in \cite{Chandru}, \cite{SVD_geometry_fields_cosmology}

\begin{align}
	l^{\mu} &= \zeta_{(0)}^A \bar{\zeta}_{(0)}^{A'} ~~~~~~~~~ n^{\mu} = \zeta_{(1)}^A \bar{\zeta}_{(1)}^{A'}\\
	m^{\mu} &= \zeta_{(0)}^A \bar{\zeta}_{(1)}^{A'} ~~~~~~~~~ \bar{m}^{\mu} = \zeta_{(1)}^A \bar{\zeta}_{(0)}^{A'}
\end{align}

\section{Computation of Contorsion spin coefficients}\label{app:Contorsion_spin interms of spinor}

We first define the product $\gamma^{\alpha}\gamma^{\beta}\gamma^{\mu}$.

\begin{equation}
	\gamma^{\alpha}\gamma^{\beta}\gamma^{\mu} =\begin{pmatrix} 0 & (\tilde{\sigma}^{\alpha})^*(\sigma^{\beta})^*(\tilde{\sigma}^{\mu})^* \\ (\sigma^{\alpha})^*(\tilde{\sigma}^{\beta})^*(\sigma^{\mu})^* & 0 \end{pmatrix}
\end{equation}
This can be expanded fully in terms of vander-warden symbols and finally it takes the form 
\begin{equation}
	\gamma^{\alpha}\gamma^{\beta}\gamma^{\mu} =2\sqrt{2}
	\begin{pmatrix}
		0 & 0 & \Bigg[\begin{split}nln -n\bar{m}m \\ -\bar{m}mn + \bar{m}nm \end{split}\Bigg] & \Bigg[\begin{split}-nl\bar{m}+n\bar{m}l \\ +\bar{m}m\bar{m}-\bar{m}nl \end{split}\Bigg] \\
		0 & 0 & \Bigg[\begin{split} -mln + m\bar{m}m\\ +lmn-lnm \end{split}\Bigg] & \Bigg[\begin{split}ml\bar{m}-m\bar{m}l \\-lm\bar{m}+lnl\end{split}\Bigg]\\
		\Bigg[\begin{split}lnl-l\bar{m}m-\\\bar{m}ml +\bar{m}lm\end{split}\Bigg] & \Bigg[\begin{split}ln\bar{m}-l\bar{m}n \\-\bar{m}m\bar{m}+\bar{m}ln\end{split}\Bigg] & 0 & 0\\
		\Bigg[\begin{split}mnl-m\bar{m}m \\ -nml+nlm\end{split}\Bigg] & \Bigg[\begin{split}mn\bar{m}-m\bar{m}n \\ -nm\bar{m}+nln\end{split}\Bigg]& 0 & 0\\
	\end{pmatrix}^{\alpha\beta\mu}
\end{equation}
We will show the explicit calculation for one Contorsion spin coefficient viz. $\rho_1$. It is given by 

\begin{equation}
	\rho_1 = -K_{(1)(3)(4)} = -l_{\mu}m_{\nu}\bar{m}_{\alpha} K^{\mu\nu\alpha} = -2i\pi l^2 [l_{\mu}m_{\nu}\bar{m}_{\alpha}] \bar{\psi}\gamma^{[\mu}\gamma^{\nu}\gamma^{\alpha]}\psi
\end{equation}
The only quantity which would give non-zero scalar product with $l_{\mu}m_{\nu}\bar{m}_{\alpha}$ is $n^{\mu}\bar{m}^{\nu}m^{\alpha}$ (This can occur in any order amongst 3 vectors because we have all the orders possible in the definition of $\gamma^{[\mu}\gamma^{\nu}\gamma^{\alpha]}$) and the product is $l_{\mu}m_{\nu}\bar{m}_{\alpha}n^{\mu}\bar{m}^{\nu}m^{\alpha}=1$. We can easily deduce that 

\begin{align}
	[l_{\mu}m_{\nu}\bar{m}_{\alpha}] \bar{\psi}\gamma^{[\mu}\gamma^{\nu}\gamma^{\alpha]}\psi &=  
	\frac{\sqrt{2}}{3}\bar{\psi} \Bigg[
	+ \begin{pmatrix}
		0&0&-1&0 \\ 0&0&0&0 \\ 0&0&0&0 \\0&0&0&0 
	\end{pmatrix}
	- \begin{pmatrix}
		0&0&1&0\\0&0&0&0 \\ 0&0&0&0 \\ 0&0&0&0
	\end{pmatrix}
	+ \begin{pmatrix}
		0&0&-1&0\\ 0&0&0&0 \\ 0&0&0&0 \\ 0&0&0&0
	\end{pmatrix} \nonumber \\
	&- \begin{pmatrix}
		0&0&0&0 \\ 0&0&0&0 \\ 0&0&0&0 \\ 0&0&-1&0
	\end{pmatrix} 
	+ \begin{pmatrix}
		0&0&0&0 \\ 0&0&0&0 \\ 0&0&0&0 \\ 0&1&0&0
	\end{pmatrix} 
	- \begin{pmatrix}
		0&0&0&0 \\ 0&0&0&0 \\ 0&0&0&0 \\ 0&-1&0&0
	\end{pmatrix}
	\Bigg] \nonumber \\
	&= \frac{\sqrt{2}}{3}\begin{pmatrix}
		Q_0 & Q_1 & \bar{P}^{0'} & \bar{P}^{1'}
	\end{pmatrix} \begin{pmatrix}
		0&0&-3&0 \\ 0&0&0&0 \\ 0&0&0&0 \\ 0&3&&0
	\end{pmatrix}\begin{pmatrix}
		P^0 \\ P^1  \\ \bar{Q}_{0'}  \\  \bar{Q}_{1'} 
	\end{pmatrix} \\
	&= \sqrt{2}[\bar{P}^{1'} P^1 - Q^1  \bar{Q}^{1'}] \\
	&= \sqrt{2}[F_2\bar{F}_2 - G_1\bar{G}_1]
\end{align}
This gives full expression for $\rho$
\begin{equation}
	\rho = -K_{(1)(3)(4)} = -2\sqrt{2}i\pi l^2 [F_2\bar{F}_2 - G_1\bar{G}_1]
\end{equation}

\section{Computation of Dynamical EM tensor on $M_4$ with torsion in NP formalism}

\begin{equation}
	\begin{split}
		\Sigma_{11}^{(NP)}(\{\}) = \frac{i\hbar c}{2 \sqrt{2}} \bigg{(} &\bar{G}_1(D+\Delta)G_1 + \bar{G}_2(D+\Delta)G_2 - (D+\Delta)\bar{G}_1G_1 - (D+\Delta)\bar{G}_2 G_2
		\\ & + \bar{F}_1(D+\Delta)F_1 + \bar{F}_2(D+\Delta)F_2 - (D+\Delta)\bar{F}_1 F_1 - (D+\Delta)\bar{F}_2F_2\bigg{)}
	\end{split}
\end{equation}

\begin{equation}
	\begin{split}
		\Sigma_{21}^{(NP)}(\{\}) = \frac{i\hbar c}{4 \sqrt{2}} \bigg{(}&
		\bar{F}_1(\delta+\delta^*) F_1 + \bar{F}_2 (\delta+\delta^*)F_2 + \bar{G}_1(\delta+\delta^*) G_1 + \bar{G}_2(\delta+\delta^*) G_2 \\ & -\bar{F}_2 (D+\Delta) F_1 - \bar{F}_1 (D+\Delta) F_2 + \bar{G}_2 (D+\Delta)G_1 + \bar{G}_1(D+\Delta) G_2 \\& -(\delta+\delta^*)\bar{F}_1 F_1 - (\delta+\delta^*)\bar{F}_2 F_2 - (\delta+\delta^*)\bar{G}_1 G_1 - (\delta+\delta^*)\bar{G}_2 G_2 
		\\ & + (D+\Delta)\bar{F}_2  F_1 + (D+\Delta)\bar{F}_1  F_2 - (D+\Delta)\bar{G}_2 G_1 - (D+\Delta)\bar{G}_1 G_2 \bigg{)}
	\end{split}
\end{equation}

\begin{equation}
	\begin{split}
		\Sigma_{31}^{(NP)}(\{\}) = \frac{i\hbar c}{4 \sqrt{2}} \bigg{(}& i\bar{F}_1(\delta-\delta^*) F_1 + i\bar{F}_2 (\delta-\delta^*)F_2 + i\bar{G}_1(\delta-\delta^*) G_1 + i\bar{G}_2(\delta-\delta^*) G_2 \\& i\bar{F}_2(D+\Delta) F_1 - i \bar{F}_1(D+\Delta) F_2 - i\bar{G}_2 (D+\Delta)G_1 + i \bar{G}_1(D+\Delta) G_2 \\& -i(\delta-\delta^*)\bar{F}_1 F_1 -i(\delta-\delta^*) \bar{F}_2 F_2 - iG_1(\delta-\delta^*) \bar{G}_1 -(\delta-\delta^*) i\bar{G}_2 G_2 \\& -i(D+\Delta)\bar{F}_2 F_1 + i(D+\Delta) \bar{F}_1 F_2 +(D+\Delta) i\bar{G}_2 G_1 -(D+\Delta) i \bar{G}_1 G_2  \bigg{)}\\
	\end{split}
\end{equation}

\begin{equation}
	\begin{split}
		\Sigma_{41}^{(NP)}(\{\}) = \frac{i\hbar c}{4 \sqrt{2}} \bigg{(}& \bar{F}_1(D-\Delta) F_1 + \bar{F}_2 (D-\Delta)F_2 + \bar{G}_1(D-\Delta) G_1 + \bar{G}_2(D-\Delta) G_2 \\& -\bar{F}_1 (D+\Delta) F_1 + \bar{F}_2(D+\Delta) F_2 + \bar{G}_1 (D+\Delta)G_1 - \bar{G}_2(D+\Delta) G_2 \\& -(D-\Delta)\bar{F}_1 F_1 - (D-\Delta)\bar{F}_2 F_2 -(D-\Delta) \bar{G}_1 G_1 - (D-\Delta)\bar{G}_2 G_2 \\& +(D+\Delta)\bar{F}_1 F_1 -(D+\Delta) \bar{F}_2 F_2 -(D+\Delta) \bar{G}_1 G_1 +(D+\Delta) \bar{G}_2 G_2 \bigg{)}\\
	\end{split}
\end{equation}

\begin{equation}
	\begin{split}
		\Sigma_{22}^{(NP)}(\{\}) = \frac{i\hbar c}{2 \sqrt{2}} \bigg{(}&-\bar{F}_2(\delta+\delta^*) F_1 - \bar{F}_1(\delta+\delta^*) F_2 + \bar{G}_2(\delta+\delta^*) G_1 + \bar{G}_1(\delta+\delta^*) G_2 \\ & +(\delta+\delta^*)\bar{F}_2F_1 +(\delta+\delta^*) \bar{F}_1 F_2 - (\delta+\delta^*)\bar{G}_2 G_1 - (\delta+\delta^*)\bar{G}_1 G_2 \bigg{)}\\
	\end{split}
\end{equation}

\begin{equation}
	\begin{split}
		\Sigma_{32}^{(NP)}(\{\}) = \frac{i\hbar c}{4 \sqrt{2}} \bigg{(}& i\bar{F}_2 (\delta+\delta^*)F_1 - i \bar{F}_1(\delta+\delta^*) F_2 - i \bar{G}_2(\delta+\delta^*) G_1 + i \bar{G}_1(\delta+\delta^*) G_2 \\& -i\bar{F}_2 (\delta-\delta^*)F_1 -i \bar{F}_1(\delta-\delta^*) F_2 + i\bar{G}_2 (\delta-\delta^*)G_1 + i\bar{G}_1(\delta-\delta^*) G_2 \bigg{)} \\ & -i(\delta+\delta^*)\bar{F}_2 F_1 + (\delta+\delta^*) i \bar{F}_1 F_2 + (\delta+\delta^*) i \bar{G}_2 G_1 -(\delta+\delta^*) i \bar{G}_1 G_2 \\& + (\delta-\delta^*)i\bar{F}_2F_1 + (\delta-\delta^*)i \bar{F}_1 F_2 -(\delta-\delta^*) i\bar{G}_2 G_1 -(\delta-\delta^*) i\bar{G}_1 G_2 \bigg{)}  
	\end{split}
\end{equation}

\begin{equation}
	\begin{split}
		\Sigma_{42}^{(NP)}(\{\}) = \frac{i\hbar c}{4 \sqrt{2}} \bigg{(}& -\bar{F}_1 (\delta+\delta^*) F_1 + \bar{F}_2(\delta+\delta^*) F_2 + \bar{G}_1(\delta+\delta^*) G_1 - \bar{G}_2 (\delta+\delta^*)G_2\\&  -\bar{F}_2(D-\Delta) F_1 -\bar{F}_1 (D-\Delta)F_2 +\bar{G}_2(D-\Delta) G_1 + \bar{G}_1(D-\Delta) G_2 \\& +(\delta+\delta^*)\bar{F}_1F_1 -(\delta+\delta^*) \bar{F}_2F_2 -(\delta+\delta^*) \bar{G}_1G_1 +(\delta+\delta^*) \bar{G}_2 G_2\\&  +(D-\Delta) \bar{F}_2F_1 + (D-\Delta) \bar{F}_1 F_2 -(D-\Delta) \bar{G}_2 G_1 -(D-\Delta)\bar{G}_1G_2	\bigg{)}
	\end{split}
\end{equation}

\begin{equation}
	\begin{split}
		\Sigma_{33}^{(NP)}(\{\}) = \frac{i\hbar c}{2 \sqrt{2}} \bigg{(}& - \bar{F}_2(\delta-\delta^*) F_1 + \bar{F}_1 (\delta-\delta^*)F_2 + \bar{G}_2(\delta-\delta^*) G_1 -  \bar{G}_1 (\delta-\delta^*)G_2	\bigg{)}\\& 
		+ (\delta-\delta^*)\bar{F}_2 F_1 -(\delta-\delta^*) \bar{F}_1 F_2 -(\delta-\delta^*) \bar{G}_2 G_1 +(\delta-\delta^*)  \bar{G}_1 G_2	\bigg{)}
	\end{split}
\end{equation}

\begin{equation}
	\begin{split}
		\Sigma_{43}^{(NP)}(\{\}) = \frac{i\hbar c}{4 \sqrt{2}} \bigg{(}& -i\bar{F}_1 (\delta-\delta^*) F_1 + i\bar{F}_2 (\delta-\delta^*)F_2 + i\bar{G}_1 (\delta-\delta^*)G_1 -i \bar{G}_2(\delta-\delta^*) G_2 \\& + i \bar{F}_2(D-\Delta) F_1 - i\bar{F}_1(D-\Delta) F_2 - i \bar{G}_2(D-\Delta) G_1 + i \bar{G}_1(D-\Delta) G_2  \\&  +i(\delta-\delta^*)\bar{F}_1 F_1 -i(\delta-\delta^*)\bar{F}_2 F_2 -i (\delta-\delta^*)\bar{G}_1 G_1 +i(\delta-\delta^*) \bar{G}_2 G_2 \\& -i(D-\Delta)\bar{F}_2 F_1 + i(D-\Delta)\bar{F}_1F_2 + i(D-\Delta) \bar{G}_2G_1 - i(D-\Delta)\bar{G}_1 G_2  \bigg{)}\\
	\end{split}
\end{equation}

\begin{equation}
	\begin{split}
		\Sigma_{44}^{(NP)}(\{\}) = \frac{i\hbar c}{2 \sqrt{2}} \bigg{(}& -\bar{F}_1(D-\Delta) F_1 + \bar{F}_2 (D-\Delta)F_2 + \bar{G}_1 (D-\Delta)G_1 - \bar{G}_2 (D-\Delta)G_2 \\& +(D-\Delta)\bar{F}_1 F_1 - (D-\Delta)\bar{F}_2F_2 -(D-\Delta) \bar{G}_1 G_1 +(D-\Delta) \bar{G}_2 G_2\bigg{)} \\
	\end{split}
\end{equation}

\end{document}